\documentclass[12pt]{article}  %fuer beidseitige Version
\usepackage{graphicx}
\usepackage{a41} %
\usepackage{epsfig} %
\usepackage{float} %
\usepackage{floatflt} %
\usepackage{amssymb} %
\usepackage{amsmath} %
\usepackage{verbatim} % 
\usepackage{cite} %
\usepackage{color} %
\allowdisplaybreaks[4]
\usepackage{eufrak}

\newcommand{\N}{\nonumber}
\newcommand{\ep}{\varepsilon}

\newcommand{\Ctil}{\tilde{C}}

\newcommand{\beq}{\begin{equation}}
\newcommand{\eeq}{\end{equation}}
\newcommand{\bea}{\begin{eqnarray}}
\newcommand{\eea}{\end{eqnarray}}

\newcommand{\GeV}{${\rm GeV}$}

\newcounter{lin}

\begin{document}
\begin{titlepage}

\begin{flushleft}
DESY 21--008   \hfill   %{\tt arXiv:xxxx.xxxx [hep-ph]} 
\\
DO--TH 21/03
\\
TTP 21--008
\\
RISC Report Series 21-06
\\
SAGEX 21--02--E
\\
May 2021 
\\
\end{flushleft}

\vspace{1cm}
\noindent
\begin{center}
{\LARGE\bf The Logarithmic Contributions to the Polarized}

\vspace*{2mm} 
{\LARGE\bf
\boldmath{$O(\alpha_s^3)$} Asymptotic Massive Wilson 
Coefficients}

\vspace*{2mm}
{\LARGE\bf 
and Operator Matrix Elements in} 

\vspace*{2mm}
{\LARGE\bf Deeply
Inelastic Scattering} \\
\end{center}
\begin{center}

\vspace{2cm}

{\large J. Bl\"umlein$^a$, A. De Freitas$^a$, M.~Saragnese$^a$, 
C.~Schneider$^c$} 

\vspace*{1mm}
{\large and 
K.~Sch\"onwald$^{a,b}$},
\vspace*{2mm}

{\it $^a$~Deutsches Elektronen Synchrotron, DESY,\\
Platanenallee 6, D--15738 Zeuthen, Germany}\\
\vspace*{2mm}

{\it $^b$~Institut f\"ur Theoretische Teilchenphysik Campus S\"ud,\\
Karlsruher Institut für Technologie (KIT) D-76128 Karlsruhe, Germany}\\
\vspace*{2mm}

{\it $^c$~
Johannes Kepler University Linz, Research Institute for Symbolic Computation (RISC),
Altenberger Stra{\ss}e 69, A--4040, Linz, Austria}

%%\today
\vspace{3cm}
\end{center}

\begin{abstract}
\noindent
We compute the logarithmic contributions to the polarized massive Wilson coefficients for deep-inelastic 
scattering in the asymptotic region $Q^2 \gg m^2$ to 3-loop order in the fixed-flavor number scheme and 
present the corresponding expressions for the polarized massive operator matrix elements needed in the 
variable flavor number scheme. The calculation is performed in the Larin scheme. For the massive operator 
matrix elements $A_{qq,Q}^{(3),\rm PS}$ and $A_{qg,Q}^{(3),\rm S}$ the complete results are presented.
The expressions are given in Mellin-$N$ space and in momentum fraction $z$-space. 
\end{abstract}
\end{titlepage}

\newpage
\sloppy

%-------------------------------------------------------------------------------------------------
\section{Introduction}
\label{sec:intro}
%-------------------------------------------------------------------------------------------------

\vspace{1mm}
\noindent
At leading twist the scaling violations of deeply--inelastic structure functions obtain corrections 
from the massless scale evolution of the parton densities, the massive Wilson coefficients, and target--mass
corrections, cf. \cite{Blumlein:2012bf}, provided that  the electroweak radiative corrections are properly removed 
\cite{Kwiatkowski:1990es,Arbuzov:1995id,Bardin:1996ch}. At low scales of the virtuality $Q^2 = - q^2$, all kind of 
other dynamical contributions, such as dynamical higher twist effects, particular small-$x$ effects, with $x$ the 
Bjorken variable, and particular hadronic contributions due to vector meson dominance etc. are present. For these 
reasons one usually applies respective cuts of the kind $Q^2 > 10~\GeV^2, W^2 = Q^2 (1-x)/x > 15~\GeV^2$ 
\cite{Alekhin:2012ig} to eliminate these effects in order to obtain clean twist--2 data for which a dedicated QCD analysis 
is going to be performed. The target mass effects can be accounted for analytically 
\cite{Blumlein:2012bf,Blumlein:1998nv,Piccione:1997zh,TM2}. Furthermore, for a precision measurement of the strong 
coupling constant $\alpha_s(M_Z^2) = g_s^2/(4\pi)$, one should use targets with no or only soft nuclear binding, 
such as the case for $p$ and $d$ targets. In the latter case still nuclear corrections have to be performed.

Under these circumstances the largest source of scaling violations next to the massless higher order corrections
are implied by the heavy quark corrections, in the form of single and two--mass corrections. Both in the unpolarized 
and polarized cases the scaling violations due to the heavy flavor Wilson coefficients turn out to be different 
from those of the massless corrections in very wide kinematic regions covered by the present experiments. Therefore, 
one can not model these effects by just adding more quasi--massless flavors in  massless higher order corrections.
In the future analysis of the polarized structure functions\footnote{For a review on the theory and phenomenology 
of polarized deep--inelastic structure functions see Ref.~\cite{Lampe:1998eu}.} at the EIC operating at high 
luminosity \cite{Boer:2011fh} these corrections are important for precision measurements of the strong coupling 
constant \cite{AS} and the charm quark mass, cf.~\cite{Alekhin:2012vu}.

In the polarized case, the leading order corrections were derived in \cite{Watson:1981ce,Vogelsang:1990ug}, the 
next-to-leading order (NLO) asymptotic corrections in \cite{Buza:1996xr,Bierenbaum:2007pn,POLl,PVFNS}, the complete 
NLO corrections in analytic form for the non-singlet and pure singlet corrections in 
\cite{Buza:1996xr,Blumlein:2016xcy,Blumlein:2019zux} and numerically in \cite{Hekhorn:2018ywm}. At 3--loop order, 
the massive operator matrix elements (OMEs) for the non-singlet $A_{qq,Q}^{(3),\rm NS}$, pure singlet 
$A_{Qq}^{(3),\rm PS}$, and the OME $A_{gq,Q}^{(3)}$ have been calculated in 
\cite{Ablinger:2014vwa,Ablinger:2019etw,Behring:2021asx}. Furthermore, two-mass corrections to  
different OMEs at 3--loop order were calculated in 
\cite{Ablinger:2017err,Ablinger:2020snj,Ablinger:2019gpu,Behring:2021asx}. The 3--loop heavy flavor non--singlet 
contributions at leading twist to the structure functions $g_1(x,Q^2)$ and $g_2(x,Q^2)$ were computed in 
\cite{Behring:2015zaa}. Likewise, in the unpolarized case the leading order corrections were obtained in 
\cite{Witten:1975bh,Babcock:1977fi,Shifman:1977yb,Leveille:1978px,Gluck:1979aw,Gluck:1980cp}
at NLO in numerical form in \cite{NLO} and in analytic form in the 
non-singlet and pure singlet cases in \cite{Buza:1995ie,Blumlein:2016xcy,Blumlein:2019qze} 
and in the asymptotic case in 
\cite{Buza:1995ie,Buza:1996wv,Bierenbaum:2007qe,Bierenbaum:2008yu,Bierenbaum:2009zt,Blumlein:2016xcy}.
At 3--loop order a series of moments has been computed in \cite{Bierenbaum:2009mv} for all massive OMEs. 
A series of 3--loop massive OMEs has been calculated in Refs.~\cite{Ablinger:2014vwa,Ablinger:2014nga,
Ablinger:2014uka,Blumlein:2012vq,AGG3,Ablinger:2014lka} and first analytic results for $A_{Qg}^{(3)}$ 
have been given in \cite{Ablinger:2017ptf}. Two-mass 3--loop corrections were calculated in 
Refs.~\cite{Ablinger:2017err,Ablinger:2017xml,Ablinger:2018brx}.

In the present paper we calculate the 3--loop logarithmic corrections to the polarized massive OMEs $A_{ij}$ both 
in Mellin $N$ and momentum fraction $z$-space in the single mass case. 
These OMEs appear as building blocks of the variable flavor number scheme (VFNS) at 3--loop oder,
through which the matching of parton distribution functions at $N_F$  and $N_F +1$ massless heavy flavors
can be obtained \cite{Buza:1996wv}. Because of the close values of the heavy quark masses $m_c$ and $m_b$ 
two-mass decoupling has been considered and implemented in the VFNS in  \cite{Ablinger:2017err,Blumlein:2018jfm, 
PVFNS}. In the present paper we will consider the heavy quark masses in the on-shell scheme.\footnote{The transformation
to the $\overline{\sf MS}$-scheme is straightforward.}
The OMEs determine also the polarized massive Wilson coefficients in the region of large virtualities 
$Q^2 \gg m_Q^2$, with $m_Q$ the corresponding heavy quark mass $m_Q = m_c (m_b)$ of charm and bottom quarks, which we derive 
for the structure function 
$g_1(x,Q^2)$. It turns out that all functions can either be expressed in terms of harmonic sums 
\cite{Vermaseren:1998uu,Blumlein:1998if} or harmonic polylogarithms \cite{Remiddi:1999ew}. The analytic continuation 
to $z$-space in Mellin-space evolution have been studied in \cite{ANCONT,Blumlein:2009ta} in the case of harmonic sums.
The expressions for the different OMEs and massive Wilson coefficients turn out to be rather long both in $N$ and $z$ 
space. As they are at present, they can be used to study the corrections at the logarithmic level and the expressions
provide the frame for the final results.

The paper is organized as follows. In Section~\ref{sec:2}, we summarize the basic formalism. It widely 
follows the representation in the unpolarized case \cite{Behring:2014eya} and resumes the representations of 
the structure functions and the relations for the single mass variable flavor number scheme. We then present 
the complete polarized massive operator matrix elements (OMEs) $A_{qq,Q}^{(3),\rm PS}$ and $A_{qg,Q}^{(3),\rm S}$ 
beyond the logarithmic terms 
in Section~\ref{sec:3}. In Section~\ref{sec:4} the polarized OME $A_{Qg}^{(3),\rm S}$ 
is given in $N$--space, followed by the expressions for the OMEs $A_{gg,Q}^{(3),\rm S}$ in Section~\ref{sec:5}.
The OMEs $A_{qq,Q}^{(3),\rm NS}, A_{Qq}^{(3),\rm PS}$ and $A_{gq,Q}^{(3),\rm S}$ have already been published 
before in Refs.~\cite{Ablinger:2014vwa,Ablinger:2019etw,Behring:2021asx} in complete form. In Sections~\ref{sec:6}, 
\ref{sec:7} and \ref{sec:8} we present the polarized 3--loop Wilson coefficients $L_q^{\rm PS}, L_g^{\rm S},
H_{Qq}^{\rm PS}$ and $H_{Qg}^{\rm S}$.
The logarithmic contributions can be represented in terms of harmonic sums and a large number of polynomials.
Section~\ref{sec:9} contains the conclusions. In Appendix~\ref{sec:A} the 
corresponding expressions for the OMEs in momentum fraction $z$-space and in Appendix~\ref{sec:B} the 
corresponding expressions 
for the polarized Wilson coefficients are presented. Here the yet unknown constant parts of the polarized massive 
OMEs $a_{Qg}^{(3)}$ and $a_{gg,Q}^{(3)}$ and the yet missing 3--loop polarized massless Wilson coefficients are 
left as symbols. In Refs.~\cite{Ablinger:2014vwa,Behring:2015zaa} the flavor non--singlet contributions were presented in the 
$\overline{\sf MS}$ scheme to 3--loop order. For a consistent treatment, we present the transformation for the massive OME
$A_{qq,Q}^{\rm NS}$ and the asymptotic massive Wilson coefficient $L_{q}^{NS}$ in the Larin scheme in Appendix~\ref{sec:C}.
%-----------------------------------------------------------------------------------------------------------------
\section{The Formalism} 
\label{sec:2}
%-----------------------------------------------------------------------------------------------------------------

\vspace*{1mm}\noindent
The explicit expressions showing the principal structure of the different massive OMEs have been derived in 
Refs.~\cite{Bierenbaum:2009mv,Behring:2014eya}, as well as for the asymptotic massive Wilson coefficients. 
The different polarized massive OMEs obey the following expansion in the strong coupling constant
%--------------------------------------------------------------------------------------------------
\begin{eqnarray}
A_{ij}^{(k)} = \delta_{ij} + \sum_{l=1}^\infty a_s^l A_{ij}^{(l),k},
\end{eqnarray}
%--------------------------------------------------------------------------------------------------
with $a_s = \alpha_s/(4\pi)$, the indices $i,j = q,g$ label the partonic channels, and $k$ denotes the different OMEs.
Here and in the following we use the shorthand notations \cite{Bierenbaum:2009mv}
%-----------------------------------------------------------------------------------------------------------------
\begin{eqnarray}
\hat{f}(N_F) &=& f(N_F+1) - f(N_F)
\\
\tilde{f}(N_F) &=& \frac{1}{N_F} f(N_F).
\end{eqnarray}
%-----------------------------------------------------------------------------------------------------------------
The heavy flavor Wilson coefficients, accounting for the single mass contributions to the structure function $g_1(x,Q^2)$, 
are given by 
%--------------------------------------------------------------------------------------------------
\begin{eqnarray}
     \label{eqWIL1}
     L_{q,(1)}^{\sf NS,Q}(N_F+1) &=& 
     a_s^2 \Bigl[A_{qq,Q}^{(2), {\sf NS}}(N_F+1) +
     \hat{C}^{(2), {\sf NS}}_{q,(1)}(N_F)\Bigr]
     \N\\
     &+&
     a_s^3 \Bigl[A_{qq,Q}^{(3), {\sf NS}}(N_F+1)
     +  A_{qq,Q}^{(2), {\sf NS}}(N_F+1) C_{q,(1)}^{(1), {\sf NS}}(N_F+1)
       \N \\
     && \hspace*{5mm}
     + \hat{C}^{(3), {\sf NS}}_{q,(1)}(N_F)\Bigr]~,  \\
      \label{eqWIL2}
      L_{q,(1)}^{\sf PS}(N_F+1) &=& 
     a_s^3 \Bigl[~A_{qq,Q}^{(3), {\sf PS}}(N_F+1)
     +  A_{gq,Q}^{(2)}(N_F+1)~~N_F\Ctil_{g,(1)}^{(1)}(N_F+1) \N \\
     && \hspace*{5mm}
     + N_F \hat{\Ctil}^{(3), {\sf PS}}_{q,(1)}(N_F)\Bigr]~,
     \\
     \label{eqWIL3}
      L_{g,(1)}^{\sf S}(N_F+1) &=& 
     a_s^2 A_{gg,Q}^{(1)}(N_F+1)N_F \Ctil_{g,(2,L)}^{(1)}(N_F+1)
     \N\\ &+&
      a_s^3 \Bigl[~A_{qg,Q}^{(3)}(N_F+1) 
     +  A_{gg,Q}^{(1)}(N_F+1)~~N_F\Ctil_{g,(2,L)}^{(2)}(N_F+1)
     \N\\ && \hspace*{5mm}
     +  A_{gg,Q}^{(2)}(N_F+1)~~N_F\Ctil_{g,(2,L)}^{(1)}(N_F+1)
     \N\\ && \hspace*{5mm}
     +  ~A^{(1)}_{Qg}(N_F+1)~~N_F\Ctil_{q,(2,L)}^{(2), {\sf PS}}(N_F+1)
     + N_F \hat{\Ctil}^{(3)}_{g,(2,L)}(N_F)\Bigr]~,
 \\ \N \\
\label{eq:WILPS}
     H_{q,(1)}^{\sf PS}(N_F+1)
     &=& a_s^2 \Bigl[~A_{Qq}^{(2), {\sf PS}}(N_F+1)
     +~\Ctil_{q,(2,L)}^{(2), {\sf PS}}(N_F+1)\Bigr]
     \\
     &+& a_s^3 \Bigl[~A_{Qq}^{(3), {\sf PS}}(N_F+1)
     +~\Ctil_{q,(2,L)}^{(3), {\sf PS}}(N_F+1) \N\\
 && %\hspace*{-20mm}
     + A_{gq,Q}^{(2)}(N_F+1)~\Ctil_{g,(2,L)}^{(1)}(N_F+1) 
     + A_{Qq}^{(2), {\sf PS}}(N_F+1)~C_{q,(2,L)}^{(1), {\sf NS}}(N_F+1) 
        \Bigr]~,       \label{eqWIL4}
         \N\\ 
\label{eq:WILS}
     H_{g,(1)}^{\sf S}(N_F+1) &=& a_s \Bigl[~A_{Qg}^{(1)}(N_F+1)
     +~\Ctil^{(1)}_{g,(2,L)}(N_F+1) \Bigr] \N\\
     &+& a_s^2 \Bigl[~A_{Qg}^{(2)}(N_F+1)
     +~A_{Qg}^{(1)}(N_F+1)~C^{(1), {\sf NS}}_{q,(2,L)}(N_F+1)\N\\ && 
     \hspace*{5mm}
     +~A_{gg,Q}^{(1)}(N_F+1)~\Ctil^{(1)}_{g,(2,L)}(N_F+1) 
     +~\Ctil^{(2)}_{g,(2,L)}(N_F+1) \Bigr]
     \N\\ &+&
     a_s^3 \Bigl[~A_{Qg}^{(3)}(N_F+1)
     +~A_{Qg}^{(2)}(N_F+1)~C^{(1), {\sf NS}}_{q,(2,L)}(N_F+1)
     \N\\ &&
     \hspace*{5mm}
     +~A_{gg,Q}^{(2)}(N_F+1)~\Ctil^{(1)}_{g,(2,L)}(N_F+1)
     \N\\ && \hspace*{5mm}
     +~A_{Qg}^{(1)}(N_F+1)\Bigl\{
     C^{(2), {\sf NS}}_{q,(2,L)}(N_F+1)
     +~\Ctil^{(2), {\sf PS}}_{q,(2,L)}(N_F+1)\Bigr\}
     \N\\ && \hspace*{5mm}
     +~A_{gg,Q}^{(1)}(N_F+1)~\Ctil^{(2)}_{g,(2,L)}(N_F+1)
     +~\Ctil^{(3)}_{g,(2,L)}(N_F+1) \Bigr]~, 
\label{eqWIL5}
\end{eqnarray}
%--------------------------------------------------------------------------------------------------
with $\beta_{0,Q} = - (4/3) \textcolor{blue}{T_F}$. The QCD color factors are given by $C_A = N_c, C_F = (N_c^2-1)/(2 
N_c), T_F = 1/2$ for $SU(N_c)$ and $N_c = 3$. $N_F$ denotes the number of massless flavors. 
Here the massive OMEs $A_{ij}$ depend on $m^2/\mu^2$ and the massless Wilson coefficients
depend on $Q^2/\mu^2$.
Note that we extended the original notion in Refs.~\cite{Buza:1996wv,Bierenbaum:2009mv}. 

The argument $(N_F+1)$ in the massive OMEs signals that these functions depend on $N_F$ massless and 
one massive flavor, while the setting of $N_F$ in the massless Wilson coefficients is a functional one.
The massless Wilson coefficients are labeled by $C_i^{(l),k}$, where $i$ refers to the parton species and $l$ to 
the expansion-order in the strong coupling constant. The massless Wilson coefficients are known to 2--loop order
\cite{vanNeerven:1991nn,Zijlstra:1993sh,Moch:1999eb,Vogt:2008yw,Blumlein:2019zux} and in the non--singlet case to 
3--loop order \cite{Vermaseren:2005qc}.

In the present paper we express all relations in the Larin scheme \cite{Larin:1993tq,Matiounine:1998re}, which is 
a consistent scheme also in the massive case. The anomalous dimensions were calculated to 2--loop order 
\cite{Mertig:1995ny,Vogelsang:1995vh} and to 3--loop order in \cite{Moch:2014sna,Behring:2019tus}.

The twist--2 contributions to the structure function $g_1(x,Q^2)$ in the single mass case for pure virtual photon 
exchange\footnote{For the structure of electroweak gauge boson exchange see e.g. \cite{Blumlein:1996vs,
Blumlein:1998nv}.} 
are given by 
%-------------------------------------------------------------------------------------------------- 
\begin{eqnarray} \frac{1}{x}g_1(x,Q^2) &=& 
      \sum_{k=1}^{N_F}e_k^2\Biggl\{ 
                   L_{q,(2,L)}^{\sf NS}\left(x,N_F+1,\frac{Q^2}{\mu^2}
                                                ,\frac{m^2}{\mu^2}\right)
                \otimes 
                   \Bigl[\Delta f_k(x,\mu^2,N_F)+\Delta f_{\overline{k}}(x,\mu^2,N_F)\Bigr]
\N\\ &&\hspace{14mm}
               +\frac{1}{N_F}L_{q,(2,L)}^{\sf PS}\left(x,N_F+1,\frac{Q^2}{\mu^2}
                                                ,\frac{m^2}{\mu^2}\right) 
                \otimes 
                   \Delta \Sigma(x,\mu^2,N_F)
\N\\ &&\hspace{14mm}
               +\frac{1}{N_F}L_{g,(2,L)}^{\sf S}\left(x,N_F+1,\frac{Q^2}{\mu^2}
                                                 ,\frac{m^2}{\mu^2}\right)
                \otimes 
                   \Delta G(x,\mu^2,N_F) 
                             \Biggr\}
\N\\
       &+&e_Q^2\Biggl[
                   H_{q,(2,L)}^{\sf PS}\left(x,N_F+1,\frac{Q^2}{\mu^2}
                                        ,\frac{m^2}{\mu^2}\right) 
                \otimes 
                   \Delta \Sigma(x,\mu^2,N_F)
\N\\ &&\hspace{7mm}
                  +H_{g,(2,L)}^{\sf S}\left(x,N_F+1,\frac{Q^2}{\mu^2}
                                           ,\frac{m^2}{\mu^2}\right)
                \otimes 
                   \Delta G(x,\mu^2,N_F)
                                  \Biggr]~.
\end{eqnarray}
%--------------------------------------------------------------------------------------------------
Here, the polarized quark and antiquark densities are denoted by $\Delta f_k$ and $\Delta f_{\bar{k}}$, 
$\Delta G$ denotes the polarized gluon density and the polarized singlet distribution is given by
%--------------------------------------------------------------------------------------------------
\begin{eqnarray}
\Delta \Sigma = \sum_{k=1}^{N_F} \left[\Delta f_k + \Delta f_{\bar{k}}\right].
\end{eqnarray}
%--------------------------------------------------------------------------------------------------
The structure function $g_2(x,Q^2)$ at twist--2 is obtained by the Wandzura--Wilczek relation 
\cite{Wandzura:1977qf,Blumlein:1996vs,Piccione:1997zh,Blumlein:1998nv,WW1}
%--------------------------------------------------------------------------------------------------
\begin{eqnarray}
g_2(x,Q^2) &=& - g_1(x,Q^2) + \int_x^1 \frac{dz}{z} g_1(z,Q^2). 
\end{eqnarray}
%--------------------------------------------------------------------------------------------------

Besides the Wilson coefficients (\ref{eqWIL1}--\ref{eqWIL5}) the massive OMEs determine the relations 
of the matching of parton distributions in the VFNS at large enough scales $\mu^2$,~\cite{Buza:1996wv,
Bierenbaum:2009mv}. Here, the PDFs for $N_F+1$ massless quarks are related to the former $N_F$ massless quarks 
process independently. 
The corresponding relations to 3--loop order read, cf. also \cite{Buza:1996wv}\footnote{Here, we 
have corrected some typographical errors in (\ref{eq:VFNS3}--\ref{eq:VFNS2}) in \cite{Buza:1996wv}, in 
accordance with the appendix of Ref.~\cite{Buza:1996wv}.}
%---------------------------------------------------------------------------------------
{\small
\begin{eqnarray}
\label{eq:VFNS1}
%-------------------------
\Delta f_k(N_F+1, \mu^2) + \Delta f_{\overline{k}}(N_F+1, \mu^2) 
&=& A_{qq,Q}^{\sf NS}\Big(N_F, \frac{\mu^2}{m^2}\Big)
\otimes \Delta \left[f_k(N_F, \mu^2) + \Delta f_{\overline{k}}(N_F, \mu^2)\right] 
\nonumber\\
&& 
+ {\tilde A_{qq,Q}^{\sf PS}\Big(N_F, \frac{\mu^2}{m^2}\Big)}\otimes
  {\Delta \Sigma(N_F, \mu^2)} 
\nonumber\\
&& + {\tilde A_{qg,Q}^{\sf  S}\Big(N_F, \frac{\mu^2}{m^2}\Big)}\otimes
  {\Delta G(N_F, \mu^2)}
\\
\label{eq:VFNS3}
%-----------------------------
  {\Delta f_{Q+\bar Q}(N_F+1, \mu^2)}
   &=& {A_{Qq}^{\sf PS}\Big(N_F, \frac{\mu^2}{m^2}\Big)}\otimes
%   \underbrace{
       {\Delta \Sigma(N_F, \mu^2)}
%\nonumber\\ &&
   + {A_{Qg}^{\sf S}\Big(N_F, \frac{\mu^2}{m^2}\Big)}
%   \underbrace{
\nonumber\\
&&
\otimes
    {\Delta G(N_F, \mu^2)}
%}_{\mbox{Gluon density}}
\\
%-------------------------
 {\Delta G(N_F+1, \mu^2)} &=&  {A_{gq,Q}^{\sf S}\Bigl(N_F,\frac{\mu^2}{m^2}\Bigr)}
\otimes {\Delta \Sigma(N_F,\mu^2)}
%\nonumber\\ &&
+ {A_{gg,Q}^S\Bigl(N_F,\frac{\mu^2}{m^2}\Bigr)} 
\nonumber\\
&& \otimes {\Delta G(N_F,\mu^2)}
\\
{\Delta \Sigma(N_F+1,\mu^2)} 
%{\sum_{k=1}^{N_F+1} \left[f_k(N_F+1,\mu^2) +
%f_{\overline{k}}(N_F+1,\mu^2)\right]} \nonumber\\
&=& \hspace*{-2mm} \left[ {A_{qq,Q}^{\sf NS}\left(N_F, \frac{\mu^2}{m^2}\right)} +
N_F
{\tilde{A}_{qq,Q}^{\sf PS} \left(N_F, \frac{\mu^2}{m^2}\right)} +
{{A}_{Qq}^{\sf PS} \left(N_F, \frac{\mu^2}{m^2}\right)} \right]
\nonumber\\ && \hspace*{-2mm}
\otimes {\Delta \Sigma(N_F,\mu^2)} \nonumber\\ &&
+\left[N_F
{\tilde{A}^{\sf S}_{qg,Q}\left(N_F, \frac{\mu^2}{m^2}\right)}
+ {{A}^{\sf S}_{Qg}\left(N_F, \frac{\mu^2}{m^2}\right)} \right]
\otimes {\Delta G(N_F,\mu^2)}.
\nonumber\\ 
\label{eq:VFNS2}
\end{eqnarray}
}
%---------------------------------------------------------------------------------------

\normalsize \noindent
The $N_F$-dependence of the OMEs is understood as functional and $\mu^2$ denotes the 
matching scale, which for the heavy-to-light transitions is normally much larger than the mass scale 
$m^2$, \cite{Alekhin:2009ni}. The corresponding matching equations in the two--mass case are given in
Ref.~\cite{Ablinger:2017err}.

The results of the calculations presented in the subsequent sections have been obtained making mutual
use of the packages {\tt HarmonicSums.m} 
\cite{Vermaseren:1998uu,Blumlein:1998if,Remiddi:1999ew,Blumlein:2003gb,Blumlein:2009ta,
HARMSUM} and {\tt Sigma.m} \cite{SIG1,SIG2}, which is based on algorithms in difference-ring theory 
\cite{DIFFRING}.

In the subsequent expressions we abbreviate the following logarithms by
%---------------------------------------------------------------------------------------
\begin{eqnarray}
\label{eq:LOGS}
L_Q = \ln\left(\frac{Q^2}{\mu^2}\right)~~~\text{and}~~~L_M = \ln\left(\frac{m^2}{\mu^2}\right)~,
\end{eqnarray}
%---------------------------------------------------------------------------------------
where we set the renormalization and factorization scales equal $\mu \equiv \mu_F = \mu_R$.

%-----------------------------------------------------------------------------------------------------------------
\section{\boldmath $A_{qq,Q}^{(3),\rm PS}$ and $A_{qg,Q}^{(3),\rm S}$} 
\label{sec:3}
%-----------------------------------------------------------------------------------------------------------------

\vspace*{1mm}\noindent
In what follows,
the OMEs $A_{qq,Q}^{(3),\rm PS}$ and $A_{qg,Q}^{(3),\rm S}$ are presented in complete form in the 
following analytically, where the $O(\ep^0)$ contribution to the unrenormalized OMEs, $a_{ij,Q}^{(3)}$, are
given separately. Up to 3--loop order the massive OMEs $A_{qq,Q}^{\rm PS}$ and $A_{qg,Q}^{\rm S}$ do not yet
receive double mass contributions. The OMEs are expressed in terms of harmonic sums 
\cite{Vermaseren:1998uu,Blumlein:1998if} $S_{\vec{a}}(N) \equiv S_{\vec{a}}$, which are defined by
%-----------------------------------------------------------------------------------------------------------------
\begin{eqnarray}
S_{b,\vec{a}}(N) &=& \sum_{k=1}^N \frac{({\rm sign}(b))^k}{k^{|b|}} S_{\vec{a}}(k),~~S_\emptyset =1, b,a_i \in \mathbb{Z} 
\backslash \{0\}.
\end{eqnarray}
%-----------------------------------------------------------------------------------------------------------------

\vspace*{1mm}\noindent
%----------------------------
\noindent
The OME $A_{qq,Q}^{PS(3)}$ is given by
%----------------------------------------------------------------------------------------------------------------
\begin{eqnarray}
A_{qq,Q}^{PS(3)} &=& 
	a_s^3 \Biggl\{
	a_{qq,Q}^{PS(3)}
	+ \textcolor{blue}{C_F N_F T_F^2} \biggl\{
                -\frac{32  L_M ^3 (N-1) (2+N)}{9 N^2 (1+N)^2}
\nonumber\\&&		
                +\frac{32 (N-1) (2+N) \big(
                        98+369 N+408 N^2+164 N^3\big)}{81 N^2 
(1+N)^5}
\nonumber\\&&
		+ L_M ^2 \biggl[
                        -\frac{32 (2+N) \big(
                                3+4 N-3 N^2+8 N^3\big)}{9 N^3 
(1+N)^3}
                        +\frac{32 (N-1) (2+N)}{3 N^2 (1+N)^2} S_1 
		\biggr]
\nonumber\\&&		
		+ L_M  \biggl[
                        -\frac{32 (2+N) P_1}{27 N^4 (1+N)^4}
                        +\frac{64 (2+N) \big(
                                3+4 N-3 N^2+8 N^3\big)}{9 N^3 
(1+N)^3} S_1 
\nonumber\\&&
                        -\frac{32 (N-1) (2+N)}{3 N^2 
(1+N)^2} S_1 ^2
                        -\frac{32 (N-1) (2+N)}{3 N^2 (1+N)^2} S_2 
		\biggr]
\nonumber\\&&		
		+\biggl[
                        -\frac{32 (N-1) (2+N) \big(
                                22+41 N+28 N^2\big)}{27 N^2 (1+N)^4}
                        -\frac{16 (N-1) (2+N) S_2 }{3 N^2 (1+N)^2}
		\biggr] S_1 
\nonumber\\&&		
                +\frac{16 (N-1) (2+N) (2+5 N)}{9 N^2 (1+N)^3} 
	(S_1 ^2 + S_2)
		-\frac{16 (N-1) (2+N)}{9 N^2 (1+N)^2} (S_1 ^3 + 2 S_3)
\nonumber\\&&		
		+\biggl[
                        \frac{16 (2+N) \big(
                                3+2 N-6 N^2+13 N^3\big)}{9 N^3 (1+N)^3}
                        -\frac{32 (N-1) (2+N) S_1 }{3 N^2 (1+N)^2}
		\biggr] \zeta_2
\nonumber\\&&		
                +\frac{32 (N-1) (2+N)}{9 N^2 (1+N)^2} \zeta_3
	\biggr\}
	\Biggr\},
\end{eqnarray}
%----------------------------------------------------------------------------------------------------------------
with
%----------------------------------------------------------------------------------------------------------------
\begin{eqnarray}
P_1=86 N^5+38 N^4+40 N^3-8 N^2-15 N-9.
\end{eqnarray}
%----------------------------------------------------------------------------------------------------------------
Here $\zeta_k,~~k \geq 2, k \in \mathbb{N}$ denotes the Riemann $\zeta$ function at integer arguments.

%----------------------------

Likewise, the OME $A_{qg,Q}^{(3)}$ is obtained by
%----------------------------------------------------------------------------------------------------------
\begin{eqnarray}
A_{qg,Q}^{(3)} &=&
	a_s^3 \Biggl\{
	a_{qg,Q}^{(3)}
	+ \textcolor{blue}{C_A N_F T_F^2} \biggl\{
                -
                \frac{8 (N-1) P_8}{81 N^5 (1+N)^5}
		+ L_M ^3 \biggl[
                        -\frac{64 (N-1)}{9 N^2 (1+N)^2}
                        +\frac{32 (N-1)}{9 N (1+N)} S_1
		\biggr]
\nonumber\\&&		
		+ L_M ^2 \biggl[
                        \frac{8 P_4}{9 N^3 (1+N)^3}
                        +\frac{32 \big(
                                1+5 N^2\big)}{9 N (1+N)^2} S_1
			-\frac{16 (N-1)}{3 N (1+N)} (S_1^2 + S_2)		
                        -\frac{32 (N-1)}{3 N (1+N)} S_{-2}		
		\biggr]
\nonumber\\&&
		+ L_M  \Biggl[
                        \frac{16 P_7}{27 N^4 (1+N)^4}
			+\biggl[
                                \frac{16 \big(
                                        -1+44 N+67 N^2+94 
N^3\big)}{27 N (1+N)^3}
                                -\frac{16 (N-1) S_2}{3 N (1+N)}
			\biggr] S_1
\nonumber\\&&			
                        -\frac{32 \big(
                                -2+5 N^2\big)}{9 N (1+N)^2} S_1^2
                        +\frac{16 (N-1)}{9 N (1+N)} S_1^3
                        -\frac{32 \big(
                                -2+6 N+5 N^2\big)}{9 N (1+N)^2} S_2
                        +\frac{32 (N-1)}{9 N (1+N)} S_3
\nonumber\\&&
                        -\frac{64 (-2+5 N)}{9 N (1+N)} S_{-2}
                        +\frac{64 (N-1)}{3 N (1+N)} S_{-3}
                        +\frac{64 (N-1)}{3 N (1+N)} S_{2,1}
		\Biggr]
\nonumber\\&&		
                -\frac{16 (N-1) \big(
                        283+584 N+328 N^2\big)}{81 N (1+N)^3} S_1
                -\frac{8 (N-1)}{3 N (1+N)^2} S_1^2
                +\frac{8 (N-1) (1+2 N)}{3 N (1+N)^2} S_2
\nonumber\\&&		
		+\biggl[
                        -\frac{8 P_3}{9 N^3 (1+N)^3}
                        -\frac{32 \big(
                                -2+5 N^2\big)}{9 N (1+N)^2} S_1
                        +\frac{8 (N-1) S_1^2}
                        {3 N (1+N)}
\nonumber\\&&			
                        +
                        \frac{8 (N-1) S_2}{3 N (1+N)}
                        +\frac{16 (N-1) S_{-2}}{3 N (1+N)}
		\biggr] \zeta_2
	+\biggl[
                        \frac{64 (N-1)}{9 N^2 (1+N)^2}
                        -\frac{32 (N-1) S_1}{9 N (1+N)}
		\biggr] \zeta_3
	\biggr\}
\nonumber\\&&	
	+ \textcolor{blue}{C_F N_F T_F^2} \biggl\{
                -\frac{(N-1) P_{10}}{81 N^6 (1+N)^6}
		+ L_M ^3 \biggl[
                        \frac{8 (N-1) P_2}{9 N^3 (1+N)^3}
                        -\frac{32 (N-1)}{9 N (1+N)} S_1
		\biggr]
\nonumber\\&&		
		+ L_M ^2 \biggl[
                        \frac{4 (N-1) P_6}{9 N^4 (1+N)^4}
                        -\frac{32 (N-1) (3+5 N)}{9 N^2 (1+N)} S_1
			+\frac{16 (N-1)}{3 N (1+N)} (S_1^2 + S_2)
		\biggr]
\nonumber\\&&		
		+ L_M  \Biggl[
                        \frac{4 P_9}{27 N^5 (1+N)^5}
			+\biggl[
                                -\frac{16 \big(
                                        -24-52 N+103 N^2\big)}{27 N^2 
(1+N)}
                                -\frac{16 (N-1)}{3 N (1+N)} S_2
			\biggr] S_1
\nonumber\\&&			
                        +\frac{16 (N-1) (3+10 N)}{9 N^2 (1+N)} S_1^2
                        -\frac{16 (N-1)}{9 N (1+N)} S_1^3
                        +\frac{16 \big(
                                -3+5 N+10 N^2\big)}{9 N^2 (1+N)} S_2
\nonumber\\&&				
                        +\frac{64 (N-1)}{9 N (1+N)} S_3
		\Biggr]
                +\frac{5248 (N-1) S_1}{81 N (1+N)}
                -\frac{896 (N-1) S_2}{27 N (1+N)}
                +\frac{160 (N-1) S_3}{9 N (1+N)}
\nonumber\\&&		
                -\frac{32 (N-1)}{3 N (1+N)} S_4
		+\biggl[
                        -\frac{4 (N-1) P_5}{9 N^4 (1+N)^4}
                        +\frac{16 (N-1) (3+10 N)}
                        {9 N^2 (1+N)} S_1
\nonumber\\&&			
                        -\frac{8 (N-1) S_1^2}{3 N (1+N)}
                        -\frac{8 (N-1) S_2}{N (1+N)}
		\biggr] \zeta_2
		+\biggl[
                        -\frac{8 (N-1) P_2}{9 N^3 (1+N)^3}
                        +\frac{32 (N-1) S_1}{9 N (1+N)}
		\biggr] \zeta_3
	\biggr\}
\Biggr\},
\end{eqnarray}
%----------------------------------------------------------------------------------------------------------
with the polynomials
%----------------------------------------------------------------------------------------------------------
\begin{eqnarray}
P_2&=&3 N^4+6 N^3-N^2-4 N+12,\\
P_3&=&6 N^5+6 N^4-67 N^3+6 N^2+43 N-18, \\
P_4&=&9 N^5+9 N^4-79 N^3+15 N^2+22 N-24,\\
P_5&=&18 N^6+54 N^5+5 N^4-20 N^3+95 N^2-132 N-108,\\
P_6&=&33 N^6+99 N^5+41 N^4-11 N^3+86 N^2-216 N-144,\\
P_7&=&99 N^7+198 N^6-410 N^5-344 N^4+128 N^3-130 N^2-39 N+90, \\
P_8&=&255 N^8+1020 N^7-532 N^6-4536 N^5-4344 N^4-1138 N^3+3 N^2+36 N
\nonumber\\&&
+108,\\
P_9&=&159 N^9+477 N^8-220 N^7-710 N^6+117 N^5-1081 N^4+2536 N^3
\nonumber\\&&
+1026 N^2-1800 N-1080,\\
P_{10}&=&1551 N^{10}+7755 N^9+10982 N^8+1910 N^7+2427 N^6+14975 N^5
\nonumber\\&&
+13952 N^4-1488 N^3-7488 N^2-6912 N-2592.
\end{eqnarray}
%----------------------------------------------------------------------------------------------------------

%----------------------------
The $O(\ep^0)$ parts of the unrenormalized massive OMEs, $a_{qq,Q}^{(3),\rm PS}$ and $a_{qg,Q}^{(3)}$ are given by
%-----------------------------------------------------------------------------------------------------------------
\begin{eqnarray}
a_{qq,Q}^{(3),\rm PS} &=& 
\textcolor{blue}{C_F T_F^2 N_F} \Biggl\{
        \frac{32 (2+N) R_2}{243 N^5 (1+N)^5}
        +\Biggl(
                -\frac{64 (2+N) R_1}{81 N^4 (1+N)^4}
                -\frac{128 (N-1) (2+N)}{9 N^2 (1+N)^2} S_2
        \Biggr) S_1
\nonumber\\ &&
        +\frac{64 (2+N) \big(
                3+2 N-6 N^2+13 N^3\big)}{27 N^3 (1+N)^3} [S_1^2+S_2]
        -\frac{128 (N-1) (2+N)}{27 N^2 (1+N)^2} [S_1^3 + 2 S_3]
\nonumber\\ &&
        +\Biggl(
                \frac{16 (2+N) \big(
                        3+2 N-6 N^2+13 N^3\big)}{9 N^3 (1+N)^3}
                -\frac{32 (N-1) (2+N)}{3 N^2 (1+N)^2} S_1
        \Biggr) \zeta_2
\nonumber\\ &&
        -\frac{224 (N-1) (2+N)}{9 N^2 (1+N)^2} \zeta_3
\Biggr\}
\\
a_{qg,Q}^{(3)} &=& 
\textcolor{blue}{C_F T_F^2 N_F} 
\nonumber\\ && \times
\Biggl\{
        \frac{112 (N-1)R_3}{9 N^3 (1+N)^3} \zeta_3
        -\frac{8 (N-1)R_6}{9 N^4 (1+N)^4}  \zeta_2
        -\frac{16 R_8}{243 N^6 (1+N)^6}
        +\frac{(N-1) (3+10 N)}{N^2 (1+N)} 
\nonumber\\ && \times
\Biggl(
                \frac{32}{81} S_1^3
                +\frac{32}{27} S_1 S_2
                +\frac{32}{9} S_1 \zeta_2
        \Biggr)
        +\frac{N-1}{N (1+N)} \Biggl(
                -\frac{8}{27} S_1^4
                -\frac{16}{9} S_1^2 S_2
                -\frac{8}{9} S_2^2
                -\frac{64}{27} S_1 S_3
\nonumber\\ &&               
 -\frac{80}{9} S_4
                -16 \Biggl(
                        \frac{1}{3} S_1^2
                        + S_2
                \Biggr) \zeta_2
                -\frac{448}{9} S_1 \zeta_3
        \Biggr)
        -\frac{32 \big(
                420+2399 N-555 N^2-2507 N^3\big)}{243 N^2 (1+N)^2} 
\nonumber\\ && \times
S_1
        -\frac{16 \big(
                -24-164 N+215 N^2\big)}{81 N^2 (1+N)} S_1^2
        -\frac{16 \big(
                -8-148 N+221 N^2\big)}{27 N^2 (1+N)} S_2
\nonumber\\ &&
        +\frac{64 (-3+5 N) (1+8 N)}{81 N^2 (1+N)} S_3
\Biggr\}
%-----
\nonumber\\ &&
+\textcolor{blue}{C_A T_F^2 N_F}  \Biggl\{
        -\frac{32 R_4}{243 N (1+N)^4} S_1
        -\frac{16 R_5}{9 N^3 (1+N)^3} \zeta_2
        -\frac{32 R_7}{243 N^5 (1+N)^5}
        +\frac{-2+5 N^2}{N (1+N)^2} 
\nonumber\\ && \times
\Biggl(
                -\frac{64}{81} S_1^3
                +\frac{64}{27} S_1 S_2
                -\frac{128}{81} S_3
                -\frac{256}{27} S_{2,1}
                -\frac{64}{9} S_1 \zeta_2
        \Biggr)
        +\frac{N-1}{N (1+N)} \Biggl(
                \Biggl(
                        -\frac{320}{27} S_3
\nonumber\\ &&
+\frac{128}{9} S_{2,1}
                \Biggr) S_1
                +\frac{8}{27} S_1^4
                -\frac{16}{9} S_1^2 S_2
                +\frac{8}{9} S_2^2
                +\frac{112}{9} S_4
                +\frac{256}{9} S_{-4}
                +\frac{256}{9} S_{3,1}
                -\frac{128}{9} S_{2,1,1}
\nonumber\\ &&
                +\Biggl(
                        \frac{16}{3} [S_2 + S_1^2]
                        +\frac{32}{3} S_{-2}
                \Biggr) \zeta_2
                +\frac{448}{9} S_1 \zeta_3
        \Biggr)
        +\frac{16 \big(
                -59-68 N+125 N^2+206 N^3\big)}{81 N (1+N)^3} S_1^2
\nonumber\\ &&
        -\frac{16 \big(
                131+26 N-509 N^2-332 N^3\big)}{81 N (1+N)^3} S_2
        -\frac{256}{9 (1+N)^2} S_3
        -\frac{64 (7-11 N) (10+11 N)}{81 N (1+N)^2} 
\nonumber\\ && \times
S_{-2}
        -\frac{256 (-2+5 N)}{27 N (1+N)} S_{-3}
        -\frac{896 (N-1)}{9 N^2 (1+N)^2} \zeta_3
\Biggr\},
\end{eqnarray}
%-----------------------------------------------------------------------------------------------------------------
with
%-----------------------------------------------------------------------------------------------------------------
\begin{eqnarray}
R_1 &=& 142 N^5+64 N^4+2 N^3-52 N^2-15 N-9,
\\
R_2 &=& 1648 N^7+2444 N^6+777 N^5-1091 N^4-106 N^3+201 N^2+72 N+27
\\
R_3 &=& 3 N^4+6 N^3-N^2-4 N+12,
\\
R_4 &=& 2228 N^4+3601 N^3+669 N^2-2191 N-1055,
\\
R_5 &=& 6 N^5+6 N^4-67 N^3+6 N^2+43 N-18,
\\
R_6 &=& 18 N^6+54 N^5+5 N^4-20 N^3+95 N^2-132 N-108,
\\
R_7 &=& 2040 N^9+6120 N^8-4816 N^7-16208 N^6-1776 N^5+7268 N^4+2443 N^3+756 N^2
\nonumber\\ &&
-225 N-810,
\\
R_8 &=& 2322 N^{11}+9288 N^{10}+5975 N^9-11499 N^8-7124 N^7+4346 N^6+522 N^5
\nonumber\\ &&
-20360 N^4-12027 N^3+5193 N^2+11880 N+4860.
\end{eqnarray}
%-----------------------------------------------------------------------------------------------------------------

\vspace*{1mm}\noindent
%-----------------------------------------------------------------------------------------------------------------
\section{\boldmath $A_{Qg}^{(3),\rm S}$} 
\label{sec:4}
%-----------------------------------------------------------------------------------------------------------------

\vspace*{1mm}\noindent
The logarithmic contributions to the massive OME $A_{Qg}^{(3),\rm S}$ are given by
%----------------------------
%-----------------------------------------------------------------------------------------------------

%-----------------------------------------------------------------------------------------------------
and the polynomials $P_i$ read
%-----------------------------------------------------------------------------------------------------
\begin{eqnarray}
P_{11}&=&-3 N^4-54 N^3-95 N^2-12 N+36,\\
P_{12}&=&N^4-94 N^3-256 N^2-161 
N+78,\\
P_{13}&=&N^4+2 N^3-5 N^2-12 N+2,\\
P_{14}&=&N^4+4 N^3-N^2-10 
N+2, \\
P_{15}&=&N^4+10 N^3+27 N^2+30 N+4,\\
P_{16}&=&N^4+17 N^3+43 N^2+33 
N+2, \\
P_{17}&=&2 N^4-4 N^3-3 N^2+20 N+12,\\
P_{18}&=&2 N^4+3 N^3-12 
N^2-23 N+6, \\
P_{19}&=&2 N^4+39 N^3+100 N^2+73 N+2,\\
P_{20}&=&3 N^4+6 N^3-N^2-4 N+12,\\
P_{21}&=&3 N^4+30 N^3+47 N^2+4 N-20,\\
P_{22}&=&3 N^4+48 N^3+123 N^2+98 N+8,\\
P_{23}&=&5 N^4+10 N^3+8 N^2+7 N+2, \\
P_{24}&=&5 N^4+13 N^3+14 N^2+16 N+6,\\
P_{25}&=&5 N^4+37 N^3+82 N^2+41 
N-48, \\
P_{26}&=&6 N^4+11 N^3-6 N^2-N-2,\\
P_{27}&=&6 N^4+12 N^3+7 
N^2+N+6, \\
P_{28}&=&9 N^4+102 N^3+245 N^2+192 N+12,\\
P_{29}&=&10 N^4+53 
N^3+92 N^2+37 N-48,\\
P_{30}&=&11 N^4-68 N^3-263 N^2-184 N+72, \\
P_{31}&=&11 N^4-26 N^3-227 N^2-286 N+48,\\
P_{32}&=&11 N^4+4 N^3-239 
N^2-304 N+240,\\
P_{33}&=&13 N^4+23 N^3+4 N^2-14 N-5,\\
P_{34}&=&13 N^4+140 N^3+365 N^2+190 N-276,\\
P_{35}&=&17 N^4+34 N^3+82 N^2+161 
N-78, \\
P_{36}&=&29 N^4+60 N^3+149 N^2+336 N+74,\\
P_{37}&=&55 N^4+86 
N^3-343 N^2-422 N+384,\\
P_{38}&=&55 N^4+182 N^3-175 N^2-542 N+240, \\
P_{39}&=&76 N^4+183 N^3+196 N^2+267 N-38,\\
P_{40}&=&97 N^4+494 N^3+1079 
N^2+898 N-408,\\
P_{41}&=&153 N^4+306 N^3+165 N^2+12 N+4,\\
P_{42}&=&183 N^4+366 N^3+305 N^2+122 N+96,\\
P_{43}&=&N^5+N^4-4 N^3+3 N^2-7 N-2,\\ 
P_{44}&=&2 N^5+6 N^4+3 N^3+11 N+2,\\
P_{45}&=&2 N^5+10 N^4+29 N^3+64 
N^2+67 N+8,\\
P_{46}&=&3 N^5+8 N^4+6 N^3+10 N^2+7 N+2,\\
P_{47}&=&8 N^5+7 
N^4-9 N^3+7 N^2+13 N+6,\\
P_{48}&=&9 N^5+9 N^4-79 N^3+15 N^2+22 N-24, \\
P_{49}&=&15 N^5+15 N^4-103 N^3+33 N^2-20 N-36, \\
P_{50}&=&18 N^5-15 
N^4-198 N^3-381 N^2-216 N+4,\\
P_{51}&=&18 N^5+47 N^4-35 N^3-141 N^2-5 
N-120, \\
P_{52}&=&40 N^5+73 N^4-142 N^3-163 N^2-150 N+54,\\
P_{53}&=&45 
N^5+45 N^4-47 N^3+27 N^2-190 N-24,\\
P_{54}&=&51 N^5+89 N^4+6 N^3-66 
N^2-104 N-72,\\
 P_{55}&=&66 N^5+336 N^4+627 N^3+415 N^2-10 N-194, \\
P_{56}&=&69 N^5+69 N^4-55 N^3+51 N^2-338 N-36,\\
 P_{57}&=&85 N^5+85 
N^4-73 N^3+197 N^2-342 N-108,\\
 P_{58}&=&103 N^5+103 N^4-79 N^3+317 
N^2-612 N-144,\\
 P_{59}&=&337 N^5+403 N^4-541 N^3-583 N^2-300 N+108, \\
P_{60}&=&436 N^5+1780 N^4+2689 N^3+2782 N^2+2167 N-134,\\
 P_{61}&=&489 
N^5+489 N^4-1187 N^3-57 N^2-742 N-144,\\
 P_{62}&=&N^6+18 N^5+63 N^4+84 
N^3+30 N^2-64 N-16,\\
 P_{63}&=&N^6+23 N^5+73 N^4+85 N^3+58 N^2+24 
N-24, \\
P_{64}&=&3 N^6+30 N^5+107 N^4+124 N^3+48 N^2+20 N+8,\\
 P_{65}&=&3 
N^6+51 N^5+153 N^4+185 N^3+160 N^2+80 N-72,\\
 P_{66}&=&4 N^6+41 N^5+126 
N^4+163 N^3+58 N^2-128 N-32,\\
 P_{67}&=&5 N^6+26 N^5+77 N^4+168 N^3+159 
N^2+19 N-22,\\
 P_{68}&=&6 N^6+75 N^5+345 N^4+719 N^3+323 N^2-696 N-96, \\
P_{69}&=&18 N^6+87 N^5+199 N^4+185 N^3+63 N^2+44 N+20, \\
P_{70}&=&23 
N^6+39 N^5-89 N^4-219 N^3-172 N^2-130 N-28,\\
 P_{71}&=&25 N^6-118 
N^5-662 N^4-500 N^3+421 N^2+186 N-264, \\
P_{72}&=&33 N^6+99 N^5+41 
N^4-11 N^3+86 N^2-216 N-144, \\
P_{73}&=&33 N^6+99 N^5+137 N^4+157 
N^3+62 N^2+8 N-16,\\
 P_{74}&=&36 N^6+48 N^5-297 N^4-977 N^3-976 N^2-362 
N+24,\\
 P_{75}&=&37 N^6+207 N^5+753 N^4+1771 N^3+1598 N^2-118 N+48, \\
P_{76}&=&57 N^6+297 N^5+567 N^4+615 N^3+468 N^2+220 N+16,\\
 P_{77}&=&80 
N^6+201 N^5-775 N^4-3495 N^3-4405 N^2-2238 N-72,\\
 P_{78}&=&94 N^6+282 
N^5+79 N^4+42 N^3+286 N^2-585 N-18, \\
P_{79}&=&129 N^6+387 N^5+509 
N^4+349 N^3+50 N^2+240 N+144,\\
 P_{80}&=&170 N^6+543 N^5+221 N^4-15 
N^3+425 N^2-864 N-288, \\
P_{81}&=&170 N^6+873 N^5+1547 N^4+951 N^3-1717 
N^2-2976 N+864,\\
P_{82}&=&243 N^6+729 N^5+923 N^4+583 N^3+14 N^2+300 
N+216,\\
P_{83}&=&321 N^6+1353 N^5+1521 N^4-713 N^3-2842 N^2-2216 
N+96,\\
P_{84}&=&333 N^6+999 N^5+1075 N^4+389 N^3-68 N^2+384 N+216, \\
P_{85}&=&633 N^6+1899 N^5+1967 N^4+697 N^3-4 N^2-48 N+8, \\
P_{86}&=&-891 
N^7-1782 N^6-3712 N^5-3058 N^4+6775 N^3
\nonumber\\&&
+7144 N^2+276 N-144, \\
P_{87}&=&9 N^7+18 N^6-124 N^5-109 N^4+199 N^3-191 N^2+138 N+72, \\
P_{88}&=&69 N^7+138 N^6-667 N^5-541 N^4+952 N^3-1277 N^2+990 N+432, \\
P_{89}&=&95 N^7+378 N^6+853 N^5+1832 N^4+2190 N^3+364 N^2-780 N-432,\\
P_{90}&=&251 N^7+1335 N^6+1745 N^5+243 N^4-529 N^3-4161 N^2-5400 
N-756,\\
P_{91}&=&N^8+427 N^7+2161 N^6+4081 N^5+3554 N^4
\nonumber\\&&
+1404 N^3+228 N^2+64 N+16,\\
P_{92}&=&2 N^8+10 N^7+22 N^6+36 N^5+29 N^4+4 N^3+33 
N^2+12 N+4,\\
P_{93}&=&2 N^8+29 N^7+135 N^6+297 N^5+333 N^4+204 N^3+28 
N^2-44 N-24,\\
P_{94}&=&3 N^8+33 N^7+149 N^6+267 N^5+196 N^4+104 N^3+64 
N^2-88 N-48,\\
P_{95}&=&12 N^8+52 N^7+60 N^6-25 N^4-2 N^3+3 N^2+8 N+4, \\
P_{96}&=&36 N^8+348 N^7+1210 N^6+2229 N^5+2168 N^4+505 N^3-424 N^2
\nonumber\\&&
-68 
N+48,\\
P_{97}&=&111 N^8+480 N^7+286 N^6-468 N^5+82 N^4+246 N^3+295 
N^2
\nonumber\\&&
+228 N+252,\\
P_{98}&=&201 N^8+840 N^7+565 N^6-699 N^5-344 N^4-645 
N^3-314 N^2
\nonumber\\&&
+324 N+360, \\
P_{99}&=&296 N^8+1184 N^7+2744 N^6+5900 
N^5+4088 N^4+476 N^3+9477 N^2
\nonumber\\&&
+4725 N+702,\\
P_{100}&=&-7299 N^{10}-39375 N^9-79900 N^8-85198 N^7-17323 N^6+129917 N^5
\nonumber\\&&
+137090 N^4+25904 N^3+12072 N^2+30672 N+8640, \\
P_{101}&=&-149 N^{10}-793 
N^9-1404 N^8-1170 N^7-1341 N^6-1221 N^5+1710 N^4
\nonumber\\&&
+2800 N^3+2256 N^2+368 N-32, \\
P_{102}&=&4 N^{10}+22 N^9+45 N^8+36 N^7-11 N^6-15 
N^5+25 N^4-41 N^3-21 N^2
\nonumber\\&&
-16 N-4, \\
P_{103}&=&10 N^{10}+62 N^9+403 
N^8+1523 N^7+2997 N^6+3197 N^5
\nonumber\\&&
+1812 N^4+478 N^3+46 N^2+24 N+8, \\
P_{104}&=&26 N^{10}+132 N^9+159 N^8-351 N^7-877 N^6+531 N^5+1820 
N^4-300 N^3
\nonumber\\&&
-252 N^2-192 N-48, \\
P_{105}&=&28 N^{10}+139 N^9+444 N^8+803 
N^7+451 N^6+3 N^5+490 N^4+219 N^3
\nonumber\\&&
+51 N^2-60 N-12, \\
P_{106}&=&435 N^{10}+2391 N^9+6946 N^8+11512 N^7+4822 N^6-7016 N^5-5369 N^4
\nonumber\\&&
-6743 N^3-2406 N^2-1764 N-216, \\
P_{107}&=&531 N^{10}+2799 N^9+4124 N^8+446 
N^7-3445 N^6-5245 N^5+4358 N^4
\nonumber\\&&
+18128 N^3-1968 N^2-10800 N-4320, \\
P_{108}&=&939 N^{10}+4893 N^9+5386 N^8-5198 N^7-10400 N^6-17636 
N^5-18137 N^4
\nonumber\\&&
+7177 N^3-21672 N^2-14112 N-4104, \\
P_{109}&=&1773 
N^{10}+9153 N^9+14204 N^8+2930 N^7-9151 N^6-8431 N^5-250 N^4
\nonumber\\&&
+13772 N^3+1920 N^2-8928 N-4320, \\
P_{110}&=&-23 N^{11}-92 N^{10}-53 N^9+322 
N^8+465 N^7-348 N^6-929 N^5-384 N^4
\nonumber\\&&
+132 N^3+102 N^2+32 N+8, \\
P_{111}&=&87 N^{12}+490 N^{11}+949 N^{10}+368 N^9-1285 N^8-2214 
N^7-1591 N^6
\nonumber\\&&
-126 N^5+644 N^4-86 N^3-268 N^2-184 N-48, \\
P_{112}&=&385 N^{12}+2182 N^{11}+4181 N^{10}+1458 N^9-5589 N^8-8414 N^7-5041 
N^6
\nonumber\\&&
-1754 N^5-760 N^4-176 N^3+152 N^2+224 N+96, \\
P_{113}&=&1623 
N^{12}+9602 N^{11}+20093 N^{10}+15520 N^9-3305 N^8-13494 N^7-5099 
N^6
\nonumber\\&&
+9414 N^5+10456 N^4+5270 N^3+1624 N^2+40 N-96.
\end{eqnarray}
%-----------------------------------------------------------------------------------------------------

%----------------------------

%-----------------------------------------------------------------------------------------------------------------
\section{\boldmath $A_{gg,Q}^{(3),\rm S}$} 
\label{sec:5}
%-----------------------------------------------------------------------------------------------------------------

\vspace*{1mm}\noindent
For the logarithmic contributions to the OME $A_{gg,Q}^{(3),\rm S}$ we obtain
%----------------------------
%--------------------------------------------------------------------------------------------------------------
\begin{eqnarray}
A_{gg,Q} &=&
	\frac{4  a_s   L_M   \textcolor{blue}{T_F} }{3}
+ a_s ^2 \Biggl\{
	\frac{16  L_M ^2  \textcolor{blue}{T_F ^2} }{9}
	+ \textcolor{blue}{C_F T_F} \biggl[
                \frac{4  L_M  P_{147}}{N^3 (1+N)^3}
                +\frac{P_{166}}{N^4 (1+N)^4}
\nonumber\\&&		
                +\frac{4  L_M ^2 (N-1) (2+N)}{N^2 (1+N)^2}
        \biggr]
+ \textcolor{blue}{C_A T_F} \biggl\{
                \frac{2 P_{155}}{27 N^3 (1+N)^3}
                + L_M  \biggl[
                        \frac{16 P_{134}}{9 N^2 (1+N)^2}
                        -\frac{80}{9} S_ 1
                \biggr]
\nonumber\\&&		
                + L_M ^2 \biggl[
                        \frac{16}{3 N (1+N)}
                        -\frac{8}{3} S_ 1
                \biggr]
                -\frac{4 (47+56 N) S_ 1}{27 (1+N)}
	\biggr\}
	\Biggr\}
+ a_s ^3 \Biggl\{
	\frac{64  L_M ^3  }{27} \textcolor{blue}{T_F ^3}
\nonumber\\&&	
	+ \textcolor{blue}{C_F T_F^2} \biggl\{
                \frac{2 P_{177}}{9 N^5 (1+N)^5}
                +\frac{80  L_M ^3 (N-1) (2+N)}{9 N^2 (1+N)^2}
                + L_M ^2 \biggl[
                        \frac{8 P_{153}}{9 N^3 (1+N)^3}
\nonumber\\&&			
                        +\frac{32 (N-1) (2+N) S_ 1}{3 N^2 (1+N)^2}
                \biggr]
                + L_M  \biggl[
                        -\frac{8 P_{173}}{27 N^4 (1+N)^4}
\nonumber\\&&			
                        +\frac{32 (N-1) (2+N) \big(
                                -6-8 N+N^2\big)}{9 N^3 (1+N)^3} S_ 1
                        +\frac{16 (N-1) (2+N) S_ 1^2}{3 N^2 (1+N)^2}
\nonumber\\&&			
                        -\frac{16 (N-1) (2+N) S_ 2}{N^2 (1+N)^2}
                \biggr]
                +\biggl[
                        -\frac{8 P_{154}}{9 N^3 (1+N)^3}
                        -
                        \frac{16 (N-1) (2+N) S_ 1}{3 N^2 (1+N)^2}
                \biggr] \zeta_2
\nonumber\\&&		
                -\frac{80 (N-1) (2+N) \zeta_3}{9 N^2 (1+N)^2}
	\biggr\}
+ \textcolor{blue}{C_F N_F T_F^2} \biggl\{
                \frac{2 P_{178}}{81 N^5 (1+N)^5}
                +\frac{64  L_M ^3 (N-1) (2+N)}{9 N^2 (1+N)^2}
\nonumber\\&&		
                + L_M  \biggl[
                        -\frac{4 P_{171}}{9 N^4 (1+N)^4}
                        -\frac{32 (N-1) (2+N) \big(
                                4+6 N+N^2\big)}{3 N^3 (1+N)^3} S_ 1
\nonumber\\&&				
                        +\frac{16 (N-1) (2+N) S_ 1^2}{N^2 (1+N)^2}
                        -\frac{80 (N-1) (2+N) S_ 2}{3 N^2 (1+N)^2}
                \biggr]
\nonumber\\&&		
                +\biggl[
                        \frac{32 (N-1) (2+N) \big(
                                22+41 N+28 N^2\big)}{27 N^2 (1+N)^4}
                        +\frac{16 (N-1) (2+N) S_ 2}{3 N^2 (1+N)^2}
                \biggr] S_ 1
\nonumber\\&&		
                -\frac{16 (N-1) (2+N) (2+5 N)}{9 N^2 (1+N)^3} S_ 1^2
                +\frac{16 (N-1) (2+N) S_ 1^3}{9 N^2 (1+N)^2}
\nonumber\\&&		
                -\frac{16 (N-1) (2+N) (2+5 N)}{9 N^2 (1+N)^3} S_ 2
                +\frac{32 (N-1) (2+N) S_ 3}{9 N^2 (1+N)^2}
                +\biggl[
                        \frac{4 P_{162}}{9 N^3 (1+N)^3}
\nonumber\\&&			
                        +\frac{16 (N-1) (2+N) S_ 1}{3 N^2 (1+N)^2}
                \biggr] \zeta_2
                -\frac{64 (N-1) (2+N) \zeta_3}{9 N^2 (1+N)^2}
	\biggr\}
+ \textcolor{blue}{C_A^2 T_F}  \biggl\{
                -\frac{4 P_{174}}{243 N^4 (1+N)^4}
\nonumber\\&&		
                + L_M ^3 \biggl[
                        -\frac{352}{27 N (1+N)}
                        +\frac{176}
                        {27} S_ 1
                \biggr]
                + L_M ^2 \Biggl[
                        -
                        \frac{2 P_{149}}{9 N^3 (1+N)^3}
                        +\biggl[
                                -\frac{8 P_{140}}{9 N^2 (1+N)^2}
\nonumber\\&&				
                                +\frac{64}{3} S_ 2
                        \biggr] S_ 1
                        -\frac{128 S_ 2}{3 N (1+N)}
                        +\frac{32}{3} S_ 3
                        +\biggl[
                                -\frac{128}{3 N (1+N)}
                                +\frac{64}{3} S_ 1
                        \biggr] S_{-2}
                        +\frac{32}{3} S_{-3}
\nonumber\\&&			
                        -\frac{64}{3} S_ {-2,1}
                \Biggr]
                + L_M  \Biggl[
                        \frac{16 S_ 2 P_{133}}{9 N^2 (1+N)^2}
                        +\frac{16 S_{-3} P_{139}}{9 N^2 (1+N)^2}
                        -\frac{32 S_{-2,1} P_{139}}{9 N^2 (1+N)^2}
                        +\frac{8 S_ 3 P_{142}}{9 N^2 (1+N)^2}
\nonumber\\&&			
                        +\frac{P_{183}}{81 (N-1) N^5 (1+N)^5 (2+N)}
                        +\biggl[
                                -\frac{4 P_{176}}{81 (N-1) N^4 
(1+N)^4 (2+N)}
\nonumber\\&&
                                +\frac{640}{9} S_ 2
                                -\frac{32}{3} S_ 3
                        \biggr] S_ 1
                        +\biggl[
                                -\frac{16 P_{163}}{9 (N-1) N^3 
(1+N)^3 (2+N)}
\nonumber\\&&
                                +\frac{32 P_{157}}{9 (N-1) N^2 
(1+N)^2 (2+N)} S_ 1
                        \biggr] S_{-2}
                        +\frac{32}{3} S_ {-2}^2
\nonumber\\&&			
                        +\biggl[
                                \frac{64 \big(
                                        -3+2 N+2 N^2\big)}{N^2 (1+N)^2}
                                -64 S_ 1
                        \biggr] \zeta_3
                \Biggr]
                -\frac{8 \big(
                        2339+4876 N+2834 N^2\big)}{243 (1+N)^2} S_ 1
                -\frac{44 S_ 1^2}{9 (1+N)}
\nonumber\\&&		
                +\frac{44 (1+2 N) S_ 2}{9 (1+N)}
                +\Biggl[
                        \frac{4 P_{160}}{27 N^3 (1+N)^3}
                        +\big(
                                \frac{16 \big(
                                        36+72 N+N^2+2 
N^3+N^4\big)}{27 N^2 (1+N)^2}
                                -
                                \frac{32}{3} S_ 2
                        \big) S_ 1
\nonumber\\&&			
                        +\frac{64 S_ 2}{3 N (1+N)}
                        -\frac{16}{3} S_ 3
                        +\biggl[
                                \frac{64}{3 N (1+N)}
                                -\frac{32}{3} S_ 1
                        \biggr] S_{-2}
                        -\frac{16}{3} S_{-3}
                        +\frac{32}{3} S_ {-2,1}
                \Biggr] \zeta_2
\nonumber\\&&		
                +\biggl[
                        \frac{352}{27 N (1+N)}
                        -\frac{176}{27} S_ 1
                \biggr] \zeta_3
	\biggr\}
	+ \textcolor{blue}{C_A N_F T_F^2} \biggl\{
                \frac{16 P_{172}}{243 N^4 (1+N)^4}
                + L_M  \biggl[
                        -\frac{16 S_ 1 P_{145}}{81 N^2 (1+N)^2}
\nonumber\\&&			
                        -\frac{4 P_{164}}{81 N^3 (1+N)^3}
                \biggr]
                + L_M ^3 \biggl[
                        \frac{128}{27 N (1+N)}
                        -\frac{64}{27} S_ 1
                \biggr]
                +\frac{32 \big(
                        283+584 N+328 N^2\big)}{243 (1+N)^2} S_ 1
\nonumber\\&&			
                +\frac{16 S_ 1^2}{9 (1+N)}
                -\frac{16 (1+2 N) S_ 2}{9 (1+N)}
                +\biggl[
                        -\frac{4 P_{137}}{27 N^2 (1+N)^2}
                        +\frac{160}{27} S_ 1
                \biggr] \zeta_2
\nonumber\\&&		
                +\biggl[
                        -\frac{128}{27 N (1+N)}
                        +\frac{64}{27} S_ 1
                \biggr] \zeta_3
	\biggr\}
+ \textcolor{blue}{C_A T_F^2} \biggl\{
                -\frac{8 P_{167}}{81 N^4 (1+N)^4}
                + L_M  \biggl[
                        -\frac{8 S_ 1 P_{141}}{9 N^2 (1+N)^2}
\nonumber\\&&			
                        -\frac{2 P_{159}}{27 N^3 (1+N)^3}
                \biggr]
                + L_M ^2 \biggl[
                        \frac{8 P_{143}}{27 N^2 (1+N)^2}
                        -\frac{640}{27} S_ 1
                \biggr]
                + L_M ^3 \biggl[
                        \frac{448}{27 N (1+N)}
                        -
                        \frac{224}{27} S_ 1
                \biggr]
\nonumber\\&&		
                +\frac{16 \big(
                        283+584 N+328 N^2\big)}{81 (1+N)^2} S_ 1
                +\frac{8 S_ 1^2}{3 (1+N)}
                -\frac{8 (1+2 N) S_ 2}{3 (1+N)}
\nonumber\\&&		
                +\biggl[
                        -\frac{4 P_{144}}{27 N^2 (1+N)^2}
                        +\frac{560}{27} S_ 1
                \biggr] \zeta_2
                +\biggl[
                        -\frac{448}{27 N (1+N)}
                        +\frac{224}{27} S_ 1
                \biggr] \zeta_3
	\biggr\}
\nonumber\\&&	
	+ \textcolor{blue}{C_F^2 T_F}  \biggl\{
                \frac{8 S_ 3 P_{136}}{3 N^3 (1+N)^3}
                +\frac{4 S_ 2 P_{151}}{N^4 (1+N)^4}
                +\frac{P_{180}}{N^6 (1+N)^6}
\nonumber\\&&		
                + L_M ^3 \biggl[
                        -\frac{4 (N-1) (2+N) \big(
                                2+3 N+3 N^2\big)}{3 N^3 (1+N)^3}
                        +\frac{16 (N-1) (2+N) S_ 1}{3 N^2 (1+N)^2}
                \biggr]
\nonumber\\&&		
                + L_M ^2 \biggl[
                        -\frac{8 (N-1) (2+N) \big(
                                -2-3 N+N^3+2 N^4\big)}{N^4 (1+N)^4}
                        +\frac{8 (N-1)^2 (2+N) (2+3 N)}{N^3 (1+N)^3} 
S_ 1
\nonumber\\&&
                        -\frac{16 (N-1) (2+N) S_ 2}{N^2 (1+N)^2}
                \biggr]
                + L_M  \Biggl[
                        -\frac{4 S_ 2 P_{138}}{N^3 (1+N)^3}
                        -\frac{2 P_{181}}{(N-1) N^5 (1+N)^5 (2+N)}
\nonumber\\&&			
                        +\biggl[
                                -\frac{8 P_{152}}{N^4 (1+N)^4}
                                +\frac{24 (N-1) (2+N) S_ 2}{N^2 
(1+N)^2}
                        \biggr] S_ 1
                        +\frac{4 \big(
                                -6-13 N+3 N^3\big)}{N^2 (1+N)^3} S_ 1^2
\nonumber\\&&				
                        -\frac{8 (N-1) (2+N) S_ 1^3}{3 N^2 (1+N)^2}
                        +\frac{16 \big(
                                14+5 N+5 N^2\big)}
                        {3 N^2 (1+N)^2} S_ 3
                        +\biggl[
                                -
                                \frac{32 \big(
                                        10+N+N^2\big)}{(N-1) N (1+N) 
(2+N)}
\nonumber\\&&
                                +\frac{256 S_ 1}{N^2 (1+N)^2}
                        \biggr] S_{-2}
                        +\frac{128 S_{-3}}{N^2 (1+N)^2}
                        -\frac{32 (N-1) (2+N) S_{2,1}}{N^2 (1+N)^2}
                        -\frac{256 S_{-2,1}}{N^2 (1+N)^2}
\nonumber\\&&			
                        -\frac{96 \big(
                                2+N+N^2\big) \zeta_3}{N^2 (1+N)^2}
                \Biggr]
                +\biggl[
                        -\frac{8 P_{132}}{N^3 (1+N)^3}
                        +\frac{8 \big(
                                -2+3 N+3 N^2\big)}{N^3 (1+N)^2} S_ 2
\nonumber\\&&				
                        -\frac{16 (N-1) (2+N) S_ 3}{3 N^2 (1+N)^2}
                        -\frac{32 (N-1) (2+N) S_{2,1}}{N^2 (1+N)^2}
                \biggr] S_ 1
\nonumber\\&&		
                +\biggl[
                        \frac{4 \big(
                                -36-22 N-6 N^2+N^3\big)}{N^3 (1+N)^2}
                        -\frac{4 (N-1) (2+N) S_ 2}{N^2 (1+N)^2}
                \biggr] S_ 1^2
\nonumber\\&&		
                +\frac{8 \big(
                        -2+3 N+3 N^2\big)}{3 N^3 (1+N)^2} S_ 1^3
                -\frac{2 (N-1) (2+N) S_ 1^4}{3 N^2 (1+N)^2}
                -\frac{2 (N-1) (2+N) S_ 2^2}{N^2 (1+N)^2}
\nonumber\\&&		
                +\frac{12 (N-1) (2+N) S_ 4}{N^2 (1+N)^2}
                -\frac{32 (2+N) S_{2,1}}{N^3 (1+N)}
                -\frac{32 (N-1) (2+N) S_{3,1}}{N^2 (1+N)^2}
\nonumber\\&&		
                +\frac{64 (N-1) (2+N) S_{2,1,1}}{N^2 (1+N)^2}
                +\biggl[
			128 \ln(2)
                        -\frac{2 P_{170}}{N^4 (1+N)^4}
\nonumber\\&&			
                        -\frac{4 (N-1) (2+N) \big(
                                -4-3 N+3 N^2\big)}{N^3 (1+N)^3} S_ 1
                        -\frac{4 (N-1) (2+N) S_ 1^2}{N^2 (1+N)^2}
\nonumber\\&&			
                        +
                        \frac{12 (N-1) (2+N) S_ 2}{N^2 (1+N)^2}
                \biggr] \zeta_2
                +\biggl[
                        -\frac{4 P_{158}}{3 N^3 (1+N)^3}
                        -\frac{16 (N-1) (2+N) S_ 1}{3 N^2 (1+N)^2}
                \biggr] \zeta_3
	\biggr\}
\nonumber\\&&	
	+\textcolor{blue}{C_A C_F T_F} \biggl\{
                -\frac{4 S_ 2 P_{148}}{N^4 (1+N)^4}
                +\frac{P_{179}}{18 N^6 (1+N)^6}
\nonumber\\&&		
                + L_M ^3 \biggl[
                        -\frac{8 (N-1) (2+N) \big(
                                -12+11 N+11 N^2\big)}{9 N^3 (1+N)^3}
                        -\frac{16 (N-1) (2+N) S_ 1}{3 N^2 (1+N)^2}
                \biggr]
\nonumber\\&&		
                + L_M ^2 \biggl[
                        -\frac{8 S_ 1 P_{150}}{3 N^3 (1+N)^3}
                        -\frac{2 P_{169}}{9 N^4 (1+N)^4}
                        -\frac{16 (N-1) (2+N) S_ 2}{N^2 (1+N)^2}
\nonumber\\&&			
                        -\frac{32 (N-1) (2+N) S_{-2}}{N^2 (1+N)^2}
                \biggr]
                + L_M  \Biggl[
                        \frac{4 S_ 2 P_{135}}{N^3 (1+N)^3}
                        +\frac{8 P_{182}}{27 (N-1) N^5 (1+N)^5 
(2+N)}
\nonumber\\&&
                        +\biggl[
                                -\frac{8 P_{175}}{9 (N-1) N^4 
(1+N)^4 (2+N)}
                                -\frac{40 (N-1) (2+N) S_ 2}{N^2 
(1+N)^2}
                        \biggr] S_ 1
\nonumber\\&&			
                        -\frac{4 \big(
                                -12-16 N+5 N^2+11 N^3\big)}{3 N^3 
(1+N)^2} S_ 1^2
                        +\frac{8 (N-1) (2+N) S_ 1^3}{3 N^2 (1+N)^2}
                        -\frac{16 \big(
                                26+5 N+5 N^2\big)}{3 N^2 (1+N)^2} S_3
\nonumber\\&&				
                        +\biggl[
                                -\frac{16 P_{146}}{(N-1) N^2 (1+N)^3 
(2+N)}
                                +\frac{32 P_{130}}{(N-1) N^2 (1+N)^2 
(2+N)}
                                 S_ 1
                        \biggr] S_{-2}
                        +
\nonumber\\&&			
                        \frac{16 \big(
                                -22+5 N+5 N^2\big)}{N^2 (1+N)^2} S_{-3}
                        +\frac{32 (N-1) (2+N) S_{2,1}}{N^2 (1+N)^2}
                        -\frac{32 \big(
                                -14+N+N^2\big) S_{-2,1}}{N^2 (1+N)^2}
\nonumber\\&&				
                        +\biggl[
                                -\frac{32 (-3+N) (4+N)}{N^2 (1+N)^2}
                                +64 S_ 1
                        \biggr] \zeta_3
                \Biggr]
                +\biggl[
                        -\frac{2 P_{165}}{9 N^2 (1+N)^5}
                        +\frac{8 \big(
                                -13+3 N^2\big)}{N^2 (1+N)^3} S_ 2
\nonumber\\&&				
                        +\frac{160 (N-1) (2+N) S_ 3}{3 N^2 (1+N)^2}
                        -\frac{64 (N-1) (2+N) S_{-2,1}}{N^2 (1+N)^2}
                \biggr] S_ 1
                +\biggl[
                        -\frac{4 P_{131}}{N^2 (1+N)^4}
\nonumber\\&&			
                        +\frac{20 (N-1) (2+N) S_ 2}{N^2 (1+N)^2}
                \biggr] S_ 1^2
                -\frac{8 \big(
                        5+4 N+N^2\big)}{3 N^2 (1+N)^3} S_ 1^3
                +\frac{2 (N-1) (2+N) S_ 1^4}{3 N^2 (1+N)^2}
\nonumber\\&&		
                +\frac{2 (N-1) (2+N) S_ 2^2}{N^2 (1+N)^2}
                -\frac{64 \big(
                        -6+8 N+7 N^2+N^3\big)}{3 N^3 (1+N)^3} S_ 3
                +\frac{36 (N-1) (2+N) S_ 4}{N^2 (1+N)^2}
\nonumber\\&&		
                +\biggl[
                        -\frac{32 (2+N) \big(
                                3+N^2\big)}{N^2 (1+N)^4}
                        +\frac{64 (N-1) (2+N) S_ 1}{N^2 (1+N)^3}
                        +\frac{32 (N-1) (2+N) S_ 1^2}{N^2 (1+N)^2}
\nonumber\\&&			
                        +\frac{32 (N-1) (2+N) S_ 2}{N^2 (1+N)^2}
                \biggr] S_{-2}
                +\biggl[
                        \frac{32 (N-1) (2+N)}{N^2 (1+N)^3}
                        +\frac{32 (N-1) (2+N) S_ 1}{N^2 (1+N)^2}
                \biggr] S_{-3}
\nonumber\\&&		
                +\frac{16 (N-1) (2+N) S_{-4}}{N^2 (1+N)^2}
                -\frac{16 (N-1) (2+N) S_{3,1}}{N^2 (1+N)^2}
                -\frac{64 (N-1) (2+N) S_{-2,1}}{N^2 (1+N)^3}
\nonumber\\&&		
                -\frac{32 (N-1) (2+N) S_{-2,2}}{N^2 (1+N)^2}
                -\frac{32 (N-1) (2+N) S_{-3,1}}{N^2 (1+N)^2}
                -\frac{16 (N-1) (2+N) S_{2,1,1}}{N^2 (1+N)^2}
\nonumber\\&&		
                +\frac{64 (N-1) (2+N) S_{-2,1,1}}{N^2 (1+N)^2}
                +\biggl[
			-64 \ln(2)
                        -\frac{4 S_ 1 P_{161}}{3 N^3 (1+N)^3}
                        +\frac{4 P_{168}}{9 N^4 (1+N)^4}
\nonumber\\&&			
                        +\frac{4 (N-1) (2+N) S_ 1^2}{N^2 (1+N)^2}
                        +\frac{12 (N-1) (2+N) S_ 2}{N^2 (1+N)^2}
                        +\frac{24 (N-1) (2+N) S_{-2}}{N^2 (1+N)^2}
                \biggr] \zeta_2
\nonumber\\&&		
                +\biggl[
                        \frac{8 P_{156}}{9 N^3 (1+N)^3}
                        +\frac{16 (N-1) (2+N) S_ 1}{3 N^2 (1+N)^2}
                \biggr] \zeta_3
	\biggr\}
	-\frac{64}{27}  \textcolor{blue}{T_F ^3} \zeta_3
+a_{gg,Q}^{(3)}
\Biggr\}.
\end{eqnarray}
%--------------------------------------------------------------------------------------------------------------
The polynomials $P_i$ read
%--------------------------------------------------------------------------------------------------------------
\begin{eqnarray}
P_{130} &=& N^4+2 N^3-7 N^2-8 N+28, \\ 
P_{131} &=& N^4+2 N^3-5 N^2-12 N+2, \\ 
P_{132} &=& 2 N^4-4 N^3-3 N^2+20 N+12, \\ 
P_{133} &=& 3 N^4+6 N^3-89 N^2-92 N+12, \\ 
P_{134} &=& 3 N^4+6 N^3+16 N^2+13 N-3, \\ 
P_{135} &=& 3 N^4+32 N^3+65 N^2-16 N-60, \\ 
P_{136} &=& 3 N^4+48 N^3+123 N^2+98 N+8, \\ 
P_{137} &=& 9 N^4+18 N^3+113 N^2+104 N-24, \\ 
P_{138} &=& 11 N^4+36 N^3+43 N^2+46 N+8, \\ 
P_{139} &=& 20 N^4+40 N^3+11 N^2-9 N+54, \\ 
P_{140} &=& 23 N^4+46 N^3+23 N^2+96 N+48, \\ 
P_{141} &=& 40 N^4+74 N^3+25 N^2-9 N+16, \\ 
P_{142} &=& 40 N^4+80 N^3+73 N^2+33 N+54, \\ 
P_{143} &=& 63 N^4+126 N^3+271 N^2+208 N-48, \\
P_{144} &=& 99 N^4+198 N^3+463 N^2+364 N-84, \\
P_{145} &=& 136 N^4+254 N^3+37 N^2-81 N+144, \\
P_{146} &=& 3 N^5+7 N^4-29 N^3-51 N^2-2 N-8, \\ 
P_{147} &=& N^6+3 N^5+5 N^4+N^3-8 N^2+2 N+4, \\ 
P_{148} &=& N^6+18 N^5+63 N^4+84 N^3+30 N^2-64 N-16, \\ 
P_{149} &=& 3 N^6+9 N^5-163 N^4-341 N^3+164 N^2-432 N-192, \\ 
P_{150} &=& 3 N^6+9 N^5+20 N^4+25 N^3-11 N^2-46 N-12, \\ 
P_{151} &=& 3 N^6+30 N^5+107 N^4+124 N^3+48 N^2+20 N+8, \\ 
P_{152} &=& 6 N^6+23 N^5-14 N^4-121 N^3-114 N^2-20 N+8, \\ 
P_{153} &=& 15 N^6+45 N^5+49 N^4-13 N^3-64 N^2+40 N+48, \\
P_{154} &=& 15 N^6+45 N^5+56 N^4+N^3-68 N^2+29 N+42, \\ 
P_{155} &=& 15 N^6+45 N^5+374 N^4+601 N^3+161 N^2-24 N+36, \\ 
P_{156} &=& 18 N^6+54 N^5+65 N^4+40 N^3-23 N^2-34 N+24, \\ 
P_{157} &=& 20 N^6+60 N^5+11 N^4-78 N^3-13 N^2+36 N-108, \\ 
P_{158} &=& 24 N^6+72 N^5+69 N^4+18 N^3+N^2+4 N+4, \\ 
P_{159} &=& 27 N^6+81 N^5-1247 N^4-2341 N^3-720 N^2+32 N-240, \\ 
P_{160} &=& 27 N^6+81 N^5+148 N^4+161 N^3+253 N^2-390 N-144, \\ 
P_{161} &=& 30 N^6+90 N^5+79 N^4+8 N^3+23 N^2+70 N+12, \\ 
P_{162} &=& 63 N^6+189 N^5+157 N^4+35 N^3+80 N^2+4 N-24, \\ 
P_{163} &=& 95 N^6+285 N^5+92 N^4-291 N^3-97 N^2+96 N-36, \\ 
P_{164} &=& 297 N^6+891 N^5-461 N^4-2119 N^3-872 N^2-96 N-432, \\ 
P_{165} &=& 233 N^7+1093 N^6+1970 N^5+1538 N^4-167 N^3
\nonumber\\&&
-2143 N^2-2412 N-288, \\ 
P_{166} &=& -15 N^8-60 N^7-82 N^6-44 N^5-15 N^4-4 N^2-12 N-8, \\ 
P_{167} &=& 3 N^8+12 N^7+2080 N^6+5568 N^5+4602 N^4+1138 N^3
\nonumber\\&&
-3 N^2-36 N-108, \\ 
P_{168} &=& 15 N^8+60 N^7+242 N^6+417 N^5+344 N^4+285 N^3
\nonumber\\&&
+185 N^2+456 N+108, \\ 
P_{169} &=& 33 N^8+132 N^7-82 N^6-840 N^5-571 N^4+564 N^3+308 N^2
\nonumber\\&&
+984 N+288, \\
P_{170} &=& 40 N^8+160 N^7+205 N^6+61 N^5+18 N^4+113 N^3+75 N^2
\nonumber\\&&
-20 N-12, \\ 
P_{171} &=& 67 N^8+268 N^7+194 N^6-508 N^5-533 N^4+480 N^3
\nonumber\\&&
+616 N^2+344 N+144, \\ 
P_{172} &=& 126 N^8+504 N^7-1306 N^6-5052 N^5-4473 N^4-1138 N^3
\nonumber\\&&
+3 N^2+36 N+108, \\ 
P_{173} &=& 219 N^8+876 N^7+1142 N^6+288 N^5-217 N^4+240 N^3
\nonumber\\&&
+410 N^2+366 N+180, \\ 
P_{174} &=& 1386 N^8+5544 N^7-11270 N^6-46284 N^5-39915 N^4
\nonumber\\&&
-9422 N^3+33 N^2+396 N+1188, \\ 
P_{175} &=& 15 N^{10}+75 N^9+2 N^8-469 N^7-506 N^6+524 N^5
\nonumber\\&&
+781 N^4+26 N^3-1192 N^2-1128 N-432, \\ 
P_{176} &=& 310 N^{10}+1748 N^9+4811 N^8+14192 N^7+24974 N^6+3194 N^5
\nonumber\\&&
-29393 N^4-16866 N^3+8694 N^2+7128 N+1944, \\ 
P_{177} &=& 391 N^{10}+1955 N^9+3622 N^8+3046 N^7+1595 N^6+1327 N^5
\nonumber\\&&
+1152 N^4+216 N^3-288 N^2-360 N-144, \\ 
P_{178} &=& 1593 N^{10}+7965 N^9+11578 N^8+1594 N^7-1379 N^6+12793 N^5
\nonumber\\&&
+17152 N^4+4432 N^3-1728 N^2-2160 N-864, \\ 
P_{179} &=& -3135 N^{12}-18810 N^{11}-42713 N^{10}-44692 N^9-22145 N^8
\nonumber\\&&
-9290 N^7-8167 N^6-4136 N^5-960 N^4+11232 N^3
\nonumber\\&&
+6720 N^2+3360 N+576, \\ 
P_{180} &=& -39 N^{12}-234 N^{11}-521 N^{10}-492 N^9-85 N^8-42 N^7-883 N^6
\nonumber\\&&
-1660 N^5-1324 N^4-492 N^3-52 N^2+48 N+16, \\ 
P_{181} &=& N^{12}+6 N^{11}-3 N^{10}-58 N^9-21 N^8+222 N^7+609 N^6
\nonumber\\&&
+1144 N^5+1122 N^4+142 N^3-180 N^2+40 N+48, \\ 
P_{182} &=& 276 N^{12}+1656 N^{11}+3334 N^{10}+869 N^9-6591 N^8-7395 N^7
\nonumber\\&&
+7452 N^6+13479 N^5+3167 N^4+7303 N^3+1110 N^2-5004 N
\nonumber\\&&
-2376, \\ 
P_{183} &=& 2493 N^{12}+14958 N^{11}+42317 N^{10}+75910 N^9+45511 N^8-60782 N^7
\nonumber\\&&
-29777 N^6+17194 N^5-130384 N^4-115536 N^3+25776 N^2
\nonumber\\&&
+24192 N+5184.
\end{eqnarray}
%--------------------------------------------------------------------------------------------------------------
Next we turn to the polarized massive Wilson coefficients in the asymptotic region $Q^2 \gg m^2$ in the single mass case.
 
%----------------------------
%-----------------------------------------------------------------------------------------------------------------
\section{The polarized Wilson coefficients \boldmath $L_q^{\rm PS}$ and $L_g^{\rm S}$}
\label{sec:6}
%-----------------------------------------------------------------------------------------------------------------

\vspace*{1mm}\noindent
The massive Wilson coefficients (\ref{eqWIL2}) and (\ref{eqWIL3}) are related to the expansion coefficients of the 
massive OMEs and the massless Wilson coefficients.
In presenting the Wilson coefficients for the structure function $g_1(x,Q^2)$ we leave the 3--loop massive OMEs, 
calculated in the previous sections
and Ref.~\cite{Ablinger:2019etw} and the 3--loop contribution to the polarized massless Wilson 
coefficients symbolic. The latter depend also on the logarithmic terms $L_Q$ (\ref{eq:LOGS}).
In the non--singlet case the complete asymptotic Wilson coefficient to the structure 
function $g_1(x,Q^2)$ has already been calculated in Ref.~\cite{Behring:2015zaa}.

For $L_{q}^{\sf PS}$ we obtain
\begin{eqnarray}
\lefteqn{L_q^{PS} = \frac{1}{2}\left[1-(-1)^N\right] }
\nonumber\\&&
	\times\Bigg\{
	a_s^3 \Biggl\{
	 \textcolor{blue}{ C_F   N_F   T_F ^2} \biggl\{
                -\frac{32 (N-1)^2 (2+N) \big(
                        22+41 N+28 N^2\big)}{27 N^3 (1+N)^4}
\nonumber\\&&
                + L_M  \Biggl[
                        -\frac{64 (N-1)^2 (2+N) (2+5 N)}{9 N^3 
(1+N)^3}
                        +(N-1) \biggl[
                                -\frac{64 (2+N) \big(
                                        3+2 N+2 N^2\big)}{9 N^3 
(1+N)^3} S_1
\nonumber\\&&
                                +\frac{64 (2+N) S_1^2}{3 N^2 (1+N)^2}
                        \biggr]
                \Biggr]
                + L_M ^2 \biggl[
                        -\frac{32 (N-1)^2 (2+N)}{3 N^3 (1+N)^2}
                        -\frac{32 (N-1) (2+N) S_1}{3 N^2 (1+N)^2}
                \biggr]
\nonumber\\&&
                +(N-1) \Biggl[
                         L_Q  \Biggl[
                                \frac{32  L_M ^2 (2+N)}{3 N^2 
(1+N)^2}
                                +\frac{32 (2+N) \big(
                                        22+41 N+28 N^2\big)}{27 N^2 
(1+N)^4}
                                + L_M  \biggl[
                                        \frac{64 (2+N) (2+5 N)}{9 N^2 
(1+N)^3}
\nonumber\\&&
                                        -\frac{64 (2+N) S_1}{3 N^2 
(1+N)^2}
                                \biggr]
                                -\frac{32 (2+N) (2+5 N)}{9 N^2 
(1+N)^3} S_1
                                +\frac{16 (2+N) S_1^2}{3 N^2 (1+N)^2}
                                +\frac{16 (2+N) S_2}{3 N^2 (1+N)^2}
                        \Biggr]
\nonumber\\&&
                        +\biggl[
                                -\frac{32 (2+N) \big(
                                        6+37 N+35 N^2+13 N^3\big)}{27 
N^3 (1+N)^4}
                                -\frac{16 (2+N) S_2}{3 N^2 (1+N)^2}
                        \biggr] S_1
\nonumber\\&&
                        +\frac{16 (2+N) \big(
                                3+4 N+7 N^2\big)}{9 N^3 (1+N)^3} 
S_1^2
                        -\frac{16 (2+N) S_1^3}{3 N^2 (1+N)^2}
                \Biggr]
                -\frac{16 (N-1)^2 (2+N)}{3 N^3 (1+N)^2} S_2
	\biggr\}
\nonumber\\&&
+A_{qq,Q}^{PS(3)}
+ \textcolor{blue}{N_F} \hat{\tilde{C}}_q^{PS(3)}\left(L_Q,N_F\right)
\Biggr\}
\Bigg\}.
\end{eqnarray}

$L_{g}^{\sf S}$ is given by
%--------------------------------------------------------------------------------
\begin{eqnarray}
\lefteqn{ L_g^S = \frac{1}{2}\left[1-(-1)^N\right] }
\nonumber\\&&
	\times\Bigg\{
	a_s^2  \textcolor{blue}{N_F   T_F ^2} \Biggl\{
        \frac{16  L_M   L_Q  (N-1)}{3 N (1+N)}
        + L_M  \biggl[
                -\frac{16 (N-1)^2}{3 N^2 (1+N)}
                -\frac{16 (N-1) S_1}{3 N (1+N)}
        \biggr]
	\Biggr\}
\nonumber\\&&
	+a_s^3 \Biggl\{
         \textcolor{blue}{N_F   T_F ^3} \biggl\{
                \frac{64  L_M ^2  L_Q  (N-1)}{9 N (1+N)}
                + L_M ^2 \biggl[
                        -\frac{64 (N-1)^2}{9 N^2 (1+N)}
                        -\frac{64 (N-1) S_1}{9 N (1+N)}
                \biggr]
	\biggr\}
\nonumber\\&&
	+ \textcolor{blue}{C_A   N_F   T_F ^2} \biggl\{
                -\frac{8 (N-1)^2 Q_9}{27 N^5 (1+N)^4}
                + L_Q  \Biggl[
                        (N-1) \Biggl[
                                \frac{8 Q_9}{27 N^4 (1+N)^4}
                                + L_M ^2 \biggl[
                                        \frac{64}{3 N^2 (1+N)^2}
\nonumber\\&&
                                        -\frac{32 S_1}{3 N (1+N)}
                                \biggr]
                                -\frac{16 (47+56 N) S_1}{27 N 
(1+N)^2}
                        \Biggr]
                        + L_M  \Biggl[
                                \frac{32 Q_5}{9 N^3 (1+N)^3}
                                +(N-1) \biggl[
                                        \frac{32 S_1^2}{3 N (1+N)}
\nonumber\\&&
                                        -\frac{32 S_2}{3 N (1+N)}
                                        -\frac{64 S_{-2}}{3 N (1+N)}
                                \biggr]
                                -\frac{64 \big(
                                        -9-2 N+3 N^2+2 N^3\big)}{9 
N^2 (1+N)^2} S_1
                        \Biggr]
                \Biggr]
                + L_M ^2 \Biggl[
                        -\frac{64 (N-1)^2}{3 N^3 (1+N)^2}
\nonumber\\&&
                        +(N-1) \biggl[
                                \frac{32 \big(
                                        -3+N^2\big) S_1}{3 N^2 (1+N)^2}
                                +\frac{32 S_1^2}{3 N (1+N)}
                        \biggr]
                \Biggr]
                +(N-1) \Biggl[
                        \frac{8 S_1 Q_{10}
                        }{27 N^4 (1+N)^4}
\nonumber\\&&
                        + L_M   L_Q ^2 \biggl[
                                \frac{64}{3 N^2 (1+N)^2}
                                -\frac{32 S_1}{3 N (1+N)}
                        \biggr]
                        +\frac{16 (47+56 N) S_1^2}{27 N (1+N)^2}
                \Biggr]
\nonumber\\&&
                + L_M  \Biggl[
                        -\frac{16 Q_{13}}{9 (N-1) N^4 (1+N)^4 (2+N)^2}
                        +(N-1) \biggl[
                                -\frac{16 S_1^3}{9 N (1+N)}
                                -\frac{64 S_{2,1}}{3 N (1+N)}
                        \biggr]
\nonumber\\&&
                        +\biggl[
                                -\frac{16 Q_8}{9 N^3 (1+N)^3 (2+N)}
                                +\frac{80 (N-1) S_2}{3 N (1+N)}
                        \biggr] S_1
                        +\frac{32 \big(
                                -9-10 N+3 N^2+7 N^3\big)}{9 N^2 
(1+N)^2} S_1^2
\nonumber\\&&
                        +\frac{32 \big(
                                3-2 N+N^2+N^3\big)}{3 N^2 (1+N)^2} 
S_2
                        +\frac{32 \big(
                                2+5 N+5 N^2\big)}{9 N (1+N) (2+N)} 
S_3
\nonumber\\&&
                        +\biggl[
                                \frac{64 Q_3}{3 (N-1) N (1+N)^2 
(2+N)^2}
                                +\frac{64 \big(
                                        2+N+N^2\big)}{3 N (1+N) 
(2+N)} S_1
                        \biggr] S_{-2}
                        -\frac{64 \big(
                                -4+N+N^2\big)}{3 N (1+N) (2+N)} 
S_{-3}
\nonumber\\&&
                        -\frac{256 S_{-2,1}}{3 N (1+N) (2+N)}
                        -\frac{32 \big(
                                2+N+N^2\big)}{N (1+N) (2+N)} \zeta_3
                \Biggr]
	\biggr\}
	+ \textcolor{blue}{C_F  N_F   T_F ^2} \biggl\{
                \frac{4 (N-1)^2 Q_{12}}{N^6 (1+N)^5}
\nonumber\\&&
                +(N-1) \Biggl[
                        \frac{4 S_1 Q_{12}}{N^5 (1+N)^5}
                        + L_M   L_Q ^2 \biggl[
                                \frac{8 Q_2}{3 N^3 (1+N)^3}
                                -\frac{32 S_1}{3 N (1+N)}
                        \biggr]
                \Biggr]
\nonumber\\&&
                + L_Q  \Biggl[
                        \frac{16  L_M 
                        ^2 (N-1)^2 (2+N)}{N^3 (1+N)^3}
                        +(N-1) \Biggl[
                                -\frac{4 Q_{12}}{N^5 (1+N)^5}
                                + L_M  \biggl[
                                        \frac{16 S_1 Q_1}{3 N^3 
(1+N)^3}
\nonumber\\&&
                                        -\frac{16 Q_6}{3 N^4 (1+N)^4}
                                        +\frac{64 S_1^2}{3 N (1+N)}
                                        -\frac{64 S_2}{3 N (1+N)}
                                \biggr]
                        \Biggr]
                \Biggr]
                + L_M ^2 \biggl[
                        -\frac{16 (N-1)^3 (2+N)}{N^4 (1+N)^3}
\nonumber\\&&
                        -\frac{16 (N-1)^2 (2+N)}{N^3 (1+N)^3} S_1
                \biggr]
                + L_M  \Biggl[
                        \frac{8 S_2 Q_4}{3 N^3 (1+N)^3}
                        -\frac{8 Q_{14}}{3 (N-1) N^5 (1+N)^5 
(2+N)^2}
\nonumber\\&&
                        +(N-1) \biggl[
                                -\frac{8 \big(
                                        -4+2 N+3 N^2
                                \big)
\big(3+2 N+3 N^2\big)}{3 N^3 (1+N)^3} S_1^2
                                -\frac{80 S_1^3}{9 N (1+N)}
                                +\frac{64 S_{2,1}}{3 N (1+N)}
                        \biggr]
\nonumber\\&&
                        +\biggl[
                                -\frac{32 Q_{11}}{3 N^4 (1+N)^4 
(2+N)}
                                +\frac{16 (N-1) S_2}{N (1+N)}
                        \biggr] S_1
                        -\frac{256 \big(
                                1+N+N^2\big)}{9 N (1+N) (2+N)} S_3
\nonumber\\&&
                        +\biggl[
                                \frac{64 Q_7}{3 (N-1) N^2 (1+N)^2 
(2+N)^2}
                                -\frac{512 S_1}{3 N (1+N) (2+N)}
                        \biggr] S_{-2}
                        -\frac{256 S_{-3}}{3 N (1+N) (2+N)}
\nonumber\\&&
                        +\frac{512 S_{-2,1}}{3 N (1+N) (2+N)}
                        +\frac{64 \big(
                                2+N+N^2\big)}{N (1+N) (2+N)} \zeta_3
                \Biggr]
	\biggr\}
+A_{qg,Q}^{(3)}
+ \textcolor{blue}{N_F} \hat{\tilde{C}}_g^{S(3)}\left(L_Q,N_F\right)   
\Biggr\}
\Bigg\},
\end{eqnarray}
%--------------------------------------------------------------------------------
with the polynomials $Q_i$ given by
%--------------------------------------------------------------------------------
\begin{eqnarray}
Q_1&=&3 N^4+2 N^3-N^2-12, \\
Q_2&=&3 N^4+6 N^3-N^2-4 N+12, \\
Q_3&=&N^5+3 N^4-3 N^3-9 N^2-8 N-8, \\
Q_4&=&9 N^5-N^4-23 N^3-15 N^2-14 N+12, \\
Q_5&=&9 N^5+9 N^4-4 N^3+15 N^2-41 N-12, \\
Q_6&=&N^6-19 N^4-22 N^3+22 N^2-34 N-36, \\
Q_7&=&N^6+3 N^5+13 N^4+21 N^3+22 N^2+12 N-24, \\
Q_8&=&13 N^6+44 N^5+71 N^4+94 N^3-90 N^2-288 N-72, \\
Q_9&=&15 N^6+45 N^5+374 N^4+601 N^3+161 N^2-24 N+36, \\
Q_{10}&=&97 N^6+161 N^5-392 N^4-807 N^3-255 N^2+24 N-36, \\
Q_{11}&=&N^8+3 N^7+11 N^6+20 N^5-15 N^4-17 N^3+49 N^2-20 N-36 
\\
Q_{12}&=&15 N^8+60 N^7+82 N^6+44 N^5+15 N^4+4 N^2+12 N+8, \\
Q_{13}&=&24 N^{10}+102 N^9+58 N^8-210 N^7-209 N^6+23 N^5+529 N^4
\nonumber\\&&
+1109 N^3+234 N^2-388 N-120, \\
Q_{14}&=&8 N^{12}+50 N^{11}+122 N^{10}+98 N^9-457 N^8-1398 N^7-1232 N^6
\nonumber\\&&
-634 N^5-793 N^4+388 N^3+1128 N^2-16 N-336.
\end{eqnarray}
%--------------------------------------------------------------------------------

%-----------------------------------------------------------------------------------------------------------------
\section{The polarized Wilson coefficients \boldmath $H_{Qq}^{(3),\rm PS}$}
\label{sec:7}
%-----------------------------------------------------------------------------------------------------------------

\vspace*{1mm}\noindent
The next Wilson coefficient is $H_{Qq}^{(3),\rm PS}$. It is given by
\begin{eqnarray}
	\lefteqn{ H_q^{PS} = \frac{1}{2} \left[ 1-(-1)^N \right] }
\nonumber\\&&
	\times \Bigg\{
	a_s^2  \textcolor{blue}{C_F   T_F}  \Biggl\{
        \frac{8 Q_{16}}{(N-1) N^4 (1+N)^4 (2+N)}
        +(2+N) \Biggl[
                \frac{8  L_M  \big(
                        1+2 N+N^3\big)}{N^3 (1+N)^3}
\nonumber\\&&			
                + L_Q  \biggl[
                        -\frac{8 \big(
                                2+N-N^2+2 N^3\big)}{N^3 (1+N)^3}
                        -\frac{8 (N-1) S_1}{N^2 (1+N)^2}
                \biggr]
                +(N-1) \biggl[
                        -\frac{4  L_M ^2}{N^2 (1+N)^2}
                        +\frac{4  L_Q ^2}{N^2 (1+N)^2}
\nonumber\\&&
                        +\frac{4 S_1^2}{N^2 (1+N)^2}
                        -\frac{12 S_2}{N^2 (1+N)^2}
                \biggr]
                +\frac{8 \big(
                        2+N-N^2+2 N^3\big)}{N^3 (1+N)^3} S_1
        \Biggr]
\nonumber\\&&
        -\frac{64}{(N-1) N (1+N) (2+N)} S_{-2}
	\Biggr\}
+a_s^3 \Biggl\{
		 \textcolor{blue}{C_F ^2  T_F} (2+N) \biggl\{
                        \frac{8 S_2 Q_{15}}{N^4 (1+N)^4}
\nonumber\\&&
                        +\frac{4 \big(
                                -1-3 N-4 N^2+4 N^3
                        \big)
\big(-2+5 N+6 N^2+9 N^3\big)}{N^6 (1+N)^5}
\nonumber\\&&
                        + L_Q  \Biggl[
                                -\frac{4 \big(
                                        2+3 N+3 N^2
                                \big)
\big(-1-3 N-4 N^2+4 N^3\big)}{N^5 (1+N)^5}
\nonumber\\&&
                                + L_M  \biggl[
                                        \frac{8 \big(
                                                2+3 N+3 N^2
                                        \big)
\big(1+2 N+N^3\big)}{N^4 (1+N)^4}
                                        -\frac{32 \big(
                                                1+2 N+N^3\big)}{N^3 
(1+N)^3} S_1
                                \biggr]
\nonumber\\&&
                                +(N-1) \Biggl[
                                         L_M ^2 \biggl[
                                                -\frac{4 \big(
                                                        2+3 N+3 
N^2\big)}{N^3 (1+N)^3}
                                                +\frac{16 S_1}{N^2 
(1+N)^2}
                                        \biggr]
                                        -\frac{8 \big(
                                                2+3 N+3 N^2\big)}
                                        {N^3 (1+N)^3} S_2
                                \Biggr]
\nonumber\\&&
                                +\biggl[
                                        \frac{16 \big(
                                                -1-3 N-4 N^2+4 
N^3\big)}{N^4 (1+N)^4}
                                        +\frac{32 (N-1) S_2}{N^2 
(1+N)^2}
                                \biggr] S_1
                        \Biggr]
\nonumber\\&&
                        + L_M  \biggl[
                                -\frac{8 \big(
                                        1+2 N+N^3
                                \big)
\big(-2+5 N+6 N^2+9 N^3\big)}{N^5 (1+N)^4}
\nonumber\\&&
                                +\frac{8 \big(
                                        -2+3 N+3 N^2
                                \big)
\big(1+2 N+N^3\big)}{N^4 (1+N)^4} S_1
                                +\frac{16 \big(
                                        1+2 N+N^3\big)}{N^3 (1+N)^3} 
S_1^2
                                -\frac{16 \big(
                                        1+2 N+N^3\big)}{N^3 (1+N)^3} 
S_2
                        \biggr]
\nonumber\\&&
                        +(N-1) \Biggl[
                                 L_M ^2 \biggl[
                                        \frac{4 \big(
                                                -2+5 N+6 N^2+9 
N^3\big)}{N^4 (1+N)^3}
                                        -\frac{4 \big(
                                                -2+3 N+3 
N^2\big)}{N^3 (1+N)^3} S_1
                                        -\frac{8 S_1^2}{N^2 (1+N)^2}
\nonumber\\&&
                                        +\frac{8 S_2}{N^2 (1+N)^2}
                                \biggr]
                                +\frac{16 S_2^2}{N^2 (1+N)^2}
                        \Biggr]
                        +\biggl[
                                -\frac{4 \big(
                                        -2+3 N+3 N^2
                                \big)
\big(-1-3 N-4 N^2+4 N^3\big)}{N^5 (1+N)^5}
\nonumber\\&&
                                -\frac{8 (N-1) \big(
                                        -2+3 N+3 N^2\big)}{N^3 
(1+N)^3} S_2
                        \biggr] S_1
                        +\biggl[
                                -\frac{8 \big(
                                        -1-3 N-4 N^2+4 N^3\big)}{N^4 
(1+N)^4}
                                -\frac{16 (N-1) S_2}{N^2 (1+N)^2}
                        \biggr] S_1^2
		\biggr\}
\nonumber\\&&
		+ \textcolor{blue}{C_F   T_F ^2} (2+N) \biggl\{
                        -\frac{32 (N-1)^2 \big(
                                22+41 N+28 N^2\big)}{27 N^3 (1+N)^4}
                        + L_M  \Biggl[
                                -\frac{64 (N-1)^2 (2+5 N)}{9 N^3 
(1+N)^3}
\nonumber\\&&
                                +(N-1) \biggl[
                                        -\frac{64 \big(
                                                3+2 N+2 N^2\big)}
                                        {9 N^3 (1+N)^3} S_1
                                        +\frac{64 S_1^2}{3 N^2 (1+N)^2}
                                \biggr]
                        \Biggr]
                        + L_M ^2 \biggl[
                                -\frac{32 (N-1)^2}{3 N^3 (1+N)^2}
\nonumber\\&&
                                -\frac{32 (N-1) S_1}{3 N^2 (1+N)^2}
                        \biggr]
                        +(N-1) \Biggl[
                                 L_Q  \Biggl[
                                        \frac{32  L_M ^2}{3 N^2 
(1+N)^2}
                                        +\frac{32 \big(
                                                22+41 N+28 
N^2\big)}{27 N^2 (1+N)^4}
\nonumber\\&&
                                        + L_M  \biggl[
                                                \frac{64 (2+5 N)}{9 
N^2 (1+N)^3}
                                                -\frac{64 S_1}{3 N^2 
(1+N)^2}
                                        \biggr]
                                        -\frac{32 (2+5 N) S_1}{9 N^2 
(1+N)^3}
                                        +\frac{16 S_1^2}{3 N^2 (1+N)^2}
                                        +\frac{16 S_2}{3 N^2 (1+N)^2}
                                \Biggr]
\nonumber\\&&
                                +\biggl[
                                        -\frac{32 \big(
                                                6+37 N+35 N^2+13 
N^3\big)}{27 N^3 (1+N)^4}
                                        -\frac{16 S_2}{3 N^2 (1+N)^2}
                                \biggr] S_1
                                +\frac{16 \big(
                                        3+4 N+7 N^2\big)}{9 N^3 
(1+N)^3} S_1^2
\nonumber\\&&
                                -\frac{16 S_1^3}{3 N^2 (1+N)^2}
                        \Biggr]
                        -\frac{16 (N-1)^2 S_2}{3 N^3 (1+N)^2}
		\biggr\}
+A_{Qq}^{PS(3)}
+\tilde{C}_q^{PS(3)}\left(L_Q,N_F+1\right)
\Biggr\}
\Bigg\},
\end{eqnarray}
where
\begin{eqnarray}
Q_{15}&=&9 N^5+6 N^4-12 N^2-8 N+1, \\
Q_{16}&=&3 N^8+10 N^7-N^6-22 N^5-14 N^4-18 N^3-30 N^2+8.
\end{eqnarray}

%-----------------------------------------------------------------------------------------------------------------
The polarized massive OME $A_{Qq}^{(3),\rm PS}$ has been calculated in Ref.~\cite{Ablinger:2019etw}.
%-----------------------------------------------------------------------------------------------------------------
\section{The polarized Wilson coefficient \boldmath $H_{Qg}^{(3),\rm S}$} 
\label{sec:8}
%-----------------------------------------------------------------------------------------------------------------

\vspace*{1mm}\noindent
Finally, $H_{Qg}^{(3),\rm S}$, (\ref{eqWIL4}), is obtained by
%-------------------------------------------------------------------------------------------

%-------------------------------------------------------------------------------------------
with the polynomials $Q_i$
%-------------------------------------------------------------------------------------------
\begin{eqnarray}
Q_{17}&=&N^4-22 N^3-79 N^2-72 N-4, \\
Q_{18}&=&N^4+3 N^3-4 N^2-8 N-4, \\
Q_{19}&=&N^4+3 N^3-2 N^2+3 N+3, \\
Q_{20}&=&N^4+17 N^3+43 N^2+33 N+2, \\
Q_{21}&=&2 N^4-N^3-24 N^2-17 N+28, \\
Q_{22}&=&3 N^4+6 N^3+2 N^2-N+6, \\
Q_{23}&=&3 N^4+6 N^3+4 N^2+N-6, \\
Q_{24}&=&3 N^4+6 N^3+11 N^2+8 N-12, \\
Q_{25}&=&3 N^4+18 N^3+47 N^2+56 N-4, \\
Q_{26}&=&3 N^4+42 N^3+71 N^2+8 N-28, \\
Q_{27}&=&4 N^4+5 N^3+3 N^2-4 N-4, \\
Q_{28}&=&7 N^4+74 N^3+171 N^2+128 N+4, \\
Q_{29}&=&9 N^4+6 N^3-55 N^2-44 N+44, \\
Q_{30}&=&9 N^4+6 N^3-35 N^2-16 N+20, \\
Q_{31}&=&9 N^4+12 N^3+8 N^2+5 N-2, \\
Q_{32}&=&9 N^4+15 N^3+17 N^2+9 N-6, \\
Q_{33}&=&9 N^4+46 N^3+93 N^2+48 N-76, \\
Q_{34}&=&11 N^4+42 N^3+47 N^2+32 N+12, \\
Q_{35}&=&25 N^4+26 N^3-3 N^2-52 N-36, \\
Q_{36}&=&29 N^4+58 N^3+5 N^2-24 N+36, \\
Q_{37}&=&43 N^4+86 N^3+107 N^2-8 N-56, \\
Q_{38}&=&45 N^4+78 N^3+53 N^2+12 N-12, \\
Q_{39}&=&85 N^4+170 N^3+61 N^2-24 N+36, \\
Q_{40}&=&111 N^4+198 N^3+135 N^2+24 N-32, \\
Q_{41}&=&763 N^4+1418 N^3+985 N^2+978 N+72, \\
Q_{42}&=&N^5+N^4-4 N^3+3 N^2-7 N-2, \\
Q_{43}&=&N^5+3 N^4-3 N^3-9 N^2-8 N-8, \\
Q_{44}&=&2 N^5+6 N^4+N^3-6 N^2+11 N+10, \\
Q_{45}&=&4 N^5-13 N^4-114 N^3-241 N^2-144 N+4, \\
Q_{46}&=&4 N^5+13 N^4-19 N^3-69 N^2-15 N-10, \\
Q_{47}&=&5 N^5+39 N^4+118 N^3+171 N^2+93 N+2, \\
Q_{48}&=&9 N^5+9 N^4-4 N^3+15 N^2-41 N-12, \\
Q_{49}&=&9 N^5+15 N^4+8 N^3+3 N^2+7 N-6, \\
Q_{50}&=&15 N^5+36 N^4-97 N^3-366 N^2-264 N+16, \\
Q_{51}&=&439 N^5+439 N^4-937 N^3-367 N^2-582 N-144, \\
Q_{52}&=&815 N^5+923 N^4-1349 N^3-491 N^2-258 N-216, \\
Q_{53}&=&N^6-7 N^4-4 N^3+16 N^2-10 N-12, \\
Q_{54}&=&2 N^6+5 N^5-22 N^4-95 N^3-114 N^2-24 N+16, \\
Q_{55}&=&2 N^6+5 N^5-3 N^4-7 N^3+2 N^2-11 N-8, \\
Q_{56}&=&11 N^6+30 N^5+9 N^4-22 N^3-10 N^2+2 N+4, \\
Q_{57}&=&13 N^6+44 N^5+71 N^4+94 N^3-90 N^2-288 N-72, \\
Q_{58}&=&15 N^6+45 N^5+374 N^4+601 N^3+161 N^2-24 N+36, \\
Q_{59}&=&15 N^6+57 N^5+47 N^4-N^3+12 N^2-38 N+4, \\
Q_{60}&=&21 N^6+45 N^5+55 N^4-13 N^3-12 N^2-8 N+8, \\
Q_{61}&=&25 N^6+85 N^5+119 N^4+75 N^3-118 N^2-102 N+44, \\
Q_{62}&=&57 N^6+153 N^5+233 N^4+163 N^3-70 N^2+120 N+144, \\
Q_{63}&=&97 N^6+161 N^5-392 N^4-807 N^3-255 N^2+24 N-36, \\
Q_{64}&=&247 N^6+795 N^5+555 N^4-71 N^3+210 N^2-360 N-432, \\
Q_{65}&=&15 N^7+28 N^6-116 N^5-453 N^4-575 N^3-221 N^2+114 N+88, \\
Q_{66}&=&21 N^7-52 N^6-86 N^5-60 N^4+17 N^3+176 N^2+24 N-40, \\
Q_{67}&=&21 N^7+220 N^6+713 N^5+1132 N^4+1010 N^3+486 N^2+38 N-52, \\
Q_{68}&=&753 N^7+1308 N^6-44 N^5-1118 N^4-1549 N^3-766 N^2-600 N-288, \\
Q_{69}&=&3479 N^7+7444 N^6-5160 N^5-13414 N^4-8111 N^3-5478 N^2
\nonumber\\&&
+1368 N+864, \\
Q_{70}&=&N^8+3 N^7+5 N^6+5 N^5-9 N^4-8 N^3+19 N^2-8 N-12, \\
Q_{71}&=&N^8+14 N^7+88 N^6+229 N^5+202 N^4+47 N^3+77 N^2+14 N+8, \\
Q_{72}&=&2 N^8+10 N^7+22 N^6+36 N^5+29 N^4+4 N^3+33 N^2+12 N+4, \\
Q_{73}&=&4 N^8-43 N^7-277 N^6-500 N^5-308 N^4+25 N^3+245 N^2+102 N-24, \\
Q_{74}&=&11 N^8+55 N^7+99 N^6+119 N^5+90 N^4+2 N^3+136 N^2+48 N+16, \\
Q_{75}&=&12 N^8+52 N^7+60 N^6-25 N^4-2 N^3+3 N^2+8 N+4, \\
Q_{76}&=&15 N^8+60 N^7+82 N^6+44 N^5+15 N^4+4 N^2+12 N+8, \\
Q_{77}&=&21 N^8+101 N^7+193 N^6+321 N^5+528 N^4+550 N^3+302 N^2+88 N+8, \\
Q_{78}&=&N^{10}+3 N^9-15 N^8-56 N^7-8 N^6+90 N^5+60 N^4+67 N^3+86 N^2
\nonumber\\&&
-12 N-24, \\
Q_{79}&=&5 N^{10}+23 N^9+31 N^8-N^7+54 N^6+268 N^5+342 N^4+98 N^3-60 N^2
\nonumber\\&&
-8 N+16, \\
Q_{80}&=&15 N^{10}+87 N^9+154 N^8+96 N^7+18 N^6-64 N^5+47 N^4+153 N^3-22 N^2
\nonumber\\&&
-12 N-8, \\
Q_{81}&=&18 N^{10}+162 N^9+740 N^8+2296 N^7+4511 N^6+5341 N^5+3593 N^4
\nonumber\\&&
+1065 N^3-130 N^2-148 N-8, \\
Q_{82}&=&24 N^{10}+102 N^9+58 N^8-210 N^7-209 N^6+23 N^5+529 N^4+1109 N^3
\nonumber\\&&
+234 N^2-388 N-120, \\
Q_{83}&=&29 N^{10}+229 N^9+620 N^8+1434 N^7+2173 N^6+505 N^5-86 N^4+712 N^3
\nonumber\\&&
+704 N^2-16 N-160, \\
Q_{84}&=&1371 N^{10}+6171 N^9+10220 N^8+5678 N^7-9493 N^6-17113 N^5-9154 N^4
\nonumber\\&&
-10864 N^3-10656 N^2+1872 N+4320, \\
Q_{85}&=&20283 N^{10}+92379 N^9+127804 N^8+11278 N^7-154181 N^6-222809 N^5
\nonumber\\&&
-170666 N^4-111392 N^3-62568 N^2+5040 N+8640, \\
Q_{86}&=&8 N^{11}+42 N^{10}+44 N^9-102 N^8-415 N^7-539 N^6-195 N^5-241 N^4
\nonumber\\&&
-414 N^3+268 N^2+104 N-96. 
\end{eqnarray}

%-----------------------------------------------------------------------------------------------------------------
\section{Conclusions}
\label{sec:9}
%-----------------------------------------------------------------------------------------------------------------

\vspace*{1mm}\noindent
We have calculated the logarithmic contributions to the massive operator matrix elements $A_{ij}$ and the massive
Wilson coefficients in the asymptotic region in the polarized case to 3--loop order, also referring to previous 
results in the literature. These quantities provide important corrections to the deep--inelastic scattering structure
functions $g_1(x,Q^2)$ and $g_2(x,Q^2)$, as well as the matching functions in the polarized variable flavor number scheme
to 3--loop order. Here the OMEs $A_{qq,Q}^{(3), \rm PS}$ and $A_{qg,Q}^{(3), \rm S}$ are new quantities, which haven
been calculated in complete form. At present, except of the massless flavor non--singlet Wilson coefficients 
\cite{Vermaseren:2005qc}, the polarized massless 3--loop Wilson coefficients are not yet known. Therefore, we have 
given them 
only symbolically in the corresponding 
expressions. Similarly, the calculation of the constant part of the unrenormalized polarized massive OMEs  $A_{Qg}^{(3)}$ and  
$A_{gg,Q}^{(3)}$ is still underway. However, the logarithmic contributions are already available for use. These contributions,
unlike the case for the non--logarithmic parts, can all be represented in terms of harmonic sums in Mellin $N$ space or
by harmonic polylogarithms in momentum fraction $z$ space. One has to finally know these expressions to 3--loop order,
because this is requested
by the experimental accuracy expected in future high--luminosity deep--inelastic scattering experiments. Likewise, the
matrix elements for the polarized variable flavor number scheme are needed to describe the decoupling of heavy flavors
at polarized hadron colliders, to give full account on these QCD corrections. Working in the Larin scheme provides 
an 
alternative to other schemes in matching the massive OMEs with the massless Wilson coefficients and polarized parton 
densities, which would also be described in this scheme. This implies a modification for their evolution, 
but maintains their
universality and leads to the same predictions of all observables. The analytic expressions for the massive 
operator matrix elements 
and Wilson coefficients in the asymptotic regions are lengthy, as expected for single differential 3--loop quantities, cf. also
\cite{Vermaseren:2005qc}. For this reason we also provide computer-readable files as attachments to this paper.

In addition to the single mass corrections, on which we have concentrated in the present paper, there are double mass
corrections due to charm and bottom quarks of the polarized massive OMEs starting with 3--loop order. They have been 
calculated for all OMEs but $A_{Qg}^{(3)}$ by now in Refs.~\cite{Ablinger:2017err,Ablinger:2019gpu,Ablinger:2020snj},
in complete analytic form and the calculation of $A_{Qg}^{(3)}$ maintaining the leading terms in the mass expansion 
$m_c^2/m_b^2$ is underway. We also presented the non--singlet OME and the associated Wilson coefficient in the Larin 
scheme since it previously has been given in the $\overline{\sf MS}$ scheme only \cite{Ablinger:2014vwa,Behring:2015zaa}.
%-----------------------------------------------------------------------------------------------------------------
\appendix
%-----------------------------------------------------------------------------------------------------------------
\section{The \boldmath $z$-Space Expressions for the massive OMEs}
\label{sec:A}
%-----------------------------------------------------------------------------------------------------------------

\vspace*{1mm}\noindent
The massive OMEs in $z$-space decompose into the three contributions $A_\delta, A_+$ and $A_{\rm reg}$. 
Their convolutions with regular functions, such as parton distribution functions, are given by
\cite{Blumlein:1989gk}
%-----------------------------------------------------------------------------------------------------------------
\begin{eqnarray}
A_\delta(z) \otimes f(z) &=& f(1)
\\
A_+(z) \otimes f(z) &=& \int_z^1 \frac{dy}{y} A_+(y) \left[f\left(\frac{z}{y}\right) - f(z)\right] - f(z) \int_0^z dy A_+(y) 
\\
A_{\rm reg}(z) \otimes f(z) &=& \int_z^1 \frac{dy}{y} A_{\rm reg}(y) f\left(\frac{z}{y}\right).
\end{eqnarray}
%-----------------------------------------------------------------------------------------------------------------
The following terms can all be expressed by harmonic polylogarithms \cite{Remiddi:1999ew}. They correspond to words
out of letters of the three-letter alphabet 
%-----------------------------------------------------------------------------------------------------------------
\begin{eqnarray}
\mathfrak{A} = \left\{f_0(x) = \frac{1}{x},f_1(x) = \frac{1}{1-x},f_{-1}(x) = \frac{1}{1+x}\right\}
\end{eqnarray}
%-----------------------------------------------------------------------------------------------------------------
and perform algebraic reductions, cf.~\cite{Blumlein:2003gb}. The harmonic polylogarithms are given by
%-----------------------------------------------------------------------------------------------------------------
\begin{eqnarray}
H_{b,\vec{a}}(x) = \int_0^x dy f_a(y) H_{\vec{a}}(y), ~~~H_\emptyset = 1, a, a_i \in \{0,1,-1\}.
\end{eqnarray}
%-----------------------------------------------------------------------------------------------------------------

With the exception of  $A_{gg,Q}^{(3)}$ and  $A_{qq,Q}^{(3),\rm NS}$, all OMEs shown in the following have only 
regular contributions.
For the OME 
$A_{qq,Q}^{(3),\rm PS}$ they read  
%-------------------------
\begin{eqnarray}
	\lefteqn{A_{qq,Q}^{PS}(z) =}
\nonumber\\&&	
	a_s^3 \Biggl\{
		 \textcolor{blue}{C_F N_F T_F^2} \biggl\{
                 L_M ^3 \biggl[
                        \frac{160}{9} (z-1)
                        -\frac{64}{9} (1+z) H_0
                \biggr]
\nonumber\\&&		
                + L_M ^2 \biggl[
                        -\frac{32}{9} (z-1) \big(
                                -1+15 H_1\big)
                        +(1+z) \Bigl(
                                -\frac{32}{3} H_0^2
                                +\frac{64}{3} H_{0,1}
                                -\frac{64}{3} \zeta_2
                        \Bigr)
\nonumber\\&&			
                        -\frac{32}{9} (7+z) H_0
                \biggr]
                + L_M  \Biggl[
                        (1+z) \Bigl(
                                -\frac{176}{9} H_0^2
                                -\frac{32}{9} H_0^3
                                +\frac{128}{3} H_{0,0,1}
                                -\frac{128}{3} H_{0,1,1}
                        \Bigr)
\nonumber\\&&			
                        -\frac{32}{27} (55+31 z) H_0
                        +(z-1) \Bigl(
                                \frac{5312}{27}-\frac{64}{9} 
H_1+\frac{160}{3} H_1^2\Bigr)
                        +\frac{64}{9} (7+z) H_{0,1}
\nonumber\\&&			
                        +\biggl[
                                -\frac{64}{9} (7+z)
                                -\frac{128}{3} (1+z) H_0
                        \biggr] \zeta_2
                \Biggr]
                +(1+z) \Bigl(
                        -\frac{64}{3} H_{0,1,1,1}
                        +\frac{448}{15} \zeta_2^2
                \Bigr)
\nonumber\\&&		
                +\frac{128}{81} (49+82 z) H_0
                +(z-1) \Bigl(
                        -\frac{10880}{81}+\frac{3712}{27} 
H_1-\frac{64}{9} H_1^2+\frac{80}{9} H_1^3\Bigr)
                -\frac{128}{27} (11+14 z) H_{0,1}
\nonumber\\&&		
                +\frac{64}{9} (2+5 z) H_{0,1,1}
                +\biggl[
                        \frac{16}{27} (103+97 z)
                        +(1+z) \Bigl(
                                \frac{176}{9} H_0
                                +\frac{16}{3} H_0^2
                                -\frac{64}{3} H_{0,1}
                        \Bigr)
\nonumber\\&&			
                        +\frac{160}{3} (z-1) H_1
                \biggr] \zeta_2
                +\biggl[
                        -\frac{32}{9} (-1+15 z)
                        +\frac{64}{9} (1+z) H_0
                \biggr] \zeta_3
	\biggr\}
	+a_{qq,Q}^{PS(3)}
\Biggr\}.
\end{eqnarray}

%-------------------------
The contributions to $A_{qg,Q}^{(3)}$ are given by
%-------------------------
\begin{eqnarray}
	\lefteqn{A_{qg,Q} =}
\nonumber\\&&	
	a_s^3 \Biggl\{
	 \textcolor{blue}{ C_F N_F T_F^2} \biggl\{
                \frac{1}{27} (58501-59018 z)
                + L_M ^3 \biggl[
                        \frac{8}{3} (-41+42 z)
\nonumber\\&&			
                        +(-1+2 z) \Bigl(
                                \frac{16}{3} H_0^2
                                -\frac{32}{9} H_1
                        \Bigr)
                        -\frac{32}{9} (13+19 z) H_0
                \biggr]
                + L_M ^2 \biggl[
                        \frac{4}{3} (-285+296 z)
\nonumber\\&&			
                        +(-1+2 z) \Bigl(
                                \frac{32}{3} H_0^3
                                +\frac{16}{3} H_1^2
                                +\frac{32}{3} H_{0,1}
                                -\frac{32}{3} \zeta_2
                        \Bigr)
                        -\frac{8}{9} (425+2 z) H_0
                        -16 (7+6 z) H_0^2
\nonumber\\&&			
                        -\frac{32}{9} (1+4 z) H_1
                \biggr]
                + L_M  \Biggl[
                        \frac{212}{9} (-69+70 z)
                        +(-1+2 z) \Bigl(
                                \frac{20}{3} H_0^4
                                -\frac{16}{9} H_1^3
                                +\frac{32}{3} H_{0,0,1}
\nonumber\\&&				
                                -\frac{32}{3} H_{0,1,1}
                                +\frac{32}{3} \zeta_3
                        \Bigr)
                        +\biggl[
                                -\frac{8}{27} (4103+1565 z)
                                +\frac{64}{3} (z-1) H_1
                                -\frac{32}{3} (-1+2 z) H_{0,1}
                        \biggr] H_0
\nonumber\\&&			
                        +\biggl[
                                \frac{4}{3} (-343+10 z)
                                +\frac{16}{3} (-1+2 z) H_1
                        \biggr] H_0^2
                        -\frac{16}{9} (50+23 z) H_0^3
                        -\frac{16}{27} (-28+131 z) H_1
\nonumber\\&&			
                        +\frac{32}{9} (-2+7 z) H_1^2
                        +\frac{64}{9} (1+4 z) H_{0,1}
                        -\frac{64}{9} (-2+7 z) \zeta_2
                \Biggr]
                +(-1+2 z) \Bigl(
                        -\frac{4}{15} H_0^5
\nonumber\\&&			
                        +\frac{5248}{81} H_1
                        -\frac{896}{27} H_{0,1}
                        +\frac{160}{9} H_{0,0,1}
                        -\frac{32}{3} H_{0,0,0,1}
                        +
                        \frac{16}{3} \zeta_2^2
                \Bigr)
                +\biggl[
                        \frac{16}{81} (7333+4195 z)
\nonumber\\&&			
                        +(-1+2 z) \Bigl(
                                \frac{896}{27} H_1
                                -\frac{160}{9} H_{0,1}
                                +\frac{32}{3} H_{0,0,1}
                        \Bigr)
                \biggr] H_0
                +\biggl[
                        \frac{8}{27} (1331+140 z)
\nonumber\\&&			
                        +(-1+2 z) \Bigl(
                                \frac{80}{9} H_1
                                -\frac{16}{3} H_{0,1}
                        \Bigr)
                \biggr] H_0^2
                +\biggl[
                        -\frac{16}{27} (-101+28 z)
                        +\frac{16}{9} (-1+2 z) H_1
                \biggr] H_0^3
\nonumber\\&&		
                +\frac{4}{9} (13+7 z) H_0^4
                +\Biggl[
                        -\frac{4}{3} (-299+305 z)
                        +(-1+2 z) \Bigl(
                                -8 H_0^3
                                -\frac{8}{3} H_1^2
                                -\frac{32}{3} H_{0,1}
                        \Bigr)
\nonumber\\&&			
                        +\biggl[
                                \frac{4}{9} (757+178 z)
                                +\frac{16}{3} (-1+2 z) H_1
                        \biggr] H_0
                        +\frac{8}{3} (34+25 z) H_0^2
                        +\frac{32}{9} (-2+7 z) H_1
                \Biggr] \zeta_2
\nonumber\\&&		
                +\biggl[
                        -\frac{8}{3} (-41+42 z)
                        +(-1+2 z) \Bigl(
                                -\frac{16}{3} H_0^2
                                +\frac{32}{9} H_1
                        \Bigr)
                        +\frac{32}{9} (13+19 z) H_0
                \biggr] \zeta_3
	\biggr\}
\nonumber\\&&	
	+ \textcolor{blue}{C_A N_F T_F^2} \biggl\{
                -\frac{8}{27} (-4028+4113 z)
                + L_M ^3 \biggl[
                        \frac{64}{3} (z-1)
                        -\frac{64}{9} (1+z) H_0
\nonumber\\&&			
                        +\frac{32}{9} (-1+2 z) H_1
                \biggr]
                + L_M ^2 \Biggl[
                        \frac{8}{3} (-65+68 z)
                        +\biggl[
                                -\frac{16}{9} (47+38 z)
                                -\frac{32}{3} (1+2 z) H_{-1}
                        \biggr] H_0
\nonumber\\&&			
                        -\frac{32}
                        {3} H_0^2
                        +
                        \frac{32}{9} (1+4 z) H_1
                        -\frac{16}{3} (-1+2 z) H_1^2
                        +\frac{32}{3} (1+2 z) H_{0,-1}
                        -\frac{32}{3} \zeta_2
                \Biggr]
\nonumber\\&&		
                + L_M  \Biggl[
                        \frac{16}{9} (-514+547 z)
                        +(-1+2 z) \Bigl(
                                \frac{16}{9} H_1^3
                                +\frac{32}{3} H_{0,0,1}
                                +32 H_{0,1,1}
                        \Bigr)
\nonumber\\&&			
                        +\biggl[
                                -\frac{32}{27} (463+211 z)
                                +(-1+2 z) \Bigl(
                                        \frac{16}{3} H_1^2
                                        -\frac{32}{3} H_{0,1}
                                \Bigr)
                                -\frac{64}{9} (2+7 z) H_{-1}
\nonumber\\&&				
                                +\frac{64}{3} (1+2 z) H_{0,-1}
                        \biggr] H_0
                        +\biggl[
                                -\frac{8}{9} (133+72 z)
                                +\frac{16}{3} (-1+2 z) H_1
                                -\frac{32}{3} (1+2 z) H_{-1}
                        \biggr] H_0^2
\nonumber\\&&			
                        +\frac{16}{9} (-5+6 z) H_0^3
                        +\biggl[
                                \frac{16}{27} (-1+95 z)
                                -\frac{64}{3} (-1+2 z) H_{0,1}
                        \biggr] H_1
                        -\frac{32}{9} (-2+7 z) H_1^2
\nonumber\\&&			
                        +\frac{64}{3} z H_{0,1}
                        +\frac{64}{9} (2+7 z) H_{0,-1}
                        -\frac{64}{3} (1+2 z) H_{0,0,-1}
                        +\biggl[
                                -\frac{64}{9} (2+3 z)
\nonumber\\&&				
                                +\frac{32}{3} (-1+2 z) H_1
                        \biggr] \zeta_2
                        +32 \zeta_3
                \Biggr]
                +\biggl[
                        \frac{8}{81} (5836+4771 z)
                        -\frac{16}{3} (-1+2 z) H_1
                \biggr] H_0
\nonumber\\&&		
                -\frac{4}{27} (-649+128 z) H_0^2
                +\frac{16}{27} (17+2 z) H_0^3
                -\frac{4}{9} (-1+2 z) H_0^4
                -\frac{16}{81} (-283+611 z) H_1
\nonumber\\&&		
                -\frac{8}{3} (z-1) H_1^2
                -\frac{16}
                {9} (3+10 z) H_{0,1}
                +\frac{32}{3} z H_{0,1,1}
                +\Biggl[
                        -\frac{8}{9} (-231+205 z)
\nonumber\\&&			
                        +\biggl[
                                \frac{8}{9} (97+100 z)
                                +\frac{16}{3} (1+2 z) H_{-1}
                        \biggr] H_0
                        -\frac{8}{3} (-3+2 z) H_0^2
                        -\frac{32}{9} (-2+7 z) H_1
\nonumber\\&&			
                        +\frac{8}{3} (-1+2 z) H_1^2
                        -\frac{16}{3} (1+2 z) H_{0,-1}
                \Biggr] \zeta_2
                +\frac{16}{3} \zeta_2^2
                +\biggl[
                        -\frac{32}{3} (-2+3 z)
                        +\frac{64}{9} (1+z) H_0
\nonumber\\&&			
                        -\frac{32}{9} (-1+2 z) H_1
                \biggr] \zeta_3
	\biggr\}
	+a_{qg,Q}^{(3)}
\Biggr\}
\end{eqnarray}

%-------------------------
and the functions $a_{qq,Q}^{(3)}(z)$ and $a_{qg,Q}^{(3)}(z)$ are 
%----------------------------------------------------------------------------------------------------------------------
\begin{eqnarray}
a_{qq,Q}^{(3),\rm PS}(z) &=& 
\textcolor{blue}{C_F T_F^2 N_F} \Biggl\{
        (1-z) \Biggl(
                \frac{85504}{243}
                -\frac{25472}{81} H_1
                +\frac{320}{27} H_1^2
                -\frac{640}{27} H_1^3
                -\frac{160}{3} H_1 \zeta_2
        \Biggr)
\nonumber\\ &&         
+(1+z) \Biggl[
                \Biggl(
                        \frac{54592}{243}
                        -64 \zeta_3
                \Biggr) H_0
                +\frac{2288}{81} H_0^2
                +\frac{176}{81} H_0^3
                +\frac{8}{27} H_0^4
                -\frac{9152}{81} H_{0,1}
\nonumber\\ &&                 
-\frac{704}{27} H_{0,0,1}
                +\frac{1408}{27} H_{0,1,1}
                -\frac{128}{9} H_{0,0,0,1}
                +\frac{256}{9} H_{0,0,1,1}
                -\frac{512}{9} H_{0,1,1,1}(z)
\nonumber\\ &&                        
                +\Biggl(
                        \frac{1232 }{27} H_0
+\frac{112}{9} H_0^2
                        -\frac{64}{3} H_{0,1}
                \Biggr) \zeta_2
                +\frac{704}{15} \zeta_2^2
        \Biggr]
        +\frac{16}{81} (617+527 z) \zeta_2
\nonumber\\ &&
        +\frac{32}{27} (-127+83 z) \zeta_3
\Biggr\}
\end{eqnarray}
%-------------------------------------
\begin{eqnarray}
a_{qg,Q}^{(3)}(z) &=& 
\textcolor{blue}{C_F T_F^2 N_F} \Biggl\{
        -
        \frac{16}{81} (-49511+50285 z)
        -(1-2 z) \Biggl(
                -\frac{8}{3} H_0^5
                +\frac{32}{27} H_0^3 H_1
                -\frac{8}{27} H_1^4
\nonumber\\ && 
                -\frac{32}{9} H_0^2 H_{0,1}
                +\frac{64}{9} H_0 H_{0,0,1}
                -\frac{64}{9} H_{0,0,0,1}
                -\frac{64}{9} H_{0,1,1,1}
        \Biggr)
        +\frac{16}{243} (107935
\nonumber\\ && 
+46408 z) H_0
        +\frac{8}{81} (21955+3853 z) H_0^2
        -\frac{8}{81} (-3997+1118 z) H_0^3
        +\frac{8}{27} (160
\nonumber\\ &&
+61 z) H_0^4
        +\Biggl(
                \frac{32}{243} (-1559+4066 z)
                +\frac{896}{81} (-5+13 z) H_0
                +\frac{64}{27} (-2+7 z) H_0^2
        \Biggr) H_1
\nonumber\\ &&         
-\frac{80}{81} (-28+71 z) H_1^2
        +\frac{64}{81} (-2+7 z) H_1^3
        +\Biggl(
                -\frac{1792}{81} (-5+13 z)
\nonumber\\ &&       
                 -\frac{128}{27} (-2+7 z) H_0
        \Biggr) 
H_{0,1}
  +\frac{128}{27} (-2+7 z) [H_{0,0,1} + H_{0,1,1}]
        +\Biggl(
                -\frac{8}{81} (-7513
\nonumber\\ &&              
+6779 z)
                -(1-2 z) \Biggl(
                        -16 H_0^3
         +\frac{32}{3} H_0 H_1
                        -\frac{16}{3} H_1^2
                        -\frac{64}{3} H_{0,1}
                \Biggr)
                +\frac{8}{9} (757+178 z) H_0
\nonumber\\ &&                
                +\frac{16}{3} (34+25 z) H_0^2
 +\frac{64}{9} (-2+7 z) H_1
        \Biggr) \zeta_2
        -\frac{608}{45} (1-2 z) \zeta_2^2
        +\Biggl(
                \frac{16}{27} (-2567
\nonumber\\ &&  
+2590 z)
                +(-1+2 z) 
\Biggl(
                        \frac{224}{3} H_0^2
                        -\frac{448}{9} H_1
                \Biggr)
                -\frac{448}{9} (13+19 z) H_0
        \Biggr) \zeta_3
\Biggr\}
%-------------------
\nonumber\\ &&
+\textcolor{blue}{C_A T_F^2 N_F} \Biggl\{
        -\frac{4352}{81} (-104+109 z)
        -(1-2 z) \Biggl(
                -\frac{32}{9} H_0^3 H_1
                -\frac{16}{9} H_0^2 H_1^2
                +\frac{32}{27} H_0 H_1^3
\nonumber\\ && 
                +\frac{8}{27} H_1^4
                +\Biggl(
                        \frac{32}{3} H_0^2
                        -\frac{64}{9} H_1^2
                \Biggr) H_{0,1}
                -\frac{64}{9} H_{0,1}^2
                +\Biggl(
                        -\frac{64}{3} H_0
                        +\frac{128}{9} H_1
                \Biggr) H_{0,0,1}
\nonumber\\ &&                
 +\Biggl(
                        \frac{64}{9} H_0
                        +\frac{128}{9} H_1
                \Biggr) H_{0,1,1}
                +\frac{64}{3} H_{0,0,0,1}
                -\frac{64}{9} H_{0,0,1,1}
                -\frac{64}{9} H_{0,1,1,1}
        \Biggr)
        +(1+2 z) 
\nonumber\\ && 
\Biggl(
                \frac{128}{27} H_{-1} H_0^3
                -\frac{128}{9} H_0^2 H_{0,-1}
                -\frac{256}{9} H_{0,0,0,-1}
                +\frac{256}{27} H_{0,0,-1} \Biggl(
                        2+3 H_0\Biggr)
        \Biggr)
\nonumber\\ && 
        +\Biggl(
                \frac{32}{243} (23638+13471 z)
                +\frac{64}{81} (70+191 z) H_{-1}
        \Biggr) H_0
        +\Biggl(
                \frac{16}{81} (3423+680 z)
\nonumber\\ &&             
+\frac{128}{27} (2+7 z) H_{-1}
        \Biggr) H_0^2
        +\frac{80}{81} (85+8 z) H_0^3
        -\frac{8}{27} (-15+26 z) H_0^4
        +\Biggl(
                -\frac{32}{243} 
\nonumber\\ &&  \times
(-1055+3283 z)
                -\frac{32}{9} (-4+11 z) H_0
                -\frac{64}{27} (-2+7 z) H_0^2
        \Biggr) H_1
        +\Biggl(
                \frac{16}{81} (-59
\nonumber\\ && 
+265 z)
                -\frac{64}{27} (-2+7 z) H_0
        \Biggr) H_1^2
        -\frac{64}{81} (-2+7 z) H_1^3
        +\Biggl(
                -\frac{128}{27} (3+26 z)
\nonumber\\ &&                
 +\frac{64}{27} (-4+17 z) H_0
                +\frac{256}{27} (-2+7 z) H_1
        \Biggr) H_{0,1}
        +\Biggl(
                -\frac{64}{81} (70+191 z)
                -\frac{128}{27} (4
\nonumber\\ &&
+11 z) H_0
        \Biggr) H_{0,-1}
        -\frac{64}{27} (-4+35 z) H_{0,0,1}
        -\frac{128}{9} (-2+3 z) H_{0,1,1}
        +\Biggl(
                -\frac{16}{81} (-2359
\nonumber\\ && 
+1311 z)
                -(1-2 z) \big(
                        -\frac{64}{9} H_0 H_1
                        +\frac{80}{9} H_1^2
                        +\frac{64}{9} H_{0,1}
                \Biggr)
                +(1+2 z) \Biggl(
                        \frac{32}{3} H_{-1} H_0
\nonumber\\ && 
         -\frac{32}{3} H_{0,-1}
                \Biggr)
                +\frac{16}{9} (97+120 z) H_0
                -\frac{16}{3} (-3+2 z) H_0^2
                -\frac{320}{27} (-2+7 z) H_1
        \Biggr) \zeta_2
\nonumber\\ &&         
+\frac{1376}{45} \zeta_2^2
        +\Biggl(
                \frac{128}{9} (-23+22 z)
                -\frac{896}{9} (1+z) H_0
                -\frac{320}{9} (1-2 z) H_1
        \Biggr) \zeta_3
\Biggr\}.
\end{eqnarray}

The single mass contributions to the OME $A_{Qg}^{(3)}$ read
%-------------------------


%-------------------------

Finally, the single mass contributions to the OME $A_{gg,Q}^{(3)}$ are given by
%-------------------------
\begin{eqnarray}
	\lefteqn{A_{gg,Q}^{(\delta)} =}
\nonumber\\&&
	\Biggl\{a_s \frac{4 L_M }{3} \textcolor{blue}{T_F}
+a_s^2 \biggl[
	\textcolor{blue}{C_F T_F} (-15+4  L_M )  
	+ \textcolor{blue}{C_A T_F} \frac{2}{9}   (5+24  L_M )  
	+\frac{16  L_M ^2  }{9} \textcolor{blue}{T_F ^2}
\biggr]
\nonumber\\&&
	+a_s^3 \Biggl\{
	\frac{64  L_M ^3  }{27} \textcolor{blue}{T_F ^3}
	+ \textcolor{blue}{C_A N_F T_F^2} \biggl[
                \frac{224}{27}
                -\frac{44  L_M }{3}
                -\frac{4 \zeta_2}{3}
        \biggr]
+ \textcolor{blue}{ C_F N_F T_F^2} \biggl[
                \frac{118}{3}
\nonumber\\&&		
                -\frac{268  L_M }{9}
                +28 \zeta_2
        \biggr]
+ \textcolor{blue}{C_A T_F^2} \biggl[
                -\frac{8}{27}
                -2  L_M 
                +\frac{56  L_M ^2}{3}
                -\frac{44}{3} \zeta_2
        \biggr]
+ \textcolor{blue}{C_F T_F^2} \biggl[
                \frac{782}{9}
\nonumber\\&&		
                -\frac{584  L_M }{9}
                +\frac{40  L_M ^2}{3}
                -\frac{40}{3} \zeta_2
        \biggr]
+ \textcolor{blue}{C_A^2 T_F}  \biggl[
                -\frac{616}{27}
                + L_M  \Bigl(
                        \frac{277}{9}
                        +\frac{16}{3} \zeta_2^2
                        +\frac{160}{9} \zeta_3
                \Bigr)
\nonumber\\&&		
                + L_M ^2 \Bigl(
                        -\frac{2}{3}
                        +\frac{16 \zeta_3}{3}
                \Bigr)
                +\Bigl(
                        4
                        -\frac{8 \zeta_3}{3}
                \Bigr) \zeta_2
        \biggr]
+ \textcolor{blue}{C_F^2 T_F}  \biggl[
                -39
                -2  L_M 
		+16 \Bigl[-5+8 \ln(2) \Bigr] \zeta_2
\nonumber\\&&		
                -32 \zeta_3
        \biggr]
+ \textcolor{blue}{ C_A C_F T_F} \biggl[
                -\frac{1045}{6}
                +\frac{736  L_M }{9}
                -\frac{22  L_M ^2}{3}
		-\frac{4}{3} \Bigl[-5+48 \ln(2) \Bigr] \zeta_2
                +16 \zeta_3
        \biggr]
\nonumber\\&&	
        -\frac{64}{27}  T_F ^3 \zeta_3
	+a_{gg,Q}^{(3),\delta}
\Biggr\} \Biggr\} \delta(1-z)
\end{eqnarray}

\begin{eqnarray}
	\lefteqn{A_{gg,Q}^{(+)} =}
\nonumber\\&&	
-a_s^2 \frac{8   \big(
	28+30  L_M+9  L_M^2\big) }{27 (z-1)} \textcolor{blue}{C_A T_F}
	+a_s^3 \Biggl\{
		\textcolor{blue}{C_A N_F T_F^2} \Biggl[
		\frac{1}{z-1}\bigg[-\frac{2176  L_M}{81}
                -\frac{64  L_M^3}{27}
\nonumber\\&&		
                -\frac{4}{243} \big(
                        -2624
                        -441 \zeta_2
                        +81 z \zeta_2
                        -144 \zeta_3
                \big)
                +\frac{32}{9} H_0
		\bigg]
                +\frac{4}{3} \zeta_2
        \Biggr]	
+ \textcolor{blue}{C_A   T_F ^2} \Biggl[
		\frac{1}{z-1}\bigg[-\frac{320  L_M}{9}
\nonumber\\&&		
                -\frac{640  L_M^2}{27}
                -\frac{224  L_M^3}{27}
                -\frac{4}{81} \big(
                        -1312
                        -717 \zeta_2
                        +297 z \zeta_2
                        -168 \zeta_3
                \big)
                +\frac{16}{3} H_0
		\bigg]
                +\frac{44}{3} \zeta_2
        \Biggr]
\nonumber\\&&	
	+ \textcolor{blue}{C_A ^2  T_F}  \biggl\{
		\frac{1}{z-1}\biggl[\frac{176  L_M^3}{27}
                -\frac{4}{243} \big(
                        5668
                        +207 \zeta_2
                        -243 z \zeta_2
                        +324 \zeta_2^2
                        +396 \zeta_3
                        -162 \zeta_2 \zeta_3
\nonumber\\&&			
                        +162 z \zeta_2 \zeta_3
                \big)
                +\Bigl(
                        -\frac{88}{9}
                        +\frac{32 H_1 \zeta_2}{3}
                \Bigr) H_0
                +\frac{8}{3} H_0^2 \zeta_2
		\biggr]
                + L_M \Biggl[
			\frac{1}{z-1}\bigg[-\frac{16}{9} H_0^2 \big(
                                10+3 H_1\big)
\nonumber\\&&				
                        +\frac{8}{81} \big(
                                -155
                                +360 \zeta_2
                                -54 \zeta_2^2
                                +54 z \zeta_2^2
                                -1044 \zeta_3
                                +180 z \zeta_3
                        \big)
                        +\Bigl(
                                -\frac{16}
                                {3}
                                -\frac{640}{9} H_1
\nonumber\\&&				
                                +\frac{32}{3} H_{0,1}
                                -\frac{64}{3} H_{0,-1}
                        \Bigr) H_0
                        -\frac{64}{3} H_{0,0,1}
                        +\frac{128}{3} H_{0,0,-1}
			\bigg]
                        -\frac{16}{3} \zeta_2^2
                        -\frac{160}{9} \zeta_3
                \Biggr]
\nonumber\\&&		
                + L_M^2 \Biggl[
			\frac{1}{z-1}\biggl[\frac{8}{9} \bigl(
                                -23
                                +12 \zeta_2
                                -6 \zeta_3
                                +6 z \zeta_3
                        \bigr)
                        -\frac{16}{3} H_0^2
                        -\frac{64}{3} H_0 H_1
			\biggr]
                        -\frac{16}{3} \zeta_3
                \Biggr]
\nonumber\\&&	
                +\Bigl(
                        -4
                        +\frac{8 \zeta_3}{3}
                \Bigr) \zeta_2
	\biggr\}
	+\textcolor{blue}{ C_A   C_F   T_F } \biggl\{
		\frac{1}{z-1}\Biggl[-\frac{2}{9} \big(
                        233
                        +210 \zeta_2
                        -288 \ln(2)  \zeta_2
\nonumber\\&&			
                        -30 z \zeta_2
                        +288 \ln(2)  z \zeta_2
                        +72 \zeta_3
                        -72 z \zeta_3
                \big)
                +\frac{8}{3}  L_M \big(
                        -5
                        +24 \zeta_3
                \big)
		-8 L_M^2
		\Biggr]
\nonumber\\&&		
		+\frac{4}{3} \Bigl[-5+48 \ln(2) \Bigr] \zeta_2
                -16 \zeta_3
	\biggr\}
+ \textcolor{blue}{C_F ^2  T_F}  \biggl\{
                16 \Bigl[
                        -5 \zeta_2
                        +8 \ln(2)  \zeta_2
                        -2 \zeta_3
                \Bigr]
\nonumber\\&&	
	-16 \Bigl[-5+8 \ln(2) \Bigr] \zeta_2
                +32 \zeta_3
	\biggr\}
+a_{gg,Q}^{(3),(+)}
\Biggr\}
\end{eqnarray}



%-------------------------

%-----------------------------------------------------------------------------------------------------------------
\section{The \boldmath $z$-Space expressions for the asymptotic massive Wilson coefficients}
\label{sec:B}
%-----------------------------------------------------------------------------------------------------------------

\vspace*{1mm}\noindent
Here we list the $z$-space contributions to the massive Wilson coefficient $L_{q}^{\sf PS}, L_{g}^{\sf S}$ and 
$H_{g}^{\sf S}$.

For $L_{q}^{\sf PS}(z)$ we obtain
\begin{eqnarray}
\lefteqn{ L_q^{PS}(z) = }
\nonumber\\&&
	a_s^3  C_F   N_F   T_F ^2 \Biggl\{
         L_M ^2 \biggl[
                \frac{32}{3} (z-1) \big(
                        12+5 H_1\big)
                +(1+z) \Bigl(
                        -\frac{32}{3} H_0^2
                        -\frac{64}{3} H_{0,1}
                        +\frac{64}{3} \zeta_2
                \Bigr)
\nonumber\\&&		
                +\frac{32}{3} (-7+z) H_0
        \biggr]
        +(z-1) \biggl[
                \frac{3296}{9}
                -\frac{32}{27} H_1 \big(
                        -125+12 H_0\big)
                +\frac{16}{9} H_1^2 \big(
                        28+15 H_0\big)
                +\frac{80}{3} H_1^3
        \biggr]
\nonumber\\&&
        + L_Q  \Biggl[
                (1+z) \Bigl(
                        \frac{64}{3} H_{0,1,1}
                        -\frac{64 \zeta_3}{3}
                \Bigr)
                + L_M ^2 \biggl[
                        -\frac{160}{3} (z-1)
                        +\frac{64}{3} (1+z) H_0
                \biggr]
\nonumber\\&&
                + L_M  \biggl[
                        \frac{64}{9} (z-1) \big(
                                -4+15 H_1\big)
                        +(1+z) \Bigl(
                                -\frac{128 H_{0,1}}{3}
                                +\frac{128 \zeta_2}{3}
                        \Bigr)
                        +\frac{128}{9} (2+5 z) H_0
                \biggr]
\nonumber\\&&
                +\frac{128}{27} (11+14 z) H_0
                +(z-1) \Bigl(
                        -\frac{3712}{27}+\frac{128}{9} 
H_1-\frac{80}{3} H_1^2\Bigr)
                -\frac{64}{9} (2+5 z) H_{0,1}
\nonumber\\&&
                +\frac{64}{9} (2+5 z) \zeta_2
        \Biggr]
        + L_M  \Biggl[
                (z-1) \biggl[
                        -\frac{64}{3}
                        -\frac{64}{9} H_1 \big(
                                32+15 H_0\big)
                        -\frac{320}{3} H_1^2
                \biggr]
\nonumber\\&&
                +(1+z) \biggl[
                        \frac{128}{9} H_{0,1} \big(
                                1+3 H_0\big)
                        -\frac{128}{3} H_{0,0,1}
                        +\frac{256}{3} H_{0,1,1}
                        -\frac{128}{3} \zeta_3
                \biggr]
                -\frac{512}{9} (1+2 z) H_0
\nonumber\\&&
                -\frac{64}
                {9} (2+5 z) H_0^2
                +\biggl[
                        \frac{64}{9} (-17+13 z)
                        -\frac{128}{3} (1+z) H_0
                \biggr] \zeta_2
        \Biggr]
        +(1+z) \Bigl(
                \frac{64}{3} H_{0,0,1,1}
                -64 H_{0,1,1,1}
\nonumber\\&&
                +\frac{32}{15} \zeta_2^2
        \Bigr)
        +\frac{128}{27} (-40+z) H_0
        -\frac{64}{27} (11+14 z) H_0^2
        +\big(
                \frac{128}{27} (-8+z)
                +\frac{64}{9} (2+5 z) H_0
        \big) H_{0,1}
\nonumber\\&&
        -\frac{64}{9} (2+5 z) H_{0,0,1}
        +\biggl[
                \frac{64}{9} (1+4 z)
                -\frac{64}{3} (1+z) H_0
        \biggr] H_{0,1,1}
        +\biggl[
                \frac{128}{27} (5+2 z)
                -\frac{64}{9} (2+5 z) H_0
\nonumber\\&&
                -\frac{160}{3} (z-1) H_1
                +\frac{64}{3} (1+z) H_{0,1}
        \biggr] \zeta_2
        +\biggl[
                -\frac{32}{9} (-17+13 z)
                +\frac{64}{3} (1+z) H_0
        \biggr] \zeta_3
\nonumber\\&&
+ A_{qq,Q}^{PS(3)}(z)
+ N_F \hat{\tilde{C}}_q^{PS(3)}(L_Q,N_F,z)
\Biggr\}.
\end{eqnarray}

For $L_{g}^{\sf S}(z)$ we obtain
\begin{eqnarray}
\lefteqn{ L_g^S(z) = }
\nonumber\\&&
	a_s^2  \textcolor{blue}{N_F   T_F ^2} \Biggl\{
        \frac{16}{3}  L_M   L_Q  (2z-1)
        + L_M  \biggl[
                -\frac{16}{3} (-3+4 z)
                +(2z-1) \Bigl(
                        -\frac{16}{3} H_0
                        -\frac{16}{3} H_1
                \Bigr)
        \biggr]
\Biggr\}
\nonumber\\&&
	+a_s^3 \Biggl\{
		\textcolor{blue}{ N_F   T_F ^3} \biggl\{
                \frac{64}{9}  L_M ^2  L_Q  (2z-1)
                + L_M ^2 \biggl[
                        -\frac{64}{9} (-3+4 z)
                        +(2z-1) \Bigl(
                                -\frac{64}{9} H_0
                                -\frac{64}{9} H_1
                        \Bigr)
                \biggr]
	\biggr\}
\nonumber\\&&
	+ \textcolor{blue}{C_A   N_F   T_F ^2} \biggl\{
                \frac{8}{27} (-3943+3928 z)
                + L_M   L_Q ^2 \biggl[
                        -64 (z-1)
                        +\frac{64}{3} (1+z) H_0
                        -\frac{32}{3} (2z-1) H_1
                \biggr]
\nonumber\\&&
                + L_M ^2 \Biggl[
                        \frac{512}{3} (z-1)
                        +(1+z) \Bigl(
                                -\frac{32}{3} H_0^2
                                -\frac{64}{3} H_{0,1}
                        \Bigr)
                        +\biggl[
                                \frac{64}{3} (-4+z)
                                +\frac{32}{3} (2z-1) H_1
                        \biggr] H_0
\nonumber\\&&
                        +\frac{32}{3} (-9+10 z) H_1
                        +\frac{32}{3} (2z-1) H_1^2
                        +32 \zeta_2
                \Biggr]
                + L_Q  \Biggl[
                        -\frac{40}{9} (-108+107 z)
\nonumber\\&&
                        + L_M ^2 \biggl[
                                -64 (z-1)
                                +\frac{64}{3} (1+z) H_0
                                -\frac{32}{3} (2z-1) H_1
                        \biggr]
                        + L_M  \Biggl[
                                -\frac{32}{3} (-22+19 z)
\nonumber\\&&
                                +\biggl[
                                        \frac{32}{9} (5+98 z)
                                        +\frac{64}{3} (2z-1) H_1
                                \biggr] H_0
                                -\frac{64}{3} (1+2 z) H_{-1}
                                 H_0
                                -
                                \frac{64}{3} (1+3 z) H_0^2
\nonumber\\&&
                                +\frac{128}{9} (-8+7 z) H_1
                                +\frac{32}{3} (2z-1) H_1^2
                                -\frac{128}{3} (1+z) H_{0,1}
                                +\frac{64}{3} (1+2 z) H_{0,-1}
                                +\frac{128}{3} \zeta_2
                        \Biggr]
\nonumber\\&&
                        +\frac{8}{27} (785+644 z) H_0
                        +\frac{16}{9} (17+2 z) H_0^2
                        -\frac{16}{9} (2z-1) H_0^3
                        -\frac{16}{27} (-47+103 z) H_1
                        -\frac{32}{3} z H_{0,1}
\nonumber\\&&
                        +\frac{32}{3} z \zeta_2
                \Biggr]
                +(2z-1) \Bigl(
                        \frac{4}{9} H_0^4
                        +\frac{32}{3} H_{0,0,0,1}
                        -\frac{64}{15} \zeta_2^2
                \Bigr)
                + L_M  \Biggl[
                        \frac{16}{9} (-325+301 z)
\nonumber\\&&
                        +\biggl[
                                \frac{32}{9} \big(
                                        -58-42 z+3 z^2\big)
                                -\frac{128}{9} (-8+7 z) H_1
                                -16 (2z-1) H_1^2
                                -\frac{32}{3} \big(
                                        -3-6 z+2 z^2\big) H_{0,1}
\nonumber\\&&
                                +\frac{64}{3} \big(
                                        -1-2 z+z^2\big) H_{0,-1}
                        \biggr] H_0
                        +\frac{32}{3} \big(
                                1+2 z+2 z^2\big) H_{-1}^2 H_0
                        +\biggl[
                                \frac{8}{9} \big(
                                        23-124 z+44 z^2\big)
\nonumber\\&&
                                +\frac{16}{3} \big(
                                        3-6 z+2 z^2\big) H_1
                        \biggr] H_0^2
                        +\frac{16}{9} (5+14 z) H_0^3
                        +\biggl[
                                \frac{16}{9} \big(
                                        -180+161 z+6 z^2\big)
\nonumber\\&&
                                +\frac{64}{3} (2z-1) H_{0,1}
                        \biggr] H_1
                        -\frac{32}{9} (-8+z) H_1^2
                        -\frac{16}{9} (2z-1) H_1^3
\nonumber\\&&
                        +\biggl[
                                -\frac{64 \big(
                                        2+6 z+12 z^2+11 z^3\big)}
                                {9 z} H_0
                                -
                                \frac{32}{3} \big(
                                        -1-2 z+z^2\big) H_0^2
                                +\frac{64}{3} (1+2 z) H_{0,1}
\nonumber\\&&
                                -\frac{64}{3} \big(
                                        1+2 z+2 z^2\big) H_{0,-1}
                        \biggr] H_{-1}
                        -\frac{32}{9} (41+62 z) H_{0,1}
                        +\frac{64 \big(
                                2+6 z+12 z^2+11 z^3\big)}{9 z} H_{0,-1}
\nonumber\\&&
                        +\frac{32}{3} \big(
                                -3+2 z+2 z^2\big) H_{0,0,1}
                        -\frac{64}{3} \big(
                                -1-2 z+z^2\big) H_{0,0,-1}
                        -\frac{32}{3} (-7+2 z) H_{0,1,1}
\nonumber\\&&
                        -\frac{64}{3} (1+2 z) H_{0,1,-1}
                        -\frac{64}{3} (1+2 z) H_{0,-1,1}
                        +\frac{64}{3} \big(
                                1+2 z+2 z^2\big) H_{0,-1,-1}
\nonumber\\&&
                        +\biggl[
                                -\frac{32}{9} \big(
                                        3-90 z+22 z^2\big)
                                -\frac{64}{3} (3+4 z) H_0
                                -\frac{64}{3} (z-1)^2 H_1
                                +\frac{32}{3} \big(
                                        -1-2 z+2 z^2\big) H_{-1}
                        \biggr] \zeta_2
\nonumber\\&&
                        -\frac{64}{3} \big(
                                2+z+2 z^2\big) \zeta_3
                \Biggr]
                +\biggl[
                        \frac{8}{27} (-2405+413 z)
                        +\frac{16}{27} (-47+103 z) H_1
                        +\frac{32}{3} z H_{0,1}
                \biggr] H_0
\nonumber\\&&
                -\frac{4}{27} (989+764 z) H_0^2
                +\frac{160}{27} (-2+z) H_0^3
                +\frac{8}{27} (-1884+1981 z) H_1
                +\frac{16}{27} (-47+103 z) H_1^2
\nonumber\\&&
                -\frac{8}{27} (785+572 z) H_{0,1}
                -\frac{32}{9} (17+5 z) H_{0,0,1}
                +\frac{64}{3} z H_{0,1,1}
                +\biggl[
                        \frac{8}{9} (293+122 z)
                        -\frac{32}
                        {9} (-17+z) H_0
\nonumber\\&&
                        -
                        \frac{16}{3} (2z-1) H_0^2
                \biggr] \zeta_2
                +\biggl[
                        -\frac{32}{9} (-17+z)
                        -\frac{32}{3} (2z-1) H_0
                \biggr] \zeta_3
	\biggr\}
	+ \textcolor{blue}{C_F   N_F   T_F ^2} \biggl\{
                4 (-889+904 z)
\nonumber\\&&
                + L_M   L_Q ^2 \biggl[
                        8 (-41+42 z)
                        +(2z-1) \Bigl(
                                16 H_0^2
                                -\frac{32}{3} H_1
                        \Bigr)
                        -\frac{32}{3} (13+19 z) H_0
                \biggr]
\nonumber\\&&
                + L_Q  \Biggl[
                        -20 (-67+70 z)
			+ L_M ^2 \Bigl[
                                -336 (z-1)
                                +48 (3+4 z) H_0
                                -16 (2z-1) H_0^2
			\Bigr]
\nonumber\\&&
                        + L_M  \Biggl[
                                -\frac{16}{3} (-472+473 z)
                                +(2z-1) \Bigl(
                                        -32 H_0^3
                                        +\frac{64}{3} H_1^2
                                        -64 H_{0,0,1}
                                        +64 \zeta_3
                                \Bigr)
\nonumber\\&&
                                +\biggl[
                                        \frac{32}{3} (156+25 z)
                                        +\frac{128}{3} (2z-1) H_1
                                \biggr] H_0
                                +\frac{16}{3} (73+52 z) H_0^2
                                -\frac{16}{3} (-109+106 z) H_1
\nonumber\\&&
                                +\frac{64}{3} (14+17 z) H_{0,1}
				+\Bigl[
                                        -64 (4+7 z)
                                        +64 (2z-1) H_0
				\Bigr] \zeta_2
                        \Biggr]
                        +32 (23+17 z) H_0
\nonumber\\&&
                        -8 (-23+9 z) H_0^2
                        +\frac{8}{3} (9+4 z) H_0^3
                        -\frac{4}{3} (2z-1) H_0^4
                \Biggr]
                + L_M ^2 \biggl[
                        928 (z-1)
\nonumber\\&&
                        +(2z-1) \Bigl(
                                \frac{16}{3} H_0^3
                                +32 H_{0,0,1}
                                -32 \zeta_3
                        \Bigr)
                        -16 (30+7 z) H_0
                        -8 (11+4 z) H_0^2
\nonumber\\&&
                        +336 (z-1) H_1
                        -48 (3+4 z) H_{0,1}
			+\Bigl[
                                48 (3+4 z)
                                -32 (2z-1) H_0
			\Bigr] \zeta_2
                \biggr]
\nonumber\\&&
                + L_M 
                 \Biggl[
                        \frac{8}{9} (-8533+8509 z)
                        +(1+z)^2 \Bigl(
                                -\frac{128}{3} H_{-1}^2 H_0
                                -\frac{256}{3} H_{0,-1,-1}
                        \Bigr)
                        +(2z-1) \Bigl(
                                \frac{28}{3} H_0^4
\nonumber\\&&
                                -\frac{80}{9} H_1^3
                                +64 H_{0,0,1,1}
                                -\frac{32}{5} \zeta_2^2
                        \Bigr)
                        +\biggl[
                                -\frac{8}{9} \big(
                                        5606+209 z+24 z^2\big)
                                +(2z-1) \Bigl(
                                        -\frac{64}{3} H_1^2
\nonumber\\&&
                                        +64 H_{0,0,1}
                                \Bigr)
                                +48 (-11+10 z) H_1
                                +\frac{32}{3} \big(
                                        -25-40 z+4 z^2\big) H_{0,1}
\nonumber\\&&
                                -\frac{128}{3} \big(
                                        -2+4 z+z^2\big) H_{0,-1}
                        \biggr] H_0
                        +\biggl[
                                -\frac{4}{9} \big(
                                        3063+462 z+32 z^2\big)
                                -\frac{16}{3} \big(
                                        -1+2 z+4 z^2\big) H_1
                        \biggr] H_0^2
\nonumber\\&&
                        -\frac{16}{9} (104+17 z) H_0^3
                        +\biggl[
                                -\frac{32}{3} \big(
                                        229-230 z+2 z^2\big)
                                -\frac{64}{3} (2z-1) H_{0,1}
                        \biggr] H_1
                        +24 (-11+10 z) H_1^2
\nonumber\\&&
                        +\biggl[
                                (1+z)^2 \Bigl(
                                        \frac{64}{3} H_0^2
                                        +\frac{256}{3} H_{0,-1}
                                \Bigr)
                                +\frac{64 \big(
                                        4+45 z+48 z^2+4 z^3\big)}{9 
z} H_0
                        \biggr] H_{-1}
                        -\frac{16}{3} (203+156 z) H_{0,1}
\nonumber\\&&
                        -\frac{64 \big(
                                4+45 z+48 z^2+4 z^3\big)}{9 z} H_{0,-1}
                        -\frac{64}{3} (z-1) (-11+2 z) H_{0,0,1}
\nonumber\\&&
                        +\frac{128}{3} \big(
                                -5+6 z+z^2\big) H_{0,0,-1}
                        -\frac{32}{3} (31+28 z) H_{0,1,1}
                        +\biggl[
                                \frac{32}
                                {9} \big(
                                        543+99 z+8 z^2\big)
\nonumber\\&&
                                +\frac{32}{3} (83+56 z) H_0
                                -96 (2z-1) H_0^2
                                +\frac{64}{3} \big(
                                        -1+2 z+2 z^2\big) H_1
                                -\frac{128}{3} (1+z)^2 H_{-1}
                        \biggr] \zeta_2
\nonumber\\&&
                        +\biggl[
                                \frac{64}{3} \big(
                                        33-4 z+4 z^2\big)
                                -128 (2z-1) H_0
                        \biggr] \zeta_3
                \Biggr]
                +(2z-1) \Bigl(
                        \frac{4}{15} H_0^5
                        +32 H_{0,0,0,0,1}
                        -32 \zeta_5
                \Bigr)
\nonumber\\&&
                +4 (-519+10 z) H_0
                -8 (69+8 z) H_0^2
                -\frac{8}{3} (32+3 z) H_0^3
                +\frac{2}{3} (-11+4 z) H_0^4
                +20 (-67+70 z) H_1
\nonumber\\&&
                -32 (23+17 z) H_{0,1}
                +16 (-23+9 z) H_{0,0,1}
                -16 (9+4 z) H_{0,0,0,1}
                +\biggl[
                        32 (23+17 z)
\nonumber\\&&
                        -16 (-23+9 z) H_0
                        +8 (9+4 z) H_0^2
                        -\frac{16}{3} (2z-1) H_0^3
                \biggr] \zeta_2
                +\biggl[
                        \frac{32}{5} (9+4 z)
                        -\frac{64}{5} (2z-1) H_0
                \biggr] \zeta_2^2
\nonumber\\&&
		+\Bigl[
                        -16 (-23+9 z)
                        +16 (9+4 z) H_0
                        -16 (2z-1) H_0^2
		\Bigr] \zeta_3
	\biggr\}
\nonumber\\&&	
+A_{qg,Q}^{(3)}
+N_F \hat{\tilde{C}}_g^{S(3)}\left(L_Q,N_F,z\right)
\Biggr\}.
\end{eqnarray}

For $H_{q}^{\sf PS}(z)$ we obtain
\begin{eqnarray}
\lefteqn{ H_q^{PS} = }
\nonumber\\&&
	a_s^2  \textcolor{blue}{C_F   T_F } \Biggl\{
        -
        \frac{4}{3} (z-1) \big(
                148+66 H_1+15 H_1^2\big)
	+ L_Q  \Bigl[
                8 (z-1) \big(
                        11+5 H_1\big)
\nonumber\\&&			
                +(1+z) \big(
                        -16 H_0^2
                        -16 H_{0,1}
                        +16 \zeta_2
                \big)
                +32 (-2+z) H_0
	\Bigr]
	+ L_M ^2 \Bigl[
                20 (z-1)
                -8 (1+z) H_0
	\Bigr]
\nonumber\\&&
	+ L_Q ^2 \Bigl[
                -20 (z-1)
                +8 (1+z) H_0
	\Bigr]
	+ L_M  \Bigl[
                8 (z-1)
                -8 (-1+3 z) H_0
                +8 (1+z) H_0^2
	\Bigr]
\nonumber\\&&
        +(1+z) \Bigl(
                \frac{16}{3} H_0^3
                -32 H_{0,0,1}
                +16 H_{0,1,1}
                +16 \zeta_3
        \Bigr)
        +\biggl[
                -\frac{256}{3} (-2+z)
                -80 (z-1) H_1
\nonumber\\&&
                -\frac{32 (1+z)^3 H_{-1}}{3 z}
                +32 (1+z) H_{0,1}
        \biggr] H_0
        +\frac{8}{3} \big(
                21+2 z^2\big) H_0^2
        +16 (-1+3 z) H_{0,1}
\nonumber\\&&
        +\frac{32 (1+z)^3 H_{0,-1}}{3 z}
        +\biggl[
                -\frac{32}{3} \big(
                        9-3 z+z^2\big)
                -32 (1+z) H_0
        \biggr] \zeta_2
	\Biggr\}
\nonumber\\&&
	+a_s^3 \Biggl\{
		\textcolor{blue}{C_F   T_F ^2} \biggl\{
                 L_M ^2 \biggl[
                        \frac{32}{3} (z-1) \big(
                                12+5 H_1\big)
                        +(1+z) \Bigl(
                                -\frac{32}{3} H_0^2
                                -\frac{64}{3} H_{0,1}
                                +\frac{64}{3} \zeta_2
                        \Bigr)
\nonumber\\&&
                        +\frac{32}{3} (-7+z) H_0
                \biggr]
                + L_Q  \Biggl[
                        (1+z) \Bigl(
                                \frac{64}{3} H_{0,1,1}
                                -\frac{64 \zeta_3}{3}
                        \Bigr)
                        + L_M ^2 \biggl[
                                -\frac{160}{3} (z-1)
                                +\frac{64}
                                {3} (1+z) H_0
                        \biggr]
\nonumber\\&&
                        + L_M  \biggl[
                                \frac{64}{9} (z-1) \big(
                                        -4+15 H_1\big)
                                +(1+z) \Bigl(
                                        -\frac{128 H_{0,1}}{3}
                                        +\frac{128 \zeta_2}{3}
                                \Bigr)
                                +\frac{128}{9} (2+5 z) H_0
                        \biggr]
\nonumber\\&&
                        +\frac{128}{27} (11+14 z) H_0
                        +(z-1) \Bigl(
                                -\frac{3712}{27}+\frac{128}{9} 
H_1-\frac{80}{3} H_1^2\Bigr)
                        -\frac{64}{9} (2+5 z) H_{0,1}
\nonumber\\&&
                        +\frac{64}{9} (2+5 z) \zeta_2
                \Biggr]
                + L_M  \Biggl[
                        (1+z) \Bigl(
                                \frac{128}{9} H_{0,1}
                                -\frac{128}{3} H_{0,0,1}
                                +\frac{256}{3} H_{0,1,1}
                                -\frac{128}{3} \zeta_3
                        \Bigr)
\nonumber\\&&
                        +\biggl[
                                -\frac{512}{9} (1+2 z)
                                -\frac{320}{3} (z-1) H_1
                                +\frac{128}{3} (1+z) H_{0,1}
                        \biggr] H_0
                        -\frac{64}{9} (2+5 z) H_0^2
\nonumber\\&&
                        +(z-1) \Bigl(
                                -\frac{64}{3}-\frac{2048}{9} 
H_1-\frac{320}{3} H_1^2\Bigr)
                        +\biggl[
                                \frac{64}{9} (-17+13 z)
                                -\frac{128}{3} (1+z) H_0
                        \biggr] \zeta_2
                \Biggr]
\nonumber\\&&
                +(1+z) \Bigl(
                        \frac{64}{3} H_{0,0,1,1}
                        -64 H_{0,1,1,1}
                        +\frac{32}{15} \zeta_2^2
                \Bigr)
                +\biggl[
                        \frac{128}{27} (-40+z)
                        +\frac{16}{9} (z-1) H_1 \big(
                                -8+15 H_1\big)
\nonumber\\&&
                        +\frac{64}{9} (2+5 z) H_{0,1}
                        -\frac{64}{3} (1+z) H_{0,1,1}
                \biggr] H_0
                -\frac{64}{27} (11+14 z) H_0^2
                +(z-1) \Bigl(
                        \frac{3296}{9}+\frac{4000}{27} H_1
\nonumber\\&&
			+\frac{448}{9} H_1^2+
                        \frac{80}{3} H_1^3\Bigr)
                +\frac{128}{27} (-8+z) H_{0,1}
                -\frac{64}{9} (2+5 z) H_{0,0,1}
                +\frac{64}{9} (1+4 z) H_{0,1,1}
\nonumber\\&&
                +\biggl[
                        \frac{128}{27} (5+2 z)
                        -\frac{64}{9} (2+5 z) H_0
                        -\frac{160}{3} (z-1) H_1
                        +\frac{64}{3} (1+z) H_{0,1}
                \biggr] \zeta_2
\nonumber\\&&
                +\biggl[
                        -\frac{32}{9} (-17+13 z)
                        +\frac{64}{3} (1+z) H_0
                \biggr] \zeta_3
	\biggr\}
	+ \textcolor{blue}{C_F ^2  T_F}  \biggl\{
		(z-1) \Bigl[
                        2656
                        +8 H_1 \big(
                                103+40 H_{0,0,1}\big)
\nonumber\\&&				
                        +\frac{8}{3} H_0^4
                        +144 H_1^2
                        -80 H_{0,1}^2
                        +160 H_{0,1,1,1}
		\Bigr]
                + L_M ^2 \biggl[
                        4 (z-1) \big(
                                92+43 H_1+10 H_1^2\big)
\nonumber\\&&
                        +(1+z) \Bigl(
                                -\frac{8}{3} H_0^3
                                +48 H_{0,0,1}
                                -32 H_{0,1,1}
                                -16 \zeta_3
                        \Bigr)
			+\Bigl[
                                -4 (51+7 z)
                                +80 (z-1) H_1
\nonumber\\&&
                                -32 (1+z) H_{0,1}
			\Bigr] H_0
                        +8 (-6+5 z) H_0^2
                        -80 z H_{0,1}
			+\Bigl[
                                80
                                +16 (1+z) H_0
			\Bigr] \zeta_2
                \biggr]
\nonumber\\&&
                + L_Q  \Biggl[
			 L_M ^2 \Bigl[
                                -4 (z-1) \big(
                                        13+20 H_1\big)
                                +(1+z) \big(
                                        8 H_0^2
                                        +32 H_{0,1}
                                        -32 \zeta_2
                                \big)
                                -16 (-2+3 z) H_0
			\Bigr]
\nonumber\\&&
                        +(z-1) \big(
                                -392
                                -288 H_1
                                -160 H_{0,1,1}
                        \big)
                        + L_M  \biggl[
                                -8 (z-1) \big(
                                        5+4 H_1\big)
\nonumber\\&&
                                +(1+z) \Bigl(
                                        -8 H_0
                                        -\frac{16}{3} H_0^3
                                        -64 H_{0,0,1}
                                        +64 \zeta_3
                                \Bigr)
                                +32 z H_0^2
                                +32 (-1+3 z) H_{0,1}
\nonumber\\&&
				+\Bigl[
                                        -32 (-1+3 z)
                                        +64 (1+z) H_0
				\Bigr] \zeta_2
                        \biggr]
                        +(1+z) \Bigl(
                                \frac{2}{3} H_0^4
                                -64 H_{0,0,0,1}
                                +128 H_{0,0,1,1}
                                -\frac{256}{5} \zeta_2^2
                        \Bigr)
\nonumber\\&&
			+\Bigl[
                                -52 (-3+z)
                                +8 (z-1) H_1 \big(
                                        13+10 H_1\big)
                                +(1+z) \big(
                                        96 H_{0,0,1}
                                        -64 H_{0,1,1}
                                \big)
\nonumber\\&&
                                -32 (-3+2 z) H_{0,1}
			\Bigr] H_0
			+\Bigl[
                                6 (3+19 z)
                                +80 (z-1) H_1
                                -32 (1+z) H_{0,1}
			\Bigr] H_0^2
                        -\frac{16}{3} z H_0^3
\nonumber\\&&
                        +8 (19+17 z) H_{0,1}
                        -16 (1+5 z) H_{0,0,1}
			+\Bigl[
                                -48 (1+5 z)
                                +(1+z) \big(
                                        -16 H_0^2
                                        +64 H_{0,1}
                                \big)
\nonumber\\&&
                                +16 (-1+3 z) H_0
                                -160 (z-1) H_1
			\Bigr] \zeta_2
			+\Bigl[
                                48 (-3+5 z)
                                -96 (1+z) H_0
			\Bigr] \zeta_3
                \Biggr]
\nonumber\\&&
                + L_M  \biggl[
                        (z-1) \Bigl(
                                312
                                -\frac{64}{3} H_0^3
                                +88 H_1
                                +16 H_1^2
                        \Bigr)
                        +(1+z) \Bigl(
                                \frac{4}{3} H_0^4
                                -160 H_{0,0,0,1}
                                +64 H_{0,0,1,1}
\nonumber\\&&
                                +\frac{288}{5} \zeta_2^2
                        \Bigr)
			+\Bigl[
                                -8 (-2+21 z)
                                +32 (z-1) H_1
                                -32 (-1+3 z) H_{0,1}
                                +64 (1+z) H_{0,0,1}
			\Bigr] H_0
\nonumber\\&&
                        +4 (7+13 z) H_0^2
                        -8 (-11+21 z) H_{0,1}
                        +32 (1+7 z) H_{0,0,1}
                        -32 (-1+3 z) H_{0,1,1}
\nonumber\\&&
			+\Bigl[
                                8 (-7+17 z)
                                -32 (3+z) H_0
                                -16 (1+z) H_0^2
			\Bigr] \zeta_2
			+\Bigl[
                                -64 (1+2 z)
                                +32 (1+z) H_0
			\Bigr] \zeta_3
                \biggr]
\nonumber\\&&
                +(1+z) \Bigl(
                        -\frac{2}{15} H_0^5
                        -80 H_{0,0,0,0,1}
                        +832 H_{0,0,0,1,1}
                        +256 H_{0,0,1,0,1}
                        -128 H_{0,0,1,1,1}
                        +80 \zeta_5
                \Bigr)
\nonumber\\&&
                +\Biggl[
                        -4 (263+195 z)
                        +(z-1) \biggl[
                                -32 H_1 \big(
                                        14+5 H_{0,1}\big)
                                -172 H_1^2
                                -\frac{80}{3} H_1^3
                        \biggr]
\nonumber\\&&
                        +(1+z) \big(
                                32 H_{0,1}^2
                                +96 H_{0,0,0,1}
                                -224 H_{0,0,1,1}
                                +64 H_{0,1,1,1}
                        \big)
                        +16 (1+10 z) H_{0,1}
\nonumber\\&&
                        -16 (-1+17 z) H_{0,0,1}
                        +160 (-2+3 z) H_{0,1,1}
                \Biggr] H_0
                +\biggl[
                        -6 (26+15 z)
                        -4 (z-1) H_1 \big(
                                43+10 H_1\big)
\nonumber\\&&
                        +(1+z) \big(
                                -48 H_{0,0,1}
                                +32 H_{0,1,1}
                        \big)
                        +80 z H_{0,1}
                \biggr] H_0^2
                +\biggl[
                        -2 (5+23 z)
                        -\frac{80}{3} (z-1) H_1
\nonumber\\&&
                        +\frac{32}{3} (1+z) H_{0,1}
                \biggr] H_0^3
		+\Bigl[
                        4 (-169+35 z)
                        -128 (1+z) H_{0,0,1}
		\Bigr] H_{0,1}
                -4 (109+33 z) H_{0,0,1}
\nonumber\\&&
                +8 (-49+13 z) H_{0,1,1}
                +32 (-8+15 z) H_{0,0,0,1}
                -144 (-1+3 z) H_{0,0,1,1}
                +\biggl[
                        4 (57+77 z)
\nonumber\\&&
                        +8 (z-1) H_1 \big(
                                43+10 H_1\big)
                        +(1+z) \Bigl(
                                \frac{8}{3} H_0^3
                                +96 H_{0,0,1}
                                -64 H_{0,1,1}
                                -96 \zeta_3
                        \Bigr)
			+\Bigl[
                                12 (5+13 z)
\nonumber\\&&
                                +160 (z-1) H_1
                                -64 (1+z) H_{0,1}
			\Bigr] H_0
                        +8 (3+z) H_0^2
                        -160 z H_{0,1}
                \biggr] \zeta_2
                +\biggl[
                        -\frac{8}{5} (-145+83 z)
\nonumber\\&&
                        +\frac{48}{5} (1+z) H_0
                \biggr] \zeta_2^2
		+\Bigl[
                        4 (207+7 z)
                        +(1+z) \big(
                                24 H_0^2
                                +128 H_{0,1}
                        \big)
                        -32 (-13+8 z) H_0
\nonumber\\&&
                        -320 (z-1) H_1
		\Bigr] \zeta_3
	\biggr\}
+A_{Qq}^{PS(3)}
+\tilde{C}^{PS(3)}(L_Q,N_F+1)
\Biggr\}.
\end{eqnarray}

Finally, $H_{}^{\sf S}(z)$ reads


%-----------------------------------------------------------------------------------------------------------------
\section{The transformation between the $\overline{\sf MS}$ and the Larin scheme in the non--singlet case}
\label{sec:C}
%-----------------------------------------------------------------------------------------------------------------

\vspace*{1mm}\noindent
In Refs.~\cite{Ablinger:2014vwa,Ablinger:2017err} the massive OME $A_{qq,Q}^{(3),\rm NS}$ has been given in the 
${\sf MS}$-scheme. This also applies to the 3--loop heavy flavor massive non--singlet Wilson coefficient in 
\cite{Behring:2015zaa},
related to the associated 3--loop massless Wilson coefficient \cite{Moch:2008fj}.
As we present here all OMEs in the Larin scheme, we also provide $A_{qq,Q}^{(3),\rm NS}$ in this scheme.
We first decompose $A_{qq,Q}^{(3),\rm NS}$ into its single and double mass pieces
%-----------------------------------------------------------------------------------------------------------------
\begin{eqnarray}
A_{qq,Q}^{(3),\rm NS} = A_{qq,Q}^{(3),\rm NS, s} + \tilde{A}_{qq,Q}^{(3),\rm NS}.
\end{eqnarray}
%-----------------------------------------------------------------------------------------------------------------
The renormalized single mass OME \cite{Ablinger:2014vwa} is given by
%-----------------------------------------------------------------------------------------------------------------
\begin{eqnarray}
A_{qq,Q}^{(2),\rm NS,L} &=& A_{qq,Q}^{(2),\rm NS,M}
+ \frac{1}{2} \left[ 
 \hat{\gamma}_{qq}^{(1),\rm NS,L} 
- \hat{\gamma}_{qq}^{(1),\rm NS,M} \right] L 
  + a_{qq,Q}^{(2),\rm NS,L} -  a_{qq,Q}^{(2),\rm NS,M}, 
\\
A_{qq,Q}^{(3),\rm NS,L,s} &=& A_{qq,Q}^{(3),\rm NS,M,s}
+ \frac{1}{2} \Biggl[
\left(
  \gamma_{qq}^{(1), \rm NS,L}
- \gamma_{qq}^{(1), \rm NS,M} \right) \beta_{0,Q}  - 
\left(
  \hat{\gamma}_{qq}^{(1), \rm NS,L}
- \hat{\gamma}_{qq}^{(1), \rm NS,M} \right) 
\nonumber\\ &&
\times (\beta_0+\beta_{0,Q}) \Biggr] L^2
+ \frac{1}{2}\Biggl[
  \hat{\gamma}_{qq}^{(2),\rm NS,L} 
- \hat{\gamma}_{qq}^{(2),\rm NS,M} 
- 4\left(
  a_{qq,Q}^{(2),\rm NS,L}
- a_{qq,Q}^{(2),\rm NS,M}
\right) 
\nonumber \\ &&
\times (\beta_0 + \beta_{0,Q} )\Biggr] L
+ 4 \left(
        \overline{a}_{qq,Q}^{(2), \rm NS,L}
        - \overline{a}_{qq,Q}^{(2), \rm NS,M}
\right) 
- \frac{\beta_{0,Q} \zeta_2}{4} 
\left[
 \gamma_{qq}^{(1),\rm NS,L}
-\gamma_{qq}^{(1),\rm NS,M} 
\right]
\nonumber \\ &&
+ \delta m_1^{(0)}
\left[
 \hat{\gamma}_{qq}^{(1),\rm NS,L}
-\hat{\gamma}_{qq}^{(1),\rm NS,M} 
\right]
+ 2 \delta m_1^{(-1)} 
\left(
          a_{qq,Q}^{(2), \rm NS,L}
        - a_{qq,Q}^{(2), \rm NS,M}
\right)
\nonumber \\ &&
+ a_{qq,Q}^{(3),\rm NS,L} - a_{qq,Q}^{(3),\rm NS,M},
\nonumber\\
\end{eqnarray}
%-----------------------------------------------------------------------------------------------------------------
where the formula applies structurally in $N$- and $z$-space.
In $z$-space products have to be understood as convolutions.
The constant part of the unrenormalized OME $A_{qq,Q}^{(3),\rm NS,L}$, $a_{qq,Q}^{(3), \rm NS,L}$, is related to the 
corresponding expression in the ${\sf M}$-scheme by
%-----------------------------------------------------------------------------------------------------------------
\begin{eqnarray}
a_{qq,Q}^{(3),\rm NS,L} &=&  a_{qq,Q}^{(3),\rm NS,M} 
+ \textcolor{blue}{C_F} \Biggl\{
        \textcolor{blue}{T_F^2} \Biggl[
                -\frac{128 W_2}{81 N^3 (1+N)^3}
                -\textcolor{blue}{N_F} \Biggl(
                        \frac{128 W_3}{81 N^3 (1+N)^3}
                        +\frac{64}{9 N (1+N)} \zeta_2
                \Biggr)
\nonumber\\ && 
                -\frac{128}{9 N (1+N)} \zeta_2
        \Biggr]
        +\textcolor{blue}{C_A T_F} \Biggl[
                \frac{32 W_1}{9 N^3 (1+N)^3} S_1
                -\frac{16 W_4}{81 N^4 (1+N)^4} 
                +\frac{64}{3 N (1+N)} S_3
\nonumber\\ &&                 
-\frac{128 \big(
                        -3+4 N+10 N^2\big)}{27 N^2 (1+N)^2} S_{-2}
                +\frac{128}{9 N (1+N)} S_{-3}
                +\frac{256}{9 N (1+N)} S_{-2,1}
\nonumber\\ &&                
 +\frac{176 \zeta_2}{9 N (1+N)}
                -\frac{128 \zeta_3}{3 N (1+N)}
        \Biggr]
\Biggr\}
+\textcolor{blue}{C_F^2 T_F} \Biggl[
        -\frac{128 \big(
                12+17 N-14 N^3+3 N^4\big)}{27 N^3 (1+N)^3} S_1
\nonumber\\ &&
        \frac{32 W_5}{27 N^4 (1+N)^4}
        +\frac{256 \big(
                -3-N+5 N^2\big)}{27 N^2 (1+N)^2} S_2
        +\frac{256 \big(
                -3+4 N+10 N^2\big)}{27 N^2 (1+N)^2} S_{-2}
\nonumber\\ && 
        -\frac{512}{9 N (1+N)} S_3
        -\frac{256}{9 N (1+N)} S_{-3}
        -\frac{512}{9 N (1+N)} S_{-2,1}
        +\frac{128}{3 N (1+N)} \zeta_3
\Biggr],
\nonumber\\\
\end{eqnarray} 
%----------------------------------------------------------------------------------------------------------------- 
with the polynomials
%----------------------------------------------------------------------------------------------------------------- 
\begin{eqnarray} 
W_1 &=& 3 N^4+6 N^3+5 N^2+2 N+2,
\\
W_2 &=& 29 N^4+38 N^3+17 N^2+8 N+6,
\\
W_3 &=& 30 N^4+50 N^3+24 N^2+4 N+3,
\\
W_4 &=& 13 N^6-113 N^5-107 N^4-255 N^3-136 N^2-294 N-144,
\\
W_5 &=& 115 N^6+198 N^5+199 N^4+28 N^3+72 N^2+16 N-12.
\end{eqnarray} 
%----------------------------------------------------------------------------------------------------------------- 
Here the expansion coefficients of the QCD-$\beta$ function \cite{BETA,Bierenbaum:2009zt} are 
%----------------------------------------------------------------------------------------------------------------- 
\begin{eqnarray} 
\beta_0 &=& \frac{11}{3} \textcolor{blue}{C_A} - \frac{4}{3} \textcolor{blue}{T_F N_F}
\\ 
\beta_1 &=& \frac{34}{3} \textcolor{blue}{C_A^2} - 4 \left(\frac{5}{3} \textcolor{blue}{C_A} + \textcolor{blue}{C_F}\right)
\textcolor{blue}{T_F N_F}
\end{eqnarray} 
%----------------------------------------------------------------------------------------------------------------- 
and the expansion parameters of the unrenormalized heavy quark mass \cite{MASS,Bierenbaum:2009zt} are 
%----------------------------------------------------------------------------------------------------------------- 
\begin{eqnarray} 
\delta m_1^{(-1)} &=& 6 \textcolor{blue}{C_F} 
\\
\delta m_1^{(0)}  &=& -4 \textcolor{blue}{C_F}.
\end{eqnarray}
%----------------------------------------------------------------------------------------------------------------- 
Furthermore, the transformation relation of the asymptotic massive Wilson coefficient is
%-----------------------------------------------------------------------------------------------------------------
\begin{eqnarray}
L_{qq,Q}^{(2),\rm NS,L} &=& L_{qq,Q}^{(2),\rm NS,M} 
  + A_{qq,Q}^{(2),\rm NS,L} -  A_{qq,Q}^{(2),\rm NS,M}
+ \hat{C}_{q}^{(2),\rm NS,L} - \hat{C}_{q}^{(2),\rm NS,M} 
\\
L_{qq,Q}^{(3),\rm NS,L} &=& L_{qq,Q}^{(3),\rm NS,M} 
  + A_{qq,Q}^{(3),\rm NS,L} -  A_{qq,Q}^{(3),\rm NS,M}
  + A_{qq,Q}^{(2),\rm NS,L} C_{q}^{(1),\rm NS,L} 
  - A_{qq,Q}^{(2),\rm NS,M} C_{q}^{(1),\rm NS,M} 
\nonumber\\ &&
  + C_{q}^{(3),\rm NS,L} - C_{q}^{(3),\rm NS,M} ,
\end{eqnarray}
%-----------------------------------------------------------------------------------------------------------------
which structurally again applies for Mellin $N$- 
and $z$-space. Here the following functions contribute
%-----------------------------------------------------------------------------------------------------------------
\begin{eqnarray}
{C}_{q}^{(1),\rm NS,L}    &=&  {C}_{q}^{(1),\rm NS,M}  - \textcolor{blue}{C_F} \frac{8}{N(N+1)}
\\
\hat{C}_{q}^{(2),\rm NS,L}    &=&  \hat{C}_{q}^{(2),\rm NS,M} 
- \textcolor{blue}{C_F T_F} 
        \frac{16 \big(
                3+N-5 N^2\big)}{9 N^2 (1+N)^2}
\\
\gamma_{qq}^{(1), \rm NS, L}  &=& \gamma_{qq}^{(1), \rm NS, M} - 2 \beta_0 z_{qq}^{(1)}
\\
a_{qq,Q}^{(2),\rm NS,L}       &=&   a_{qq,Q}^{(2),\rm NS,M} - 16 \textcolor{blue}{C_F T_F} \frac{5 N^2 - N -3}{9 N^2 (N+1)^2}
\\
\bar{a}_{qq,Q}^{(2),\rm NS,L} &=&   \bar{a}_{qq,Q}^{(2),\rm NS,M} - \textcolor{blue}{C_F T_F} \Biggl[
\frac{8(9+12N+10N^2+26N^3+28N^4)}{27 N^3 (N+1)^3} 
\nonumber\\ &&
+ \frac{8 \zeta_2}{3 N (N+1)}\Biggr]
\\
{\gamma}_{qq}^{(2), \rm NS, L}  &=& {\gamma}_{qq}^{(2), \rm NS, M} + 2 \beta_0\left[\left(z_{qq}^{(1)}\right)^2 - 2 
z_{qq}^{(2), \rm NS}\right] - 2 \beta_1 z_{qq}^{(1)}. 
\end{eqnarray}
%-----------------------------------------------------------------------------------------------------------------
The functions $z_{ij}^{(k)}$ are given in Refs.~\cite{Matiounine:1998re,Moch:2014sna,Behring:2019tus}.
%-----------------------------------------------------------------------------------------------------------------
\begin{eqnarray}
        z_{qq}^{(1)} &=&
- \frac{8
\textcolor{blue}{C_F}}{N(N+1)},
\\
%-----
z_{qq}^{(2),\text{NS}}
&=&
\textcolor{blue}{C_F T_F N_F}
                \frac{16 \big(-3-N+5 N^2\big)}{9 N^2 (1+N)^2}
+ \textcolor{blue}{C_A C_F} \Biggl\{
        -\frac{4 W_6}{9 N^3 (1+N)^3}
        -\frac{16}{N (1+N)} S_{-2}
\Biggr\}
\nonumber \\ &&
+
\textcolor{blue}{C_F^2}
\Biggl\{
          \frac{8\big(2+5 N+8 N^2+N^3+2 N^4\big)}{N^3 (1+N)^3}
        + \frac{16(1+2 N)}{N^2 (1+N)^2} S_1
        + \frac{16}{N (1+N)} S_2
\nonumber \\ &&
        + \frac{32}{N (1+N)} S_{-2}
\Biggr\},
\end{eqnarray}
%-----------------------------------------------------------------------------------------------------------------
with
%-----------------------------------------------------------------------------------------------------------------
\begin{eqnarray}
W_6 = 103 N^4 + 140 N^3 + 58 N^2 + 21 N + 36.
\end{eqnarray}
%-----------------------------------------------------------------------------------------------------------------

The corresponding expressions in $z$ space read
%-----------------------------------------------------------------------------------------------------------------
\begin{eqnarray}
a_{qq,Q}^{(3),\rm NS,L}(z) &=& a_{qq,Q}^{(3),\rm NS,M}(z) + 
\textcolor{blue}{C_F} \Biggl\{
        \textcolor{blue}{T_F^2} \Biggl[
                \textcolor{blue}{N_F} \Biggl(
                        (1-z) \Biggl(
                                -
                                \frac{1280}{27}-\frac{640}{81} H_0-\frac{64}{27} H_0(z)^2\big)
\nonumber\\ && 
                        -\frac{64 (1-z)}{9} \zeta_2
                \Biggr)
                + (1-z) \Biggl(
                        -\frac{3712}{81}-\frac{1280}{81} H_0 - \frac{128}{27} H_0^2\Biggr)
                -\frac{128}{9} (1-z) \zeta_2
        \Biggr]
\nonumber\\ &&
        +\textcolor{blue}{C_A T_F} \Biggl[
                (1-z) \Biggl(
                        -\frac{3664}{81}
                        +\frac{352}{9} H_1
                        +\frac{32}{3} H_0^2 H_1
                        -\frac{64}{3} H_0 H_{0,1}
                \Biggr)
\nonumber\\ && 
                + (1+z) \Biggl(
                        \frac{1280}{27} H_{-1} H_0
                        +\frac{64}{9} H_{-1} H_0^2
                        -\frac{1280}{27} H_{0,-1}
                        -\frac{128}{9} H_{0,0,-1}
                        +\frac{256}{9} H_{0,1,-1}
\nonumber\\ &&              
         +\frac{256}{9} H_{0,-1,1}
                        -\frac{128}{9} H_{0,1} \big(
                                -1+2 H_{-1}\big)
                \Biggr)
                -\frac{32}{81} (200+7 z) H_0
                -\frac{16}{27} (47+53 z) H_0^2
\nonumber\\ &&                 
-\frac{128}{27} H_0^3
                +\frac{256}{9} H_{0,0,1}
                +\Biggl(
                        -\frac{16}{27} (-89+57 z)
                        -\frac{64}{9} (-1+3 z) H_0
\nonumber\\ &&               
         +\frac{256}{9} (1+z) H_{-1}
                \Biggr) \zeta_2
                +\frac{128}{3} (-2+z) \zeta_3
        \Biggr]
\Biggr\}
+\textcolor{blue}{C_F^2 T_F}  \Biggl[
        (1-z) \Biggl(
                \frac{4480}{27}
\nonumber\\ &&                
 -\frac{896}{9} H_1
                -\frac{1280}{27} H_0 H_1
                -\frac{256}{9} H_0^2 H_1
                +\frac{256}{9} H_0 H_{0,1}
        \Biggr)
        + (1+z) \Biggl(
                -\frac{2560}{27} H_{-1} H_0
\nonumber\\ &&                
 -\frac{128}{9} H_{-1} H_0^2
                +\frac{2560}{27} H_{0,-1}
                +\frac{256}{9} H_{0,0,-1}
                -\frac{512}{9} H_{0,1,-1}
                -\frac{512}{9} H_{0,-1,1}
\nonumber\\ &&               
 +\frac{128}{9} H_{0,1} \big(
                        -3+4 H_{-1}\big)
        \Biggr)
        -\frac{32}{27} (-112+3 z) H_0
        +\frac{256}{27} (4+7 z) H_0^2
\nonumber\\ &&        
+\frac{64}{27} (1+3 z) H_0^3
        -\frac{512}{9} H_{0,0,1}
        +\Biggl(
                \frac{256}{9} H_0
                + (1+z) \Biggl(
                        -\frac{128}{27}-\frac{512}{9} H_{-1}\Biggr)
        \Biggr) \zeta_2
\nonumber\\ &&
        -\frac{256}{9} (-4+z) \zeta_3
\Biggr]
\end{eqnarray}
%-----------------------------------------------------------------------------------------------------------------
and
%-----------------------------------------------------------------------------------------------------------------
\begin{eqnarray}
{C}_{q}^{(1),\rm NS,L}(z)     &=&  {C}_{q}^{(2),\rm NS,M}(z) - 8 \textcolor{blue}{C_F} (1 - z),
\\
\hat{C}_{q}^{(2),\rm NS,L}    &=&  \hat{C}_{q}^{(2),\rm NS,M} + \frac{16}{9} \textcolor{blue}{C_F T_F} (1 - z) (5 + 3 H_0),
\\
\gamma_{qq}^{(1), \rm NS, L}  &=& \gamma_{qq}^{(1), \rm NS, M}(z) + 16 \beta_0 \textcolor{blue}{C_F} (1 - z),
\\
a_{qq,Q}^{(2),\rm NS,L}       &=&   a_{qq,Q}^{(2),\rm NS,M} - \frac{16}{9} \textcolor{blue}{C_F T_F} (1 - z) (5 + 3 H_0),
\\
\bar{a}_{qq,Q}^{(2),\rm NS,L} &=&   \bar{a}_{qq,Q}^{(2),\rm NS,M} - (1-z) \textcolor{blue}{C_F T_F} 
                \left( 
                        \frac{224}{27} 
                        + \frac{8}{3} \zeta_2
                        + \frac{40}{9} H_0
                        + \frac{4}{3} H_0^2
                \right),
\\
\hat{\gamma}_{qq}^{(2), \rm NS, L}  &=& \hat{\gamma}_{qq}^{(2), \rm NS, M} 
+
        \textcolor{blue}{C_A C_F T_F}
        \biggl[
                -\frac{19904}{27}
                -\biggl(
                         \frac{256}{9} (10-z)
                        -\frac{256}{3} (1+z) H_{-1}
                \biggr) H_0
                \nonumber \\ &&
                -\frac{128}{3} H_0^2
                -\frac{256}{3} (1+z) H_{0,-1}
                +\frac{256}{3} \zeta_2
        \biggr]
        + \textcolor{blue}{C_F^2 T_F}
        \biggl[
                \frac{1472}{3}
                +\biggl(
                         \frac{128}{3} (5+6 z)
                         \nonumber \\ &&
                        -\frac{512}{3} (1+z) H_{-1}
                \biggr) H_0
                +\frac{128}{3} (1+z) H_0^2
                -\frac{256}{3} (1-z) H_0 H_1
                +\frac{512}{3} (1+z) H_{0,-1}
                \nonumber \\ &&
                -\frac{256}{3} (1+z) \zeta_2
        \biggr]
        + \textcolor{blue}{C_F T_F^2 ( 1 - 2 N_F)} (1-z)
        \biggl[
                \frac{1280}{27}
                +\frac{256}{9} H_0
        \biggr].
\end{eqnarray}
%-----------------------------------------------------------------------------------------------------------------

For the double mass contributions \cite{Ablinger:2017err} we obtain
%-----------------------------------------------------------------------------------------------------------------
\begin{eqnarray}
\tilde{A}_{qq,Q}^{(3),\rm NS,L} &=& \tilde{A}_{qq,Q}^{(3),\rm NS,M} + 
        \frac{1}{2} \beta_{0,Q} \left( \hat{\gamma}_{qq}^{(1), \rm NS, L} - \hat{\gamma}_{qq}^{(1), \rm NS, M} \right) \left( L_1^2 + L_2^2 \right)
        \nonumber \\ &&
        + 4 \bigl( a_{qq,Q}^{(2),\rm NS,L} 
        - a_{qq,Q}^{(2),\rm NS,M} \bigr) \left( L_1 + L_2 \right)
        + 8 \beta_{0,Q} \left( \bar{a}_{qq,Q}^{(2),\rm NS,L} - \bar{a}_{qq,Q}^{(2),\rm NS,M} \right)
        \nonumber \\ &&
        +\tilde{a}_{qq,Q}^{(3),\rm NS,L} - \tilde{a}_{qq,Q}^{(3),\rm NS,M},
\end{eqnarray}
%-----
with
%-----
\begin{eqnarray}
\tilde{a}_{qq,Q}^{(3),\rm NS,L}(N) &=& \tilde{a}_{qq,Q}^{(3),\rm NS,M}(N) + 
        C_F T_F^2 
        \Biggl\{
                -\frac{256}{9N(N+1)}\left( L_1^2 + L_1 L_2 + L_2^2 + \zeta_2 \right)
                \nonumber \\ && \hspace{-0.9cm}
                +\frac{256(3+N+5N^2)}{27N^2(N+1)^2}\left( L_1 + L_2 \right)
                -\frac{256(6+8N+17N^2+38N^3+29N^4)}{81N^3(N+1)^3}
        \Biggr\},
\\ 
        \tilde{a}_{qq,Q}^{(3),\rm NS,L}(z) &=& \tilde{a}_{qq,Q}^{(3),\rm NS,M}(z) + 
        C_F T_F^2 (1-z)
        \Biggl\{
                -\frac{256}{9}\left( L_1^2 + L_1 L_2 + L_2^2 + \zeta_2 \right)
                \nonumber \\ && \hspace{-0.9cm}
                -\frac{256}{27}(5+3H_0)\left( L_1 + L_2 \right)
                - \frac{256}{27} H_0^2
                - \frac{2560}{81} H_0
                - \frac{7424}{81}
        \Biggr\}.
\end{eqnarray}
%-----------------------------------------------------------------------------------------------------------------

\vspace*{4mm} \noindent
{\bf Acknowledgment.} We would like thank A.~Behring for discussions. This project has received funding from the 
European Union’s Horizon 2020 research and innovation programme under the Marie Sk\l{}odowska–Curie grant agreement 
No. 764850, SAGEX and from the Austrian Science Fund (FWF) grant SFB F50 (F5009-N15). The larger formulae have been 
typesetted using {\tt SigmaToTeX} \cite{SIG1,SIG2}.

%\newpage
%%%%%%%%%%%%%%%%%%%%%%%%%%%%%%%%%%%%%%%%%%%%%%%%%%%%%%%%%%%%%%%%%%%%%%%%%%%%%%%%
%
%       Bibliography
%
%%%%%%%%%%%%%%%%%%%%%%%%%%%%%%%%%%%%%%%%%%%%%%%%%%%%%%%%%%%%%%%%%%%%%%%%%%%%%%%%

%%%%%%%%%%%%%%%%%%%%%%%%%%%%%%%%%%%%%%%%%%%%%%%%%%%%%%%%%%%%%%%%%%%%%%%%%%%%%%%%

%%%%%%%%%%%%%%%%%%%%%%%%%%%%%%%%%%%%%%%%%%%%%%%%%%%%%%%%%%%%%%%%%%%%%%%%%%%%%%%
\end{document}